\documentclass[12pt, openany]{book}

\usepackage[a4paper, margin=1in]{geometry} 
\usepackage{parskip} 
\usepackage{amsmath, amssymb} 
\usepackage{enumitem} 
\usepackage{graphicx} 
\usepackage{hyperref}
\hypersetup{
  colorlinks=true,
  linkcolor=blue!50!black,
  citecolor=blue!50!black,
  urlcolor=blue!50!black
}
\usepackage{fancyhdr} 
\usepackage{titlesec} 
\usepackage{amsthm} 
\usepackage{xparse}
\usepackage{stmaryrd}
\usepackage[most]{tcolorbox}
\usepackage{nicematrix}
\usepackage{siunitx}
\usepackage{xspace}
\usepackage{cite}
\usepackage{upgreek}
\usepackage{subcaption}

\usepackage{mathrsfs}

\def\0{\phantom{0}} \def\+{\phantom{+}}

\setlist[itemize]{left=10pt,label=--}

\usepackage{footnote}
\usepackage{etoolbox}
\usepackage{tikz}
\usepackage{listofitems} 
\usetikzlibrary{arrows.meta} 
\usepackage[outline]{contour} 
\usepackage{pgfplots}
\pgfplotsset{compat=1.17}
\contourlength{1.4pt}

\pgfmathdeclarefunction{relu}{1}{%
  \pgfmathparse{max(0,#1)}%
}

\pgfmathdeclarefunction{sigmoid}{1}{%
  \pgfmathparse{1 / (1 + exp(-#1))}%
}

\tikzset{>=latex} 
\usepackage{xcolor}
\colorlet{myred}{red!80!black}
\colorlet{myblue}{blue!80!black}
\colorlet{mygreen}{green!60!black}
\colorlet{myorange}{orange!70!red!60!black}
\colorlet{mydarkred}{red!30!black}
\colorlet{mydarkblue}{blue!40!black}
\colorlet{mydarkgreen}{green!30!black}
\tikzstyle{node}=[thick,circle,draw=myblue,minimum size=22,inner sep=0.5,outer sep=0.6]
\tikzstyle{node in}=[node,green!20!black,draw=mygreen!30!black,fill=mygreen!25]
\tikzstyle{node hidden}=[node,blue!20!black,draw=myblue!30!black,fill=myblue!20]
\tikzstyle{node convol}=[node,orange!20!black,draw=myorange!30!black,fill=myorange!20]
\tikzstyle{node out}=[node,red!20!black,draw=myred!30!black,fill=myred!20]
\tikzstyle{connect}=[thick,mydarkblue]
\tikzstyle{connect arrow}=[-{Latex[length=4,width=3.5]},thick,mydarkblue,shorten <=0.5,shorten >=1]
\tikzset{
  node 1/.style={node in},
  node 2/.style={node hidden},
  node 3/.style={node out},
}
\def\nstyle{int(\lay<\Nnodlen?min(2,\lay):3)} 

\makeatletter
\DeclareFontFamily{OMX}{MnSymbolE}{}
\DeclareSymbolFont{MnLargeSymbols}{OMX}{MnSymbolE}{m}{n}
\SetSymbolFont{MnLargeSymbols}{bold}{OMX}{MnSymbolE}{b}{n}
\DeclareFontShape{OMX}{MnSymbolE}{m}{n}{
    <-6>  MnSymbolE5
   <6-7>  MnSymbolE6
   <7-8>  MnSymbolE7
   <8-9>  MnSymbolE8
   <9-10> MnSymbolE9
  <10-12> MnSymbolE10
  <12->   MnSymbolE12
}{}
\DeclareFontShape{OMX}{MnSymbolE}{b}{n}{
    <-6>  MnSymbolE-Bold5
   <6-7>  MnSymbolE-Bold6
   <7-8>  MnSymbolE-Bold7
   <8-9>  MnSymbolE-Bold8
   <9-10> MnSymbolE-Bold9
  <10-12> MnSymbolE-Bold10
  <12->   MnSymbolE-Bold12
}{}

\let\llangle\@undefined
\let\rrangle\@undefined
\DeclareMathDelimiter{\llangle}{\mathopen}%
                     {MnLargeSymbols}{'164}{MnLargeSymbols}{'164}
\DeclareMathDelimiter{\rrangle}{\mathclose}%
                     {MnLargeSymbols}{'171}{MnLargeSymbols}{'171}
\makeatother

\titleformat{\section}{\normalfont\Large\bfseries}{\thesection}{1em}{}
\titleformat{\subsection}{\normalfont\large\bfseries}{\thesubsection}{1em}{}

\newlist{boxitemize}{itemize}{1}
\setlist[boxitemize,1]{label=--,topsep=7pt,parsep=7pt,itemsep=0pt,left=10pt}

\newlist{boxexitemize}{enumerate}{1}
\setlist[boxexitemize,1]{label=(\alph*),topsep=7pt,parsep=7pt,itemsep=0pt,left=10pt}

\newlist{boxdescription}{description}{1}
\setlist[boxdescription,1]{topsep=7pt,parsep=7pt,itemsep=0pt,left=10pt}

\newlist{exitemize}{enumerate}{1}
\setlist[exitemize,1]{label=(\alph*)}

\definecolor{thmbackcolor}{HTML}{E9E9F3}
\definecolor{thmframecolor}{HTML}{CBC9D7}

\definecolor{defbackcolor}{HTML}{E9E9F3}
\definecolor{defframecolor}{HTML}{CBC9D7}

\definecolor{exmpbackcolor}{HTML}{ECF6ED}
\definecolor{exmpframecolor}{HTML}{D6E0DA}

\definecolor{exbackcolor}{HTML}{F9E5E5}
\definecolor{exframecolor}{HTML}{E6D6D6}

\definecolor{solbackcolor}{HTML}{F1F0EB}
\definecolor{solframecolor}{HTML}{E1E0DD}

\newtcolorbox[auto counter, number within=chapter, number freestyle={\noexpand\thechapter.\noexpand\arabic{\tcbcounter}}]{mytheorem}[2][]{%
  enhanced,
  colback=defbackcolor,
  colframe=defframecolor,
  coltitle = black,
  arc=2pt,
  fonttitle=\bfseries,
  breakable,before skip=15pt,after skip=15pt,
  title=Theorem~\thetcbcounter: #2, #1
}

\newtcolorbox[auto counter, number within=chapter, number freestyle={\noexpand\thechapter.\noexpand\arabic{\tcbcounter}}]{myproposition}[2][]{
  enhanced,
  colback=defbackcolor,
  colframe=defframecolor,
  coltitle = black,
  arc=2pt,
  fonttitle=\bfseries,
  breakable,before skip=15pt,after skip=15pt,
  title=Proposition~\thetcbcounter: #2, #1
}

\newtcolorbox[auto counter, number within=chapter, number freestyle={\noexpand\thechapter.\noexpand\arabic{\tcbcounter}}]{mylemma}[2][]{%
  enhanced,
  colback=defbackcolor,
  colframe=defframecolor,
  coltitle = black,
  arc=2pt,
  fonttitle=\bfseries,
  breakable,before skip=15pt,after skip=15pt,
  title=Lemma~\thetcbcounter: #2, #1
}

\newtcolorbox[auto counter, number within=chapter, number freestyle={\noexpand\thechapter.\noexpand\arabic{\tcbcounter}}]{mydefinition}[2][]{%
  enhanced,
  colback=defbackcolor,
  colframe=defframecolor,
  coltitle = black,
  arc=2pt,
  fonttitle=\bfseries,
  breakable,before skip=15pt,after skip=15pt,
  title=Definition~\thetcbcounter: #2, #1
}

\newtcolorbox[auto counter, number within=chapter, number freestyle={\noexpand\thechapter.\noexpand\arabic{\tcbcounter}}]{myexercise}[2][]{%
  enhanced,
  colback=exbackcolor,
  colframe=exframecolor,
  coltitle = black,
  arc=2pt,
  fonttitle=\bfseries,
    breakable,before skip=15pt,after skip=15pt,
  title=Exercise~\thetcbcounter: #2, #1
}

\newtcolorbox[auto counter, number within=chapter, number freestyle={\noexpand\thechapter.
\noexpand\arabic{\tcbcounter}}]{mysolution}[2][]{%
  enhanced,
  colback=solbackcolor,
  colframe=solframecolor,
  coltitle = black,
  arc=2pt,
  fonttitle=\bfseries,
    breakable,before skip=15pt,after skip=15pt,
  title=Solution:
}

\newtcolorbox[auto counter, number within=chapter, number freestyle={\noexpand\thechapter.\noexpand\arabic{\tcbcounter}}]{myremark}[2][]{%
  enhanced,
  colback=solbackcolor,
  colframe=solframecolor,
  coltitle = black,
  arc=2pt,
  fonttitle=\bfseries,
    breakable,before skip=15pt,after skip=15pt,
  title=Remark~\thetcbcounter: #2, #1
}

\newtcolorbox[auto counter, number within=chapter, number freestyle={\noexpand\thechapter.\noexpand\arabic{\tcbcounter}}]{myexample}[2][]{%
  enhanced,
  colback=exmpbackcolor,
  colframe=exmpframecolor,
  coltitle = black,
  arc=2pt,
  fonttitle=\bfseries,
    breakable,before skip=15pt,after skip=15pt,
  title=Example~\thetcbcounter: #2, #1
}

\title{Verification of Neural Networks\\[3ex]\large Lecture Notes}
\author{Benedikt Bollig\\[3ex]
Université Paris-Saclay, CNRS, ENS Paris-Saclay, LMF\\Gif-sur-Yvette, France}

\date{}

\begin{document}
\maketitle



\newcommand{\Perm}[1]{S_{#1}}
\newcommand{\perm}{\pi}
\renewcommand{\epsilon}{\varepsilon}
\renewcommand{\phi}{\varphi}

\newcommand{\Reals}{\mathbb{R}}
\newcommand{\Rationals}{\mathbb{Q}}
\newcommand{\Naturals}{\mathbb{N}}
\newcommand{\posNaturals}{\mathbb{N}_+}
\newcommand{\Integers}{\mathbb{Z}}

\newcommand{\df}{=}
\newcommand{\fcomp}{\circ}

\newcommand{\REF}{\text{REF}\xspace}

\newcommand{\LRA}{\text{LRA}\xspace}
\newcommand{\NNSL}{\text{NNL}\xspace}
\newcommand{\existsNNL}{{\exists}\text{NNL}\xspace}
\newcommand{\forallNNL}{{\forall}\text{NNL}\xspace}
\newcommand{\existsLRA}{{\exists}\text{LRA}\xspace}

\newcommand{\sfin}{\mathsf{in}}
\newcommand{\sfout}{\mathsf{out}}

\newcommand{\varx}{x}
\newcommand{\vary}{y}
\newcommand{\varz}{z}
\newcommand{\vectvarx}{\vect{x}}
\newcommand{\vectvary}{\vect{y}}
\newcommand{\vectvarz}{\vect{z}}

\newcommand{\twosquares}{\mathrel{\scalebox{0.7}{$\Box\!\Box$}}}
\newcommand{\twosquaresbelow}{\mathrel{\rotatebox[origin=c]{90}{\scalebox{0.7}{$\Box\!\Box$}}}}

\newcommand{\vectconc}[2]{{#1}\ensuremath{\twosquaresbelow}{#2}}
\newcommand{\matrixconc}[2]{{#1}\ensuremath{\twosquares}{#2}}

\newcommand{\vertconc}[2]{{#1}\ensuremath{\twosquaresbelow}{#2}}
\newcommand{\vertconcwo}{\ensuremath{\twosquaresbelow}}
\newcommand{\horizconc}[2]{{#1}\ensuremath{\twosquares}{#2}}

\newcommand{\heaviside}{\textup{heaviside}}

\newcommand{\realx}{x}
\newcommand{\realy}{y}
\newcommand{\realz}{z}

\newcommand{\nrealx}{r}
\newcommand{\nrealy}{s}
\newcommand{\nrealz}{p}


\newcommand{\States}{Q}
\newcommand{\Trans}{\Delta}
\newcommand{\Final}{F}
\newcommand{\init}{\iota}

\newcommand{\BA}{\mathcal{A}}
\newcommand{\comma}{\mathord{\bullet}}
\newcommand{\commaord}{\,\mathord{\bullet}\,}
\newcommand{\binsign}{s}
\newcommand{\binint}{u}
\newcommand{\binfract}{v}
\newcommand{\dec}[1]{\mathit{dec}(#1)}
\newcommand{\sign}[1]{\mathit{sign}(#1)}
\newcommand{\WF}[1]{\textup{WF}^{#1}}
\newcommand{\wfBA}[1]{\BA^{#1}_{\textup{wf}}}
\newcommand{\tsize}{k}
\newcommand{\wtuple}{\vect{w}}

\newcommand{\lastsymbol}[1]{\mathit{last}(#1)}

\newcommand{\BAconst}[3]{\BA^{#1}_{#2=\mathsf{const}(#3)}}
\newcommand{\BAmult}[4]{\BA^{#1}_{#2=\mathsf{mult}(#3,#4)}}
\newcommand{\BAadd}[4]{\BA^{#1}_{#2=\mathsf{add}(#3,#4)}}
\newcommand{\BAle}[3]{\BA^{#1}_{#2 \le #3}}
\newcommand{\BAeq}[3]{\BA^{#1}_{#2 = #3}}
\newcommand{\BAproj}[2]{\mathit{proj}_{\le #2}(#1)}
\newcommand{\BAcup}[2]{#1 \cup #2}
\newcommand{\BAcap}[2]{#1 \cap #2}
\newcommand{\BAcompl}[1]{\overline{#1}}
\newcommand{\Ind}{J}
\newcommand{\letter}{u}

\newcommand{\state}{q}
\newcommand{\statex}{p}
\newcommand{\statey}{q}

\newcommand{\runlabel}[1]{\mathit{label}(#1)}

\newcommand{\Formulas}{\mathcal{F}}
\newcommand{\FSet}{\mathfrak{A}}
\newcommand{\NNLrestr}[1]{\NNSL[#1]}
\newcommand{\Formrestr}[2]{#1[#2]}

\newcommand{\SAT}[1]{\mathsf{SAT}(#1)}
\newcommand{\UnivSAT}[1]{\mathsf{UnivSAT}(#1)}



\renewcommand{\implies}{\;\Rightarrow\;}
\renewcommand{\iff}{\;\Leftrightarrow\;}


\newcommand{\argmax}{\textup{argmax}}
\newcommand{\softmax}{\textup{softmax}}
\newcommand{\ReLU}{\textup{ReLU}}
\newcommand{\NLReLU}{\textup{NLReLU}}
\newcommand{\sigmoid}{\sigma}
\renewcommand{\ln}{\textup{ln}}
\newcommand{\idactivation}{\textup{id}}

\newcommand{\val}{\mathit{val}}
\newcommand{\frestr}[2]{#1\big|_{#2}}

\newcommand{\manhattan}[2]{\mathord{\parallel}#1,#2\mathord{\parallel}_\mathsf{Manhattan}}


\newcommand{\vect}[1]{{\bf #1}}
\newcommand{\vectone}[2]{#1_{#2}}
\newcommand{\weight}{a}
\newcommand{\wmatrix}[1]{{\bf A}}
\newcommand{\wmatrixone}[2]{\wmatrix{#1}[#2]}
\newcommand{\wmatrixtwo}[3]{\wmatrix{#1}[#2,#3]}
\newcommand{\bias}[1]{{\bf #1}}
\newcommand{\biasone}[2]{#1[#2]}


\newcommand{\NN}{\mathcal{N}}
\newcommand{\nlayers}{\ell}
\newcommand{\activation}{f}
\newcommand{\Layer}{\mathscr{L}}
\newcommand{\layer}{k}

\newcommand{\run}{\rho}

\newcommand{\indim}[1]{\mathit{in}(#1)}
\newcommand{\outdim}[1]{\mathit{out}(#1)}
\newcommand{\nndim}[1]{\mathit{dim}(#1)}

\newcommand{\lfunction}[1]{\llbracket #1 \rrbracket}
\newcommand{\nnfunction}[1]{\llbracket #1 \rrbracket}

\newcommand{\NNLplus}{\NNSL^\ast}
\newcommand{\NNLrelu}{\NNSL[\ReLU]}
\newcommand{\existsNNLrelu}{\existsNNL[\ReLU]}
\newcommand{\basicexistsNNLrelu}{\textup{REACH}}
\newcommand{\forallNNLrelu}{\forallNNL[\ReLU]}


\chapter*{Preface}

These lecture notes provide an introduction to the verification of neural networks from a theoretical perspective. We discuss feed-forward neural networks, recurrent neural networks, attention mechanisms, and transformers, together with specification languages and algorithmic verification techniques.

They are based on a lecture series given in the academic year 2023/2024 as
part of the MPRI 2.8 course ``Advanced Techniques of Verification''.
I thank the students of the course for their questions and comments during the lectures, which helped improve the presentation of the material.

The neural-network drawings presented in Sections 3.1 and 3.2 are adapted from Izaak Neutelings \cite{Neutelings2021}, available at tikz.net, and used under the CC BY-SA 4.0 license. All other figures are original unless stated otherwise.

Any comments, suggestions, errata, etc.\ are very welcome.
Please send them by email to \texttt{bollig@lmf.cnrs.fr}.

\tableofcontents
\newpage

\chapter{Introduction}

AI-based systems, particularly neural networks, are playing an important role in our daily lives. Neural networks are used for image and speech recognition, in autonomous cars, medical diagnostics, anomaly detection, financial and weather forecasting, etc. As they are increasingly used in safety-critical applications (e.g., in autonomous cars or medical diagnostics), there are of course high safety requirements for AI-based components. However, neural networks are black-box models with a rather opaque structure where small changes can have significant effects. Another reason to ask: Can we give formal guarantees for a given neural network? Or, in other words, can we \emph{verify} a neural network?

Formal methods are well-suited to provide answers here. They include techniques that first mathematically model systems and requirements specifications and then algorithmically determine the compatibility of system and specification. Formal methods are often associated with verifying systems written in some programming language. Now, neural networks are not a program in this sense (they are written or trained by a machine). However, like ``programs,'' they follow a precise sequence of instructions and are, in principle, amenable to formal verification.

While the specification of programs is often natural (think of terms like termination or deadlock freedom), it may appear unclear what correctness means for neural networks. For example, when is an image classifier that is supposed to classify animals correct? Probably, when it recognizes a dog as such, a cat as a cat, and so on. But to apply formal methods, we must formalize correctness in a precise mathematical sense. Now, to write down when a picture represents a dog and when a cat, is naturally incredibly challenging. And if we could, we probably would not need a neural network anymore. But there are many other desirable properties that we \emph{can} formalize. For example, when an image classifier is robust, that is, when small changes in a given image do not entirely change the classification. Or when a neural network is fair, that is, when it does not take sensitive features into account in order to assign credits or jobs.

Neural networks often serve as building blocks of larger systems and may act as a controller,
i.e., choose an action to be performed in
a given state. We are then in the setting of \emph{reactive systems}.
Verifying them is particularly challenging, as specifications usually combine
temporal properties with arithmetic expressions such as
``whenever a system variable $\realx$
reaches a critical threshold $\gamma$, i.e., $\realx \ge \gamma$,
then there is a time point in the near future when it falls back
below the threshold, i.e., $\realx < \gamma$.''

Applying formal methods to neural networks is an exciting new field with many interesting developments.
We refer here to various survey papers and lecture notes \cite{Leucker20,UrbanM2021,albarghouthi-book,ZhangXWH2022,BolligLN22} that we recommend for further reading.
We do not aim to provide optimal algorithms. The goal of this lecture is to give a sense of what verification of a neural network means and to explore its theoretical possibilities and limitations. However, one should remember that scalability is an essential criterion for verification methods for neural networks, which can have millions of parameters.

\chapter{Preliminaries}

In this chapter, we recall some standard concepts from linear algebra and automata theory.
Linear algebra allows one to describe neural networks in a compact, elegant manner.
We will use automata-based techniques (among others) to address their verification.

\section{Sets, Functions, Vectors, Matrices}

\paragraph{Sets and Functions.}

By $\Naturals = \{0,1,2,\ldots\}$, we denote the set of natural numbers,
and by $\posNaturals = \{1,2,\ldots\}$ the set of positive natural numbers.
The set of real numbers is denoted $\Reals$, and the set of rational numbers
by $\Rationals$.
As part of an input to a decision problem or
of an object like a matrix, we implicitly assume
that a rational number is effectively given
in terms of binary encodings of its numerator and denominator.
For $x, y \in \Reals$, we let $[x,y] = \{z \in \Reals \mid x \le z \le y\}$,
$(x,y] = \{z \in \Reals \mid x < z \le y\}$, and so forth.

For functions $f: A \to B$ and $g: B \to C$, we denote by $g \fcomp f: A \to C$ the composition of $f$ and $g$, defined by $(g \fcomp f)(a) = g(f(a))$ for all $a \in A$. Moreover, given $A' \subseteq A$, we let $\frestr{f}{A'}: A' \to B$ denote the \emph{restriction} of $f$ to the domain $A'$.

For a finite set $A$, the number of elements in $A$ is denoted by $|A|$.

\paragraph{Vectors and Matrices.}

Let $m,n \in \posNaturals$.
For a vector $\vect{x} \in \Reals^n$ and $i \in \{1,\ldots,n\}$, we let
$\vectone{\realx}{i}$ refer to the $i$-th component of $\vect{\realx}$, i.e., $\vect{\realx} = (\realx_1,\ldots,\realx_n)^\top$. If $n=1$, we may simply write $\realx$ for $\vect{\realx}$. Similarly, $\vect{\realy} = (\realy_1,\ldots,\realy_n)^\top$, $\vect{\realx}' = (\realx_1',\ldots,\realx_n')^\top$, and so on.
Moreover, for a matrix $\wmatrix{W} \in \Reals^{n \times m}$, $i \in \{1,\ldots,n\}$, and $j \in \{1,\ldots,m\}$,
we let $\weight_{i,j}$ refer to the element of $\wmatrix{W}$ in the $i$-th row and $j$-th column, i.e.,
\[
\wmatrix{W} = \begin{pmatrix}
    \weight_{1,1} & \weight_{1,2} & \ldots & \weight_{1,m} \\
    \weight_{2,1} & \weight_{2,2} & \ldots & \weight_{2,m} \\
    \vdots  & \vdots  & \ddots & \vdots  \\
    \weight_{n,1} & \weight_{n,2} & \ldots & \weight_{n,m}
\end{pmatrix} \in \Reals^{n \times m}\,.
\]
Similarly, $\wmatrix{W}' = \left(\weight_{i,j}'\right)_{i,j}$ and so forth.

For $\activation = (\activation_1,\ldots,\activation_m)$
with $\activation_1,\ldots,\activation_m: \Reals \rightarrow \Reals$ and $\vect{\realx} \in \Reals^m$, we define \[\activation(\vect{\realx}) = (\activation_1(\vectone{\realx}{1}), \ldots, \activation_m(\vectone{\realx}{m}))^\top \in \Reals^m\,.\]

The \emph{vertical concatenation}
of matrices $\vect{A} \in \Reals^{m \times n}$ and
$\vect{B} \in \Reals^{k \times n}$ is defined
by $\vertconc{\vect{A}}{\vect{B}} = \vect{C} \in \Reals^{(m+k) \times n}$
where $c_{i,j} = a_{i,j}$ for all $i \in \{1,\ldots,m\}$
and $j \in \{1,\ldots,n\}$, and
$c_{i,j} = b_{i-m,j}$ for all $i \in \{m+1,\ldots,m+k\}$
and $j \in \{1,\ldots,n\}$.
The \emph{horizontal concatenation}
of $\vect{A} \in \Reals^{m \times n}$ and
$\vect{B} \in \Reals^{m \times k}$ is defined accordingly
by $\horizconc{\vect{A}}{\vect{B}} = \vect{C} \in \Reals^{m \times (n+k)}$
where $c_{i,j} = a_{i,j}$ for all $i \in \{1,\ldots,m\}$
and $j \in \{1,\ldots,n\}$, and
$c_{i,j} = b_{i,j-n}$ for all $i \in \{1,\ldots,m\}$
and $j \in \{n+1,\ldots,n+k\}$.
The special case of vectors is defined analogously.
In particular,
the vertical concatenation of $\vect{x} \in \Reals^m$ and $\vect{y} \in \Reals^n$ is
$\vectconc{\vect{x}}{\vect{y}} = (x_1,\ldots,x_m,y_1,\ldots,y_n)^\top \in \Reals^{m+n}$.

For a given $m \in \posNaturals$ (which we suppose to be clear from the context), we let
\[
\argmax:
\begin{cases}
\Reals^m \to 2^{\{1,\ldots,m\}}\\
\vect{\realx} \mapsto \{i \in \{1,\ldots,m\} \mid \realx_i = \max(\vect{\realx})\} \,.
\end{cases}
\]

For $m \in \posNaturals$, the set of permutations $\perm: \{1,\ldots,m\} \to \{1,\ldots,m\}$ is denoted $\Perm{m}$. We extend $\perm$ to $\perm: \Reals^m \to \Reals^m$ letting $\perm(\vect{\realx}) = (\vectone{\realx}{\perm(1)},\ldots,\vectone{\realx}{\perm(m)})$.
This convention reads the coordinates of $\vect{\realx}$ in the order
$\perm(1),\ldots,\perm(m)$; using the inverse convention would give
equivalent definitions below after renaming the permutation.

Below we define two properties that play an important role in the realm of neural networks:

\begin{mydefinition}{Permutation Equivariance and Invariance}
Let $m \in \posNaturals$ and $A$ be a set.
\begin{boxitemize}
\item A function $f: \Reals^m \to A$ is called \emph{permutation invariant} if, for all $\vect{\realx} \in \Reals^m$ and $\perm \in \Perm{m}$, we have $f(\vect{\realx}) = f(\perm(\vect{\realx}))$.

\item A function $f: \Reals^m \to \Reals^m$ is called \emph{permutation equivariant} if, for all $\vect{\realx} \in \Reals^m$ and $\perm \in \Perm{m}$, we have $f(\perm(\vect{\realx})) = \perm(f(\vect{\realx}))$.
\end{boxitemize}
\end{mydefinition}

\newcommand{\extdelta}{\hat\delta}

\section{Languages and Büchi Automata}

Automata are a useful tool in verification and for deciding arithmetic theories such as Presburger arithmetic and linear real arithmetic \cite{Haase2020}. In this course, we will need to decide linear real arithmetic and, to do so, rely on Büchi automata, which are devices that run over infinite words (or strings). Later on, these infinite words will represent real numbers.

An \emph{alphabet} is a nonempty set (possibly infinite).
Let $\Sigma$ be an alphabet. A \emph{finite word} over $\Sigma$ of length $n \in \Naturals$ is a finite sequence $w = \letter_1\ldots \letter_n$ with $\letter_1,\ldots,\letter_n \in \Sigma$.
We denote the length $n$ of $w$ by $|w|$.
If $n=0$, then $w$ is the empty word, denoted by $\epsilon$.
An infinite word over $\Sigma$ is a countably infinite sequence $w=\letter_1\letter_2\letter_3\ldots$ with $\letter_1,\letter_2,\ldots \in \Sigma$.
The set of finite words over $\Sigma$ is denoted by $\Sigma^\ast$,
the set of nonempty finite words over $\Sigma$ by $\Sigma^+$ (i.e., $\Sigma^+ = \Sigma^\ast \setminus \{\epsilon\}$),
and the set of
infinite words by $\Sigma^\omega$.

We will often deal with mappings
of the form $\delta: Q \times \Sigma \to Q$ where $Q$ is a (possibly infinite) set.
This mapping can be inductively extended to
$\extdelta: Q \times \Sigma^\ast \to Q$ letting
$\extdelta(q, \epsilon) = q$
and, for $w \in \Sigma^\ast$ and $\letter \in \Sigma$, $\extdelta(q, w \cdot \letter) =
\delta(\extdelta(q, w), \letter)$.
Abusing notation, we usually still write $\delta$ instead
of $\extdelta$.

In the remainder of this section, we consider finite alphabets.

\begin{mydefinition}{Büchi Automaton}
Let $\Sigma$ be a finite alphabet.
A \emph{Büchi automaton} over $\Sigma$ is a tuple $\BA = (\States,\Trans,\init,\Final)$ where
\begin{boxitemize}
\item $\States$ is a finite set of \emph{states},
\item $\Trans \subseteq \States \times \Sigma \times \States$ is the set of \emph{transitions}
\item $\init \in \States$ is the \emph{initial state}, and
\item $\Final \subseteq \States$ is the set of \emph{final states}.
\end{boxitemize}
\end{mydefinition}

Büchi automaton $\BA$ recognizes a language $L(\BA) \subseteq \Sigma^\omega$ as follows.
A \emph{run} of $\BA$ is an infinite sequence $\run = \state_0 \letter_1 \state_1 \letter_2 \state_2 \ldots \in \States(\Sigma\States)^\omega$ such that $\state_0 = \init$ and, for all $i \in \posNaturals$,
$(\state_{i-1},\letter_i,\state_i) \in \Trans$. The \emph{label} of $\run$ is defined
as $\runlabel{\run} = \letter_1\letter_2\letter_3\ldots \in \Sigma^\omega$. Run $\run$ is
called \emph{accepting} if it sees some final state infinitely often, i.e.,
the set $\{i \in \Naturals \mid \state_i \in \Final\}$ is infinite. Finally,
the \emph{language recognized by} $\BA$ is defined by
\[L(\BA) = \{\runlabel{\run} \mid \run \textup{ is an accepting run of } \BA\} \subseteq \Sigma^\omega\,.\]

\begin{myexample}{}
The figure below depicts a Büchi automaton $\BA=(\States,\Trans,\init,\Final)$ over the alphabet $\Sigma=\{0,1\}$.
We have $\States = \{\state_0,\state_1\}$, $\init = \state_0$, and $\Final = \{\state_1\}$.
The set of transitions $\Trans$ includes $(\state_0,0,\state_0)$, $(\state_0,1,\state_1)$, etc.
The language $L(\BA)$ is the set of words from $\Sigma^\omega$ in which letter $1$
occurs infinitely often.\medskip
\begin{center}
\includegraphics[scale=0.5]{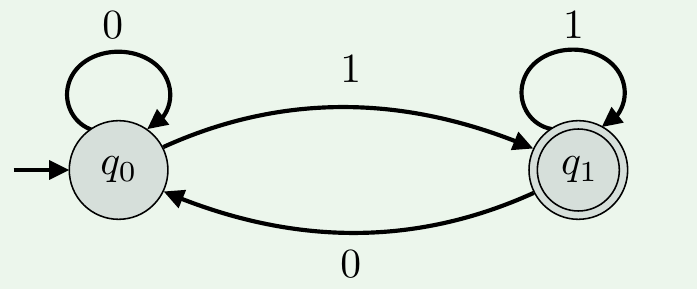}
\end{center}
\end{myexample}

Büchi automata enjoy several useful closure and decidability properties:

\begin{mytheorem}[label=thm:baclosure]{Closure Properties of Büchi Automata}
Let $\Sigma$ be a finite alphabet and let $\BA,\BA_1,\BA_2$ be Büchi automata over $\Sigma$.
We can effectively construct a Büchi automaton
\begin{boxexitemize}
\item $\BAcup{\BA_1}{\BA_2}$ over $\Sigma$ such that $L(\BAcup{\BA_1}{\BA_2}) = L(\BA_1) \cup L(\BA_2)$;

\item $\BAcap{\BA_1}{\BA_2}$ over $\Sigma$ such that $L(\BAcap{\BA_1}{\BA_2}) = L(\BA_1) \cap L(\BA_2)$;

\item $\BAcompl{\BA}$ over $\Sigma$ such that $L(\BAcompl{\BA}) = \Sigma^\omega \setminus L(\BA)$.
\end{boxexitemize}
\end{mytheorem}

\begin{mytheorem}[label=thm:BAemptiness]{Büchi Automata Emptiness}
The following problem is decidable:
\begin{boxdescription}
\item[Input:] A finite alphabet $\Sigma$ and a Büchi automaton over $\Sigma$.
\item[Question:] Is $L(\BA)$ nonempty?
\end{boxdescription}
The complexity is linear in the number of states and transitions.
\end{mytheorem}
For more background on languages and automata,
we refer the reader to \cite{Thomas97}.

\chapter{Feed-Forward Neural Networks}

In this chapter, we define feed-forward neural networks and their verification problem.
We also present a specification language for neural networks. It is based
on linear real arithmetic, which encompasses, as we will see, many pertinent properties
of neural networks.


\section{Definition of Neural Networks}

A neural network is a stack of layers. Every layer transforms an input vector into an output vector.
The more layers we have, the deeper is the network. In practice, the term \emph{deep learning} is used when several such transformations are composed. A layer and its computation are depicted in Figure~\ref{fig:layer}. It consists of $m$ input nodes and $n$ output nodes, and it maps an input vector $\vect{\realx} = (\realx_1,\ldots,\realx_m)^\top \in \Reals^m$ to an output vector $\vect{\realy} = (\realy_1,\ldots,\realy_n)^\top \in \Reals^n$. Here, intermediate values $\realz_i$ are computed as
a linear combination $b_i + \sum_j a_{i,j} \cdot \realx_j$.
Thus, the transformation induced by a layer can be written in terms of matrix multiplications. 

\begin{figure}[h]
\centering
\begin{tikzpicture}[x=2.7cm,y=1.6cm]
  \def\NI{5} 
  \def\NO{4} 
  \def\yshift{0.4} 
  
  \foreach \i [evaluate={\c=int(\i==\NI); \y=\NI/2-\i-\c*\yshift; \index=(\i<\NI?int(\i):"m");}]
              in {1,...,\NI}{ 
    \node[node in,outer sep=0.6] (NI-\i) at (0,\y) {$\realx_{\index}$};
  }
  
  \foreach \i [evaluate={\c=int(\i==\NO); \y=\NO/2-\i-\c*\yshift; \index=(\i<\NO?int(\i):"n");}]
    in {\NO,...,1}{ 
    \ifnum\i=1 
      \node[node hidden]
        (NO-\i) at (1,\y) {$\realz_{\index}$};
      \foreach \j [evaluate={\index=(\j<\NI?int(\j):"m");}] in {1,...,\NI}{ 
        \draw[connect,white,line width=1.2] (NI-\j) -- (NO-\i);
        \draw[connect] (NI-\j) -- (NO-\i)
          node[pos=0.50] {\contour{white}{$\weight_{1,\index}$}};
      }
    \else 
      \node[node,blue!20!black!80,draw=myblue!20,fill=myblue!5]
        (NO-\i) at (1,\y) {$\realz_{\index}$};
      \foreach \j in {1,...,\NI}{ 
        \draw[connect,myblue!20] (NI-\j) -- (NO-\i);
      }
    \fi
  }
  
  \path (NI-\NI) --++ (0,1+\yshift) node[midway,scale=1.2] {$\vdots$};
  \path (NO-\NO) --++ (0,1+\yshift) node[midway,scale=1.2] {$\vdots$};
  
  \def\agr#1{{\realx_{#1}}}
  \node[right,scale=0.9] at (0.32,1.7)
  	{$+b_1$};
  \node[right,scale=0.9] at (1.3,-0.8)
    {$\begin{aligned}
      {
      \begin{pmatrix}
        \realz_{1} \\[0.3em]
        \realz_{2} \\
        \vdots \\
        \realz_{n}
      \end{pmatrix}}
      &=
      \color{black}
      \begin{pmatrix}
        \weight_{1,1} & \weight_{1,2} & \ldots & \weight_{1,m} \\
        \weight_{2,1} & \weight_{2,2} & \ldots & \weight_{2,m} \\
        \vdots  & \vdots  & \ddots & \vdots  \\
        \weight_{n,1} & \weight_{n,2} & \ldots & \weight_{n,m}
      \end{pmatrix}
      {
      \begin{pmatrix}
        \realx_{1} \\[0.3em]
        \realx_{2} \\
        \vdots \\
        \realx_{m}
      \end{pmatrix}}
      +
      \begin{pmatrix}
        b_{1} \\[0.3em]
        b_{2} \\
        \vdots \\
        b_{n}
      \end{pmatrix}
      \\[0.5em]
      \vect{\realz} &= 
           \wmatrix{W} \cdot {\vect{\realx}}+\mathbf{b}
            \\[0.5em]
      \vect{\realy} &= 
           f(\vect{\realz})
    \end{aligned}$};
  
\end{tikzpicture}
\caption{A neural network layer with activation function $f: \Reals^n \to \Reals^n$\label{fig:layer}}
\end{figure}

\begin{mydefinition}{Layer}
Let $m,n \in \posNaturals$. A \emph{feed-forward layer} with input dimension $m$ and output dimension $n$ is a triple $\Layer=(\wmatrix{W}, \bias{b}, \activation)$ where $\wmatrix{W} \in \Rationals^{n \times m}$ is the \emph{weight matrix}, $\vect{b} \in \Rationals^n$ is the \emph{bias vector}, and $\activation: \Reals^n \rightarrow \Reals^n$ is the \emph{activation function}.
\end{mydefinition}

We let $\indim{\Layer} \df m$ denote the \emph{input dimension} of $\Layer$.
Moreover, $\outdim{\Layer} \df n$ is the \emph{output dimension}, which is usually referred to as the \emph{number of neurons} of $\Layer$.
In other words, the weights of the $i$-th output neuron are given by the $i$-th row of $\wmatrix{W}$.
We also let $\nndim{\Layer} = (m,n)$ (note the difference from the dimension of the matrix $\wmatrix{W}$).
Layer $\Layer$ defines the function
\[
\lfunction{\Layer}: 
\begin{cases}
\Reals^{\indim{\Layer}} \to \Reals^{\outdim{\Layer}}\\
\vect{\realx} \mapsto \activation (\wmatrix{W} \cdot \vect{\realx} + \vect{b})\,.
\end{cases}
\]

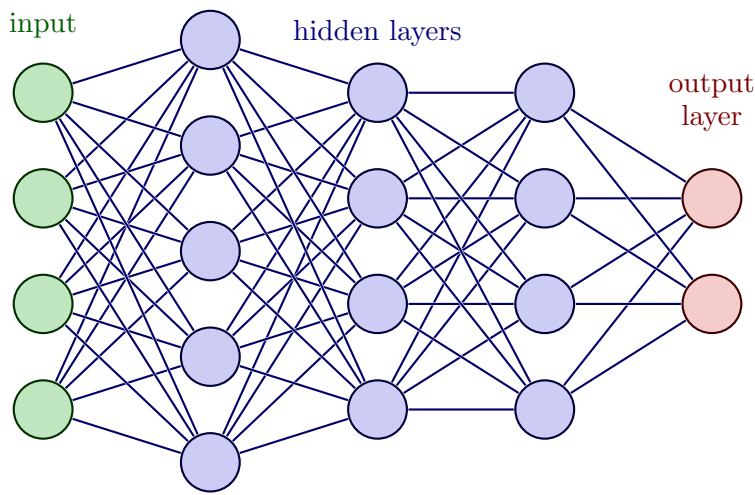
\begin{figure}
\centering
\begin{tikzpicture}[x=2.2cm,y=1.4cm]
  \readlist\Nnod{4,5,4,4,2} 
  
  \foreachitem \N \in \Nnod{ 
    \def\lay{\Ncnt} 
    \pgfmathsetmacro\prev{int(\Ncnt-1)} 
    \foreach \i [evaluate={\y=\N/2-\i; \x=\lay; \n=\nstyle;}] in {1,...,\N}{ 
      
      \node[node \n] (N\lay-\i) at (\x,\y) {};
      
      \ifnum\lay>1 
        \foreach \j in {1,...,\Nnod[\prev]}{ 
          \draw[connect,white,line width=1.2] (N\prev-\j) -- (N\lay-\i);
          \draw[connect] (N\prev-\j) -- (N\lay-\i);
        }
      \fi 
      
    }
  }
  
  \node[above=5,align=center,mygreen!60!black] at (N1-1.90) {\small input};
  \node[above=2,align=center,myblue!60!black] at (N3-1.90) {\small hidden layers};
  \node[above=10,align=center,myred!60!black] at (N\Nnodlen-1.90) {\small output\\[-0.2em]\small layer};
  
\end{tikzpicture}
\caption{Structure of a neural network with 4 layers (+ 1 input layer)\label{fig:nn}}
\end{figure}

Let us turn to neural networks. A \emph{(feed-forward) neural network} is a sequence of feed-forward layers such that neighboring layers have compatible input/output dimensions. The idea is that the output of one layer is the input to the next layer.
Thus, the function defined by a neural network is the composition of the functions defined by its layers.
The neural network illustrated in Figure~\ref{fig:nn} consists of 4 layers (we omit weights and activation functions).

\begin{mydefinition}{Feed-forward Neural Network}
A \emph{feed-forward neural network} is a sequence $\NN = (\Layer^{(1)},\ldots,\Layer^{(\nlayers)})$
of layers $\Layer^{(k)} = (\wmatrix{W}^{(\layer)}, \bias{b}^{(\layer)}, \activation^{(\layer)})$ such that
$\outdim{\Layer^{(\layer)}} = \indim{\Layer^{(\layer+1)}}$ for all $\layer \in \{1,\ldots,\nlayers-1\}$.
\end{mydefinition}

Since, in this chapter, we only talk about feed-forward neural networks (rather than recurrent neural networks or graph neural networks), we just say \emph{neural network}.

Note that, in the literature, the number of layers usually considers the collection of input values as a separate layer,
which indeed makes sense when looking at the graph representation of a neural network (cf.\ Example~\ref{exmp:nn}).
Thus, while $\NN = (\Layer^{(1)},\ldots,\Layer^{(\nlayers)})$ has $\nlayers$ layers,
the ``graph representation'' exhibits $\nlayers+1$ layers.

Similarly to a layer, $\indim{\NN} \df \indim{\Layer^{(1)}}$ is the \emph{input dimension},
and $\outdim{\NN} \df \outdim{\Layer^{(\nlayers)}}$ the output dimension of $\NN$.
In addition, we let $\nndim{\NN} \df (\indim{\NN},\outdim{\NN})$.
The neural network computes
$\nnfunction{\NN}: \Reals^{\indim{\NN}} \to \Reals^{\outdim{\NN}}$
defined as the function composition \[\nnfunction{\NN} \df \lfunction{\Layer^{(\nlayers)}} \fcomp \ldots \fcomp \lfunction{\Layer^{(1)}}\,.\]

We will switch between a graph representation and matrix view at discretion.
Moreover, we will consider $\Reals^m$ and $\Reals^n$ as sets of vectors or sets of tuples whatever is more convenient. In particular, we may simply write $\nnfunction{\NN}(\realx_1,\ldots,\realx_m) = (\realy_1, \ldots,\realy_n)$
instead of $\nnfunction{\NN}((\realx_1,\ldots,\realx_m)^\top) = ((\realy_1,\ldots,\realy_n)^\top)$.

Two neural networks $\NN_1=(\Layer_1^{(1)},\ldots,\Layer_1^{(\nlayers_1)})$ and $\NN_2=(\Layer_2^{(1)},\ldots,\Layer_2^{(\nlayers_2)})$ such that $\outdim{\NN_1} = \indim{\NN_2}$ can be concatenated, and we let \[\NN_1 \cdot \NN_2 = (\Layer_1^{(1)},\ldots,\Layer_1^{(\nlayers_1)},\Layer_2^{(1)},\ldots,\Layer_2^{(\nlayers_2)})\,.\] Note that $\nndim{\NN_1 \cdot \NN_2} = (\indim{\NN_1},\outdim{\NN_2})$.

\paragraph{Activation Functions.}
There are different types of activation functions used in practice. They usually depend on the concrete task at hand (e.g., classification vs.\ regression) and the type of the layer (internal vs.\ output layer), but also on training-related issues such as the choice of the loss function (a precise discussion is, however, not in the scope of this course). Often, a layer $\Layer=(\wmatrix{W}, \bias{b}, \activation)$ has a \emph{local activation function} $\activation$ in the following sense: There is $g: \Reals \to \Reals$ such that, for all $\vect{\realx} \in \Reals^{n}$, we have $\activation(\vect{\realx}) = (g(\realx_1),\ldots,g(\realx_n))^\top$. The definitions and graphs of some common local activation functions are given in Figure~\ref{fig:activations}
(the activation function $\NLReLU$ was recently studied in \cite{abs-1908-03682}).
Their extensions to $\Reals^{n} \to \Reals^{n}$ are denoted in the same way ($\sigmoid$, $\tanh$, and $\ReLU$), where we always assume that $n$ is clear from the context. Another example of a local activation function is the identity function, which we denote by $\idactivation: \Reals^n \to \Reals^n$ (again assuming that $n$ is understood).
An instance of a \emph{global activation function} is
\[
\softmax:
\begin{cases}
\Reals^{n} \to \Reals^{n}\\
\vect{\realx} \mapsto (\realy_1,\ldots,\realy_n) \textup{ where } \realy_i = \displaystyle{\frac{e^{\realx_i}}{\sum_{j=1}^{n} e^{\realx_j}}}
\end{cases}
\]
It ``squeezes'' or normalizes a vector so that it represents a probability distribution.
It is, therefore, frequently used in classification tasks.

An important class of neural networks employs ReLU activation functions (or the identity function, mostly in output layers):

\begin{mydefinition}{Standard and ReLU Neural Network}
A neural network $\NN = (\Layer^{(1)},\ldots,\Layer^{(\nlayers)})$ with layers
$\Layer^{(k)} = (\wmatrix{W}^{(\layer)}, \bias{b}^{(\layer)}, \activation^{(\layer)})$ is called a
\emph{ReLU neural network} if, for all $k \in \{1,\ldots,\nlayers\}$, we have $\activation^{(\layer)} \in \{\idactivation, \ReLU\}$.
%
\end{mydefinition}

\begin{figure}[t]
\centering
\begin{tikzpicture}
  \begin{axis}[
    domain=-3:3,           
    samples=100,           
    axis lines=middle,      
    xlabel={$x$},          
    ylabel={$g(x)$},       
    ymin=-1,
    legend pos=north west, 
    legend entries={ReLU, NLReLU, ~sigmoid ($\sigma$)~, tanh},
    legend style={font=\footnotesize, xshift=-30pt},
  ]
  
  \addplot[blue, thick] {relu(x)};

  \addplot[orange, thick] {ln(1 + relu(x))};
  
  \addplot[red, thick] {sigmoid(x)};
  
  \addplot[teal, thick] {tanh(x)};  
  
  \end{axis}
  \node[anchor=west] at (6.9,5.7) {$\textup{ReLU}(x) = \max(0, x)$}; 
    \node[anchor=west] at (6,3.7) {$\textup{NLReLU}(x) = \ln(1 + \textup{ReLU}(x))$}; 
  \node[anchor=west] at (-3.2,2.1) {$\sigma(x) = \displaystyle{\frac{1}{1 + e^{-x}}}$}; 
  \node[anchor=west] at (-3.9,0.1) {$\textup{tanh}(x) = \displaystyle{\frac{e^x - e^{-x}}{e^x + e^{-x}}}$}; 

\end{tikzpicture}
\caption{(Local) activation functions given by $g: \Reals \to \Reals$ \label{fig:activations}}
\end{figure}

\paragraph{Classification vs.\ Regression.}

As far as feed-forward neural networks are concerned, two predominant classes of tasks are \emph{classification} and \emph{regression}. In a regression problem, the goal is to predict a continuous, numerical value or quantity. In other words, one is trying to find a relationship between input features and the output, which is a real-valued number. Corresponding tasks can be price prediction, weather forecast, or estimating health indicators.
In a classification problem, objects are assigned a category or label among $n$ labels. In that case, frequently, the last layer has output dimension $n$ and uses a softmax activation function, which works well with the categorical cross-entropy loss function during training. When $\NN$ acts as a classifier with $\nnfunction{\NN}: \Reals^m \to \Reals^n$, element $\vect{\realx} \in \Reals^m$ is assigned category $\min(\argmax(\nnfunction{\NN}(\vect{\realx}))) \in \{1,\ldots,n\}$. In the case of binary classification (i.e., in presence of two classes to choose from), it is also common to have $n=1$ and $\sigmoid$ as activation function in the last layer, and to choose one class or the other depending on whether $\nnfunction{\NN}(\vect{\realx}) \in [0,1]$ exceeds a given threshold $\gamma \in [0,1]$.
In the case of binary classification, possible applications are object detection, fraud detection, medical diagnostics, etc.

\begin{myexample}[label=exmp:nn]{Neural Networks}
Below are two simple neural networks:\\ 
\begin{center}
\begin{tikzpicture}[x=2.2cm,y=1.4cm,scale=1.4]
  \readlist\Nnod{2,3,1} 
  
  \foreachitem \N \in \Nnod{ 
    \def\lay{\Ncnt} 
    \pgfmathsetmacro\prev{int(\Ncnt-1)} 
    \foreach \i [evaluate={\y=\N/2-\i; \x=\lay; \n=\nstyle;}] in {1,...,\N}{ 
      
      \node[node \n] (N\lay-\i) at (\x,\y) {};
      
      \ifnum\lay>1 
        \foreach \j in {1,...,\Nnod[\prev]}{ 
          \draw[connect,line width=1.2] (N\prev-\j) -- (N\lay-\i);
          \draw[connect] (N\prev-\j) -- (N\lay-\i);
        }
      \fi 
      
    }
  }
  
  
  \node[fill=exmpbackcolor, circle, inner sep=2pt, minimum size=0.1cm] at (2.5,0) {\scalebox{0.9}{$1$}};
  \node[fill=exmpbackcolor, circle, inner sep=2pt, minimum size=0.1cm] at (2.5,-0.5) {\scalebox{0.9}{$1$}};
  \node[fill=exmpbackcolor, circle, inner sep=2pt, minimum size=0.1cm] at (2.46,-1) {\scalebox{0.9}{$-1$}};
  
  \node[fill=exmpbackcolor, circle, inner sep=2pt, minimum size=0.1cm] at (1.69,0.35) {\scalebox{0.9}{$1$}};
  \node[fill=exmpbackcolor, circle, inner sep=2pt, minimum size=0.1cm] at (1.64,0.04) {\scalebox{0.9}{$-1$}};
  \node[fill=exmpbackcolor, circle, inner sep=2pt, minimum size=0.1cm] at (1.69,-0.32) {\scalebox{0.9}{$0$}};
  \node[fill=exmpbackcolor, circle, inner sep=2pt, minimum size=0.1cm] at (1.69,-0.66) {\scalebox{0.9}{$1$}};
  \node[fill=exmpbackcolor, circle, inner sep=2pt, minimum size=0.1cm] at (1.69,-1.02) {\scalebox{0.9}{$0$}};
  \node[fill=exmpbackcolor, circle, inner sep=2pt, minimum size=0.1cm] at (1.64,-1.36) {\scalebox{0.9}{$-1$}};

  \node[inner sep=2pt, minimum size=0.1cm] at (1,0.7) {(a)};

  \node[inner sep=2pt, minimum size=0.1cm] at (1,0) {$\realx_1$};
  \node[inner sep=2pt, minimum size=0.1cm] at (1,-1) {$\realx_2$};

  \node[inner sep=2pt, minimum size=0.1cm] at (2,0.5) {\scalebox{0.6}{$\ReLU$}};
  \node[inner sep=2pt, minimum size=0.1cm] at (2,-0.5) {\scalebox{0.6}{$\ReLU$}};
  \node[inner sep=2pt, minimum size=0.1cm] at (2,-1.5) {\scalebox{0.6}{$\ReLU$}};
  
  \node[inner sep=2pt, minimum size=0.1cm] at (3,-0.5) {\scalebox{0.8}{$\textup{id}$}};

\end{tikzpicture}
~~~~
\begin{tikzpicture}[x=2.2cm,y=1.4cm,scale=1.4]
  \readlist\Nnod{2,3,1} 
  
  \foreachitem \N \in \Nnod{ 
    \def\lay{\Ncnt} 
    \pgfmathsetmacro\prev{int(\Ncnt-1)} 
    \foreach \i [evaluate={\y=\N/2-\i; \x=\lay; \n=\nstyle;}] in {1,...,\N}{ 
      
      \node[node \n] (N\lay-\i) at (\x,\y) {};
      
      \ifnum\lay>1 
        \foreach \j in {1,...,\Nnod[\prev]}{ 
          \draw[connect,line width=1.2] (N\prev-\j) -- (N\lay-\i);
          \draw[connect] (N\prev-\j) -- (N\lay-\i);
        }
      \fi 
      
    }
  }
  
  
  \node[fill=exmpbackcolor, circle, inner sep=2pt, minimum size=0.05cm] at (2.5,0) {\scalebox{0.6}{$0.24$}};
  \node[fill=exmpbackcolor, circle, inner sep=2pt, minimum size=0.05cm] at (2.5,-0.5) {\scalebox{0.6}{$0.84$}};
  \node[fill=exmpbackcolor, circle, inner sep=2pt, minimum size=0.05cm] at (2.5,-1) {\scalebox{0.6}{$0.97$}};
 
  \node[fill=exmpbackcolor, circle, inner sep=2pt, minimum size=0.05cm] at (1.72,0.35) {\scalebox{0.6}{$0.33$}};
  \node[fill=exmpbackcolor, circle, inner sep=2pt, minimum size=0.05cm] at (1.72,0.04) {\scalebox{0.6}{$0.20$}};
  \node[fill=exmpbackcolor, circle, inner sep=2pt, minimum size=0.05cm] at (1.69,-0.32) {\scalebox{0.6}{$-0.10$}};
  \node[fill=exmpbackcolor, circle, inner sep=2pt, minimum size=0.05cm] at (1.72,-0.66) {\scalebox{0.6}{$1.13$}};
  \node[fill=exmpbackcolor, circle, inner sep=2pt, minimum size=0.05cm] at (1.72,-1.02) {\scalebox{0.6}{$1.03$}};
  \node[fill=exmpbackcolor, circle, inner sep=2pt, minimum size=0.05cm] at (1.69,-1.38) {\scalebox{0.6}{$-1.03$}};

  \node[inner sep=2pt, minimum size=0.1cm] at (1,0.7) {(b)};

  \node[inner sep=2pt, minimum size=0.1cm] at (1,0) {$\realx_1$};
  \node[inner sep=2pt, minimum size=0.1cm] at (1,-1) {$\realx_2$};

  \node[inner sep=2pt, minimum size=0.1cm] at (2,0.5) {\scalebox{0.6}{$\ReLU$}};
  \node[inner sep=2pt, minimum size=0.1cm] at (2,-0.5) {\scalebox{0.6}{$\ReLU$}};
  \node[inner sep=2pt, minimum size=0.1cm] at (2,-1.5) {\scalebox{0.6}{$\ReLU$}};
  
  \node[inner sep=2pt, minimum size=0.1cm] at (3,-0.5) {\scalebox{0.8}{$\textup{id}$}};
  \node[inner sep=2pt, minimum size=0.1cm] at (3,-0.17) {\scalebox{0.6}{$+0.07$}};

\end{tikzpicture}
\end{center}
~\\[-1ex]
\begin{boxexitemize}
\item[(a)] We have $\NN = (\Layer^{(1)}, \Layer^{(2)})$ where
$\Layer^{(k)} = (\wmatrix{W}^{(\layer)}, \bias{b}^{(\layer)}, \activation^{(\layer)})$.
Function $\activation^{(1)}$ is the ReLU activation function and $\activation^{(2)}$ is the identity.
Moreover, we have $\bias{b}^{(1)} = (0, 0, 0)^\top$, $\bias{b}^{(2)} = (0)$,
\[
\wmatrix{W}^{(1)} =
\begin{pmatrix}
    1 & -1 \\
    0 & \+1 \\
    0 & -1
\end{pmatrix} \in \Reals^{3 \times 2}
\textup{\,, and }
\wmatrix{W}^{(2)} =
\begin{pmatrix}
    1 & 1 & -1
\end{pmatrix} \in \Reals^{1 \times 3}\,.
\]
Note that $\NN$ computes the maximum function, i.e., $\nnfunction{\NN}(\realx_1, \realx_2) = \max(\vectone{\realx}{1}, \vectone{\realx}{2}) = \max(\vectone{\realx}{1} - \vectone{\realx}{2}, 0) + \vectone{\realx}{2}$ for all $(\realx_1, \realx_2) \in \Reals^2$.

\item[(b)] This neural network has the same structure and activation functions as in (a), but different weights:
We have $\bias{b}^{(1)} = (0, 0, 0)^\top$, $\bias{b}^{(2)} = (0.07)$,
\[
\wmatrix{W}^{(1)} =
\begin{pmatrix}
    \+0.33 & \+0.2\0 \\
    -0.1\0 & \+1.13 \\
    \+1.03 & -1.03
\end{pmatrix}
 \in \Reals^{3 \times 2}
\textup{\,, and }
\wmatrix{W}^{(2)} =
\begin{pmatrix}
    0.24 & 0.84 & 0.97
\end{pmatrix} \in \Reals^{1 \times 3}\,.
\]
It is, however, less clear what $\NN$ computes. For example, we have
$\nnfunction{\NN}(3,2) = 3.049$, $\nnfunction{\NN}(4,9) = 9.026$, and
$\nnfunction{\NN}(4,93) = 92.79$.

\end{boxexitemize}
The first neural network is taken from \cite{HertrichBSS23}, where the authors study the question how many layers are needed to compute certain functions in a ReLU neural network.
\end{myexample}

\paragraph{Convolutional Neural Networks.}

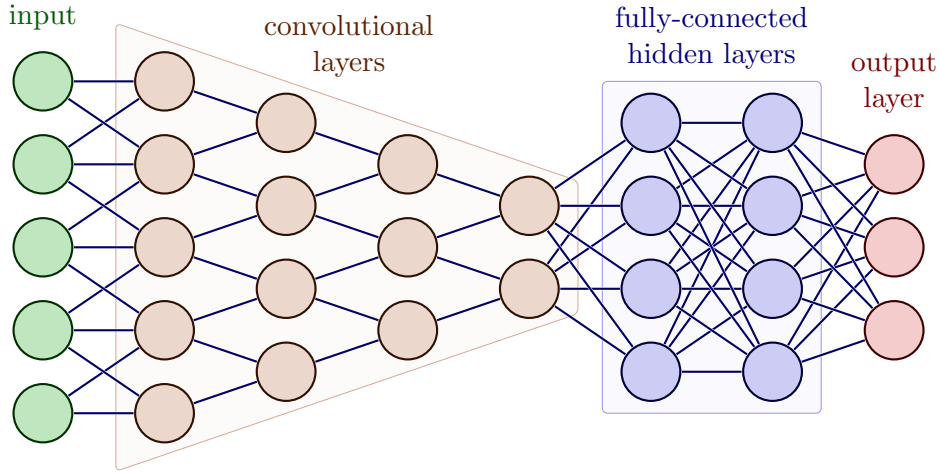
\begin{figure}
\centering
\begin{tikzpicture}[x=1.6cm,y=1.1cm]
  \large
  \readlist\Nnod{5,5,4,3,2,4,4,3} 
  \def\NC{6} 
  \def\nstyle{int(\lay<\Nnodlen?(\lay<\NC?min(2,\lay):3):4)} 
  \tikzset{ 
    node 1/.style={node in},
    node 2/.style={node convol},
    node 3/.style={node hidden},
    node 4/.style={node out},
  }
  
  \draw[myorange!40,fill=myorange,fill opacity=0.02,rounded corners=2]
    (1.6,-2.7) --++ (0,5.4) --++ (3.8,-1.9) --++ (0,-1.6) -- cycle;
  \draw[myblue!40,fill=myblue,fill opacity=0.02,rounded corners=2]
    (5.6,-2.0) rectangle++ (1.8,4.0);
  \node[right=19,above=3,align=center,myorange!60!black] at (3.1,1.8) {\small convolutional\\[-0.3em]\small layers};
  \node[above=3,align=center,myblue!60!black] at (6.5,1.9) {\small fully-connected\\[-0.3em]\small hidden layers};
  
  \foreachitem \N \in \Nnod{ 
    \def\lay{\Ncnt} 
    \pgfmathsetmacro\prev{int(\Ncnt-1)} 
    \foreach \i [evaluate={\y=\N/2-\i+0.5; \x=\lay; \n=\nstyle;}] in {1,...,\N}{ 
      
      \node[node \n,outer sep=0.6] (N\lay-\i) at (\x,\y) {};
      
      \ifnum\lay>1 
        \ifnum\lay<\NC 
          \foreach \j [evaluate={\jprev=int(\i-\j); \cconv=int(\Nnod[\prev]>\N); \ctwo=(\cconv&&\j>0);
                       \c=int((\jprev<1||\jprev>\Nnod[\prev]||\ctwo)?0:1);}]
                       in {-1,0,1}{
            \ifnum\c=1
              \ifnum\cconv=0
                \draw[connect,white,line width=1.2] (N\prev-\jprev) -- (N\lay-\i);
              \fi
              \draw[connect] (N\prev-\jprev) -- (N\lay-\i);
            \fi
          }
          
        \else 
          \foreach \j in {1,...,\Nnod[\prev]}{ 
            \draw[connect,white,line width=1.2] (N\prev-\j) -- (N\lay-\i);
            \draw[connect] (N\prev-\j) -- (N\lay-\i);
          }
        \fi
      \fi 
      
    }
  }
  
  \node[above=3,align=center,mygreen!60!black] at (N1-1.90) {\small input};
  \node[above=3,align=center,myred!60!black] at (N\Nnodlen-1.90) {\small output\\[-0.3em]\small layer};
  
\end{tikzpicture}
\caption{A convolutional neural network (CNN)\label{fig:cnn}}
\end{figure}

Note that the neural networks that we defined above are fully connected:
every neuron is connected to all neurons in the previous layer.
In particular, all the weights/parameters in a weight matrix are, in principle, trainable and
can be adjusted by the learning algorithm. There are other types of neural networks, such as
convolutional neural networks (CNNs), that relax this condition.
A CNN is illustrated in Figure~\ref{fig:cnn}.
Certain neurons in CNNs share weights and
they are sparse in the sense that some of the connections to preceding neurons are missing.
A CNN can be captured in terms of our definition by setting the corresponding weights to $0$.
However, one should have in mind that, when considering training algorithms,
one has to make a distinction between trainable parameters and those that cannot be modified.
When defining neural networks as graphs, this is simple, as one would just omit the corresponding edges.

\section{A Specification Language for Neural Networks}
\label{sec:NNSL}

In this section, we will discuss what it means for a neural network to be \emph{correct}. Here, we mean correctness in a strict mathematical sense. This, in turn, requires a formal specification to be given. Below are some examples of specifications. After discussing them, we will see a formal specification language that encompasses all of them.

\begin{exitemize}
\item $\nnfunction{\NN}: \Reals^m \to \Reals$ computes the maximum function.
\item $\nnfunction{\NN}: \Reals^m \to \Reals^m$ implements a sorting algorithm.
\item $\nnfunction{\NN}: \Reals^m \to \Reals^n$ is permutation invariant.
\item $\nnfunction{\NN}: \Reals^m \to \Reals^m$ is permutation equivariant.
\item $\nnfunction{\NN}: \Reals^m \to \Reals^n$ such that, for given sets $R \subseteq \Reals^m$ and $K \subseteq \{1,\ldots,m\}$, the following holds: for all $\vect{\realx},\vect{\realx}' \in R$ satisfying $\realx_i = \realx_i'$ for all $i \in K$, we have $\argmax(\nnfunction{\NN}(\vect{\realx}))=\argmax(\nnfunction{\NN}(\vect{\realx}'))$.
\item $\nnfunction{\NN}: \Reals^m \to \Reals^n$ such that, for a given set $R \subseteq \Reals^m$ and $\epsilon > 0$, the following holds: for all $\vect{\realx} \in R$ and $\vect{\realx}' \in \Reals^m$ such that $\manhattan{\vect{\realx}}{\vect{\realx}'} \le \epsilon$, we have $\argmax(\nnfunction{\NN}(\vect{\realx}))=\argmax(\nnfunction{\NN}(\vect{\realx}'))$.\footnote{For $\vect{\realx}, \vect{\realx}' \in \Reals^m$, the \emph{Manhattan distance} of $\vect{\realx}$ and $\vect{\realx}'$ is defined as $\manhattan{\vect{\realx}}{\vect{\realx}'} = \sum_{i=1}^m |\realx_i - \realx_i'|$.}
\item $\nnfunction{\NN_1}, \nnfunction{\NN_2}: \Reals^m \to \Reals^n$ such that, for a given set $R \subseteq \Reals^m$, the following holds: for all $\vect{\realx} \in R$, we have $\argmax(\nnfunction{\NN_1}(\vect{\realx}))=\argmax(\nnfunction{\NN_2}(\vect{\realx}))$.
\end{exitemize}

\begin{myexercise}[label=ex:specification_discussion]{}
For all of the above properties, discuss whether they are typically relevant for machine-learning tasks and provide corresponding examples. Can you come up with other relevant specifications? Are there specifications that imply others?
\end{myexercise}

\begin{mysolution}{\ref{ex:specification_discussion}}
\begin{boxexitemize}
\item Computing the maximum function is not a typical machine-learning task: it does not require learning from data to discover unknown or hidden patterns (though, in principle, the task can be addressed by learning a machine-learning model).

\item The same discussion as for (a) applies to the sorting problem.

\item Here, we aggregate multiple inputs into one or several values. Suppose every input value is the age of a particular person in a group of $m$ people, and that the neural network makes a prediction on the number of votes that a particular candidate may receive in the upcoming election. The prediction should not depend on the order of the input values. In other words, permuting the input values should result in the same output value.

\item Similarly to the previous case, suppose that each input represents the probability of a medical staff member being infected, and the output determines, for every medical staff member, whether they should be tested or quarantined. Again, the decision per agent should not depend on the order of the input values. That is, permuting the input values should result in the same output values permuted.

\item Suppose the neural network takes features of a person as input, such as age, gender, salary, etc., and the neural network's task is to decide whether a loan is granted. This decision should be independent of sensitive features like gender or ethnicity. Such a property is referred to as a fairness property. The property thus considers that the features in $\{1,\ldots,m\} \setminus K$ are sensitive.

\item We can think of $\NN$ as an image classifier. If $R$ represents a set of images, e.g., the set the classifier was trained on, we would like $\NN$ to be robust in the sense that small perturbations on the input images do not change the predictions. This would be particularly important when $\NN$ is used in an autonomous car to detect traffic signs.

\item The formula states that neural networks $\NN_1$ and $\NN_2$ are equivalent on an input set $R$. That is, for every input in $R$, they yield the same index/class from $\{1,\ldots,n\}$. Suppose that $\NN_2$ is much smaller than $\NN_1$. If the property is satisfied, we could safely replace $\NN_1$ with the more efficient neural network~$\NN_2$.
\end{boxexitemize}
\end{mysolution}

\begin{myexercise}{}
Let $n=3$. For each of the specifications (a) and (b), provide a neural network satisfying it.
\end{myexercise}

\begin{myexercise}{}
Consider the neural networks from Example~\ref{exmp:nn} and the properties (a)--(f) above.
For every combination of a neural network $\NN$ and a property $\phi$, verify whether $\phi$ is a
suitable specification for $\NN$ (syntactically) and, if so, whether $\NN$ satisfies $\phi$.
Can the two neural networks from the example be considered equivalent? And what would a
corresponding specification look like?
\end{myexercise}

\newcommand{\Var}{\mathcal{X}}
\newcommand{\const}{a}
\newcommand{\consta}{a}
\newcommand{\constb}{b}
\newcommand{\constc}{c}
\newcommand{\constd}{d}
\newcommand{\quant}{\theta}

The formal specification language for neural networks will be based on \emph{linear real arithmetic}, a decidable logic that allows one to combine logical connectives with linear expressions. It is defined over an infinite countable set of variables $\Var = \{\varx,\vary,\varx_1,\varx_2,\ldots\}$ that range over the real numbers.\footnote{Note that we use $x$ etc.\ to denote both real numbers and variables. Variables will be \emph{interpreted} as real numbers so that denoting them in the same way makes sense. However, it is important to keep in mind that variables and real numbers are different objects.}

\begin{mydefinition}{Linear Real Arithmetic}
Formulas from LRA (linear real arithmetic) are given by the following grammar:
\[
\begin{array}{lrcl}
\textup{terms} & t & ::= & \consta \cdot \varx ~\mid~ \constb ~\mid~ t + t\\[1ex]
\textup{formulas} & \phi & ::= & t \le t ~\mid~ \neg \phi ~\mid~ \phi \vee \phi ~\mid~ \exists \varx.\phi
\end{array}
\]
where $\varx \in \Var$ and $\consta, \constb \in \Rationals$.
\end{mydefinition}

\newcommand{\Int}{I}
\newcommand{\termsem}[2]{#2(#1)}
\newcommand{\modfunct}[3]{#1[#2 \mapsto #3]}

An occurrence of a variable is \emph{free} in an LRA formula $\phi$ if
it is not in the scope of a quantifier $\exists$/$\forall$.
A variable is called free in $\phi$ if it has some free occurrence in $\phi$.
Given a tuple of variables $\vectvarx = (\varx_1,\ldots,\varx_n)$,
we may write $\phi(\vectvarx)$ or $\phi(\varx_1,\ldots,\varx_n)$ for a formula $\phi$ whose
free variables are among $\varx_1,\ldots,\varx_n$ (though not all of them need to have a free
occurrence). A \emph{sentence} is a formula without free variables.

We define some common abbreviations such as
$\varx$ for $1 \cdot \varx$,
$t_1 = t_2$ for $t_1 \le t_2 \wedge t_2 \le t_1$,
$t_1 < t_2$ for $t_1 \le t_2 \wedge \neg (t_1 = t_2)$,
$\phi \wedge \psi$ for $\neg(\neg \phi \vee \neg \psi)$,
$\phi \implies \psi$ for $\neg \phi \vee \psi$,
$\phi \iff \psi$ for $(\phi \implies \psi) \wedge (\psi \implies \phi)$,
$\forall \varx.\phi$ for $\neg \exists \varx.\neg \phi$, etc.
We will also write $\exists x_1,\ldots,x_n.\phi$ instead of
$\exists x_1.\exists x_2.\ldots \exists x_n.\phi$ and so forth.

Moreover, for (tuples of) variables $x$, $\vectvary = (\vary_1,\ldots,\vary_n)$,
$\vectvary' = (\vary_1',\ldots,\vary_n')$, and a set $K \subseteq \{1,\ldots,n\}$,
we define the following abbreviations:
\begin{align*}
x \in \vectvary &~\equiv~ \bigvee_{i = 1}^n x = y_i\\[1ex]
x = \max(\vectvary) &~\equiv~ x \in \vectvary \;\wedge\; \displaystyle\bigwedge_{i=1}^n \vary_i \le x\\[1ex]
\argmax(\vectvary) = K &~\equiv~
\displaystyle\bigwedge_{i \in K} y_i = \max(\vectvary) ~\wedge~
\displaystyle\bigwedge_{i \in \{1,\ldots,n\} \setminus K} \neg(y_i = \max(\vectvary))
\\[1ex]
\argmax(\vectvary) = \argmax(\vectvary') &~\equiv~ \bigvee_{K \,\subseteq\, \{1,\ldots,n\}}
\left(
\argmax(\vectvary) = K ~\wedge~ \argmax(\vectvary') = K
\right)
\end{align*}
The latter definition takes into account that $\argmax$ returns the set of indices
carrying the maximal value in a tuple/vector.


\begin{myexample}[label=exmp:sentence]{}
Consider the formulas
\begin{align*}
\phi_1(\varx,\vary) &= \varx < \vary \implies \exists \varz.(\varx < \varz \;\wedge\; \varz < \vary)\\[1ex]
\phi_2(\varx,\vary) &= \varx < \vary \implies ((\varx < 0.5 \cdot \varx + 0.5 \cdot \vary) \;\wedge\;
(0.5 \cdot \varx + 0.5 \cdot \vary < \vary))
\end{align*}
Both, $\phi_1$ and $\phi_2$ have free variables $\varx$ and $\vary$.
The formulas $\forall \varx.\forall \vary.\phi_1(\varx,\vary)$ and $\forall \varx.\forall \vary.\phi_2(\varx,\vary)$ are sentences,
as they do not have any free variables.
\end{myexample}

The semantics of LRA formulas is defined inductively.
To evaluate formulas with free variables, such as $\phi(x,y) = (x \le 0.2 \cdot y)$, we need
to assign values to $x$ and $y$. This is done by an \emph{interpretation
function} $\Int: \Var \to \Reals$. The above formula $\phi$ is evaluated to true iff $\Int(x) \le 0.2 \cdot \Int(y)$.
Towards the semantics, we will first assign to each term $t$
a real number $\termsem{t}{\Int} \in \Reals$ inductively as follows:
\begin{itemize}
\item $\termsem{\consta \cdot x}{\Int} = \consta \cdot \Int(x)$
\item $\termsem{\constb}{\Int} = \constb$
\item $\termsem{t_1 + t_2}{\Int} = \termsem{t_1}{\Int} + \termsem{t_2}{\Int}$
\end{itemize}

Now, models of formulas $\phi$ are interpretation functions:
\begin{itemize}
\item $\Int \models t_1 \le t_2$ if $\termsem{t_1}{\Int} \le \termsem{t_2}{\Int}$
\item $\Int \models \neg \phi$ if $\Int \not\models \phi$
\item $\Int \models \phi \vee \psi$ if $\Int \models \phi$ or $\Int \models \psi$
\item $\Int \models \exists \varx.\phi$ if there is $\nrealx \in \Reals$ such that $\modfunct{\Int}{\varx}{\nrealx} \models \phi$
\end{itemize}
Here, $\modfunct{\Int}{\varx}{\nrealx}$ is the interpretation function that coincides with $\Int$ on
all variables apart from $\varx$, while $\varx$ is mapped to $\nrealx$.

We say that formula $\phi$ is \emph{satisfiable} if
there is an interpretation function $\Int$ such that $\Int \models \phi$.
Note that, to evaluate a formula (a term), it is enough to know
the interpretation of the free variables (variables, respectively)
that occur in it.
Therefore, given a formula $\phi(x_1,\ldots,x_n)$ and $\nrealx_1,\ldots,\nrealx_n \in \Reals$,
we write $\models \phi(\nrealx_1,\ldots,\nrealx_n)$ if $\Int \models \phi(x_1,\ldots,x_n)$
for some interpretation $\Int$ such that $\Int(x_i) = \nrealx_i$ for all $i \in \{1,\ldots,n\}$.
In the particular case where $\phi$ is a sentence,
satisfiability is independent of an interpretation function.
That is, we either have one of the following:
\begin{itemize}
\item $\Int \models \phi$ for all interpretation functions $\Int$
\item $\Int \not\models \phi$ for all interpretation functions $\Int$
\end{itemize}
In the former case, we write $\models \phi$, and we say that $\phi$ is \emph{true}. In the latter case, $\phi$ is \emph{false}.

\begin{myexample}{}
We continue Example~\ref{exmp:sentence}.
The sentences $\forall \varx.\forall \vary.\phi_1(\varx,\vary)$
and $\forall \varx.\forall \vary.\phi_2(\varx,\vary)$ are both true.
On the other hand, $\forall \varx.\forall \vary.\exists \varz.(\varx < \varz \;\wedge\; \varz < \vary)$ is false.
\end{myexample}

\begin{mydefinition}{Satisfiability Problem}
For a class of formulas $\Formulas$, the decision problem $\SAT{\Formulas}$ is defined as follows:
\begin{boxdescription}
\item[Input:] A formula $\phi \in \Formulas$.
\item[Question:] Is $\phi$ satisfiable?
\end{boxdescription}
\end{mydefinition}

Note that free variables in the given formula $\phi$ are implicitly interpreted as existential variables, as the question is whether \emph{there is} a suitable interpretation function. Moreover, if $\phi$ is a sentence, then the problem amounts to asking if $\phi$ is true.

The definition of $\SAT{\Formulas}$ applies to all classes of formulas $\Formulas$ that we consider in this lecture, as they will all be based on interpretation functions of the form $\Int: \Var \to \Reals$.

\begin{mytheorem}[label=thm:declra]{}
The problem $\SAT{\LRA}$ is decidable.
\end{mytheorem}

Before we prove this theorem, we give our specification language for neural networks. It is basically LRA, but with an additional predicate that allows us to talk about the input-output relation induced by a neural network.

\begin{mydefinition}{Neural Network Logic}
Formulas from \NNSL (neural network logic) are given by the following grammar:
\[
\begin{array}{rcl}
t & ::= & \consta \cdot \varx ~\mid~ \constb ~\mid~ t + t\\[1ex]
\phi & ::= & t \le t ~\mid~ \neg \phi ~\mid~ \phi \vee \phi ~\mid~ \exists \varx.\phi ~\mid~ {\NN(\varx_1,\ldots,\varx_m) = (\vary_1,\ldots,\vary_n)}
\end{array}
\]
where $\NN$ is a neural network with input dimension $m \in \posNaturals$ and output dimension $n \in \posNaturals$, $\varx,\varx_1,\ldots,\varx_m,\vary_1,\ldots,\vary_n \in \Var$, and $\consta, \constb \in \Rationals$.
\end{mydefinition}

Note that, in the formula $\NN(\varx_1,\ldots,\varx_m) = (\vary_1,\ldots,\vary_n)$, the variables $\varx_1,\ldots,\varx_m$ and $\vary_1,\ldots,\vary_n$ are free. As \NNSL is an extension of LRA, it only remains to define the semantics of $\NN(\varx_1,\ldots,\varx_m) = (\vary_1,\ldots,\vary_n)$:
\[
\Int \models \NN(\varx_1,\ldots,\varx_m) = (\vary_1,\ldots,\vary_n) \textup{~ if ~} \nnfunction{\NN}(\Int(\varx_1), \ldots, \Int(\varx_m)) =  (\Int(\vary_1), \ldots, \Int(\vary_n))
\]


Accordingly, given an \NNSL sentence $\phi$, we write $\models \phi$ (and say that $\phi$ is true) if $\Int \models \phi$ for some/all $\Int$. Given an \NNSL formula $\phi$ containing the neural networks $\NN_1,\ldots,\NN_k$, we may
write $\phi[\NN_1,\ldots,\NN_k]$ instead of just $\phi$ to highlight that $\phi$ talks about $\NN_1,\ldots,\NN_k$.\footnote{We could have defined \NNSL formulas using ``neural network variables'' so that models interpret these variables as neural networks, but this would cause some notational overhead.}

For a set of NNL formulas $\Formulas$ and a set of activation functions $\FSet$, we denote by
$\Formrestr{\Formulas}{\FSet}$ the set of formulas $ \phi \in \Formulas$ such that every neural network
occurring in $\phi$ uses only activation functions from $\{\idactivation\} \mathrel{\cup} \FSet$.
To simplify notation further, we may just write a list of functions instead of a set.
For example, $\NNLrestr{\ReLU}$ is the set of NNL formulas whose neural networks use the identity
function or $\ReLU$ in their layers. Similarly, $\NNLrestr{\ReLU,\sigmoid,\tanh}$ admits
activation functions from $\{\idactivation\} \cup \{\ReLU,\sigmoid,\tanh\}$,
while $\NNLrestr{\emptyset}$ admits only the identity function.

Now, let us define some concrete \NNSL specifications:

\begin{myexercise}[label=ex:inj_sur]{}
Let $\NN$ be a neural network with $\indim{\NN} = \outdim{\NN} = 2$. Write \NNSL sentences $\phi_1[\NN]$ and $\phi_2[\NN]$ such that the following hold:
\begin{boxitemize}
\item $\nnfunction{\NN}$ is surjective iff $\models \phi_1[\NN]$
\item $\nnfunction{\NN}$ is injective iff $\models \phi_2[\NN]$
\end{boxitemize}
\end{myexercise}

\begin{mysolution}{\ref{ex:inj_sur}}
\begin{boxitemize}
\item $\phi_1[\NN] = \forall y_1,y_2.\exists x_1,x_2.\,\NN(x_1,x_2) = (y_1,y_2)$
\item $\phi_2[\NN] = \forall x_1,x_2,x_1',x_2',y_1,y_2.
\left(
\begin{array}{c}
\NN(x_1,x_2) = (y_1,y_2) ~\wedge~ \NN(x_1',x_2') = (y_1,y_2)\\[0.8ex]
\implies\\[1ex]
x_1 = x_1' ~\wedge~ x_2 = x_2'\\[0.8ex]
\end{array}
\right)$
\end{boxitemize}
\end{mysolution}

\begin{myexercise}{}
Write an \NNSL sentence for the XOR function: A given neural network $\NN$ should compute
a function $\Reals^2 \to \Reals$ that, for every input $(x_1,x_2) \in \{0,1\}^2$, outputs
the truth value $x_1 \oplus x_2$.
\end{myexercise}

\begin{myexercise}{}
Write \NNSL sentences for the properties (a)--(g) given at the beginning of Section~\ref{sec:NNSL}:
For each property $P$ among (a)--(f), define an \NNSL sentence $\phi[\NN]$ such that $\models \phi[\NN]$ iff $\NN$ satisfies $P$. For property (g), write an \NNSL sentence $\phi[\NN_1,\NN_2]$ such that $\models \phi[\NN_1, \NN_2]$ iff $\NN_1,\NN_2$ satisfy (g).
\end{myexercise}

\begin{mysolution}{}
For a (definable) set $R \subseteq \Reals^m$, let $\phi_R(\varx_1,\ldots,\varx_m)$ be an LRA formula such that, for all $\nrealx_1,\ldots,\nrealx_m \in \Reals$, we have $(\nrealx_1,\ldots,\nrealx_m) \in R$ iff $\models \phi_R(\nrealx_1,\ldots,\nrealx_m)$.
\begin{boxexitemize}\itemsep=2ex
\item
$\forall \varx_1,\ldots,\varx_m,\vary.
\left(\NN(\varx_1,\ldots,\varx_m) =
\vary ~\implies~ \vary = \max(\varx_1,\ldots,\varx_m)\right)$
\item 
$\forall \varx_1,\ldots,\varx_m,\vary_1,\ldots,\vary_m.
\left(
\begin{array}{c}
\NN(\varx_1,\ldots,\varx_m) = (\vary_1,\ldots,\vary_m)\\[0.8ex]
\implies\\[1ex]
\displaystyle\bigwedge_{1 \le i < j \le m} \vary_i \le \vary_j
\;\wedge\;
\displaystyle\bigvee_{\perm \in \Perm{m}}
\bigwedge_{i = 1}^m x_i = y_{\perm(i)}
\end{array}
\right)$

\item $\forall \varx_1,\ldots,\varx_m.\exists\vary_1,\ldots,\vary_n.
\displaystyle\bigwedge_{\perm \in \Perm{m}} \NN(\varx_{\perm(1)},\ldots,\varx_{\perm(m)}) = (\vary_1,\ldots,\vary_n)$

\item $\forall \varx_1,\ldots,\varx_m,\vary_1,\ldots,\vary_m.
\left(
\begin{array}{c}
\NN(\varx_1,\ldots,\varx_m) = (\vary_1,\ldots,\vary_m)\\[0.8ex]
\implies\\[1ex]
\displaystyle\bigwedge_{\perm \in \Perm{m}}
\NN(\varx_{\perm(1)},\ldots,\varx_{\perm(m)}) = (\vary_{\perm(1)},\ldots,\vary_{\perm(m)})\\[0.8ex]
\end{array}
\right)$

\item \[
\forall \vectvarx,\vectvarx',\vectvary,\vectvary'.
\left(
\begin{array}{@{}c@{}}
\NN(\vectvarx) = \vectvary \,\wedge\, \NN(\vectvarx') = \vectvary'
\,\wedge\, \phi_R(\vectvarx) \,\wedge\, \phi_R(\vectvarx') \,\wedge\,  \displaystyle\bigwedge_{i \in K} \varx_i = \varx_i'\\[0.8ex]
\implies\\[1ex]
\argmax(\vectvary) = \argmax(\vectvary')\end{array}
\right)
\]

\item 
\[
\forall \vectvarx,\vectvarx',\vectvary,\vectvary'.
\left(
\begin{array}{@{}c@{}}
\left(
\begin{array}{@{}r@{\;}l@{}}
& \NN(\vectvarx) = \vectvary \;\wedge\; \NN(\vectvarx') = \vectvary'
\;\wedge\; \phi_R(\vectvarx)\\[1ex]
\wedge & \exists \varz_1,\ldots,\varz_m.
\left(
\begin{array}{@{}c@{}}
\displaystyle\bigwedge_{i=1}^m
\left(
\begin{array}{@{}c@{}}
(\varx_i \le \varx_i' \implies \varz_i = \varx_i' - \varx_i)\\[1ex]
\wedge\\[1ex]
(\varx_i' < \varx_i \implies \varz_i = \varx_i - \varx_i')
\end{array}
\right)\\[3ex]
\wedge~ \varz_1 + \ldots + \varz_m \le \epsilon
\end{array}
\right)
\end{array}
\right)\\[7ex]
\implies\\[1ex]
\argmax(\vectvary) = \argmax(\vectvary')
\end{array}
\right)
\]

\item $\forall \vectvarx,\vectvary,\vectvary'.
\left(
\begin{array}{c}
\NN_1(\vectvarx) = \vectvary \;\wedge\; \NN_2(\vectvarx) = \vectvary' \;\wedge\; \phi_R(\vectvarx)\\[1ex]
\implies\\[1ex]
\argmax(\vectvary) = \argmax(\vectvary')
\end{array}
\right)$
\end{boxexitemize}
\end{mysolution}

An NNL formula $\phi[\NN_1,\ldots,\NN_k]$ is considered to represent a neural network specification.
Verifying $\NN_1,\ldots,\NN_k$ amounts to deciding whether $\phi[\NN_1,\ldots,\NN_k]$ is satisfiable.
We will assume that $\NN_1,\ldots,\NN_k$ are all ReLU neural networks, i.e., $\phi \in \NNLrelu$.
To establish decidability of verification, due to Theorem~\ref{thm:declra},
it is then enough to show that $\phi$ can be translated into an equivalent LRA sentence.

\newcommand{\nnltrans}[1]{\tilde{#1}}
\newcommand{\polyred}{\le_{\textup{poly}}}

\begin{mydefinition}{}
Let $\Formulas_1$ and $\Formulas_2$ be classes of formulas (whose semantics depends on interpretation
functions $\Int: \Var \to \Reals$). We write $\Formulas_1 \le \Formulas_2$ if
there is an algorithm that translates every
$\phi(\varx_1,\ldots,\varx_n) \in \Formulas_1$ into
$\nnltrans{\phi}(\varx_1,\ldots,\varx_n) \in \Formulas_2$ such that, for all
interpretation functions $\Int$, we have $\Int \models \phi$ iff $\Int \models \nnltrans{\phi}$.

\medskip
If the translation can be done in polynomial time, then we write $\Formulas_1 \polyred \Formulas_2$.
\end{mydefinition}

\begin{myproposition}[label=prop:nntoformula]{}
We have $\NNLrelu \polyred \LRA$.
\end{myproposition}

\begin{proof}
We only need to consider formulas involving neural networks.
To do so, we translate every neural network $\NN$, say with input dimension $m$ and output dimension $n$,
into an LRA formula $\phi_\NN(\varx_1,\ldots,\varx_m,\vary_1,\ldots,\vary_n)$ such that
\begin{align}
\label{eq:nntoformula}
\begin{split}
& \textup{for all }
\nrealx_1,\ldots,\nrealx_m,\nrealy_1,\ldots,\nrealy_n \in \Reals\textup{, we have}\\
& \nnfunction{\NN}(\nrealx_1,\ldots,\nrealx_m) = (\nrealy_1,\ldots,\nrealy_n)
\textup{ iff } \models \phi_\NN(\nrealx_1,\ldots,\nrealx_m,\nrealy_1,\ldots,\nrealy_n)
\end{split}
\end{align}

We proceed by induction and first suppose that $\NN = (\Layer)$ has one layer $\Layer = (\wmatrix{W}, \bias{b}, \activation)$. If $\activation = \idactivation$, we set
\[
{\phi}_\NN(x_1,\ldots,x_m,y_1,\ldots,y_n) ~=~
\displaystyle\bigwedge_{i = 1}^n
y_i = b_i + \sum_{j=1}^m a_{i,j} \cdot x_j
\,.
\]
If $\activation = \ReLU$, we set
\[
{\phi}_\NN(x_1,\ldots,x_m,y_1,\ldots,y_n) ~=~
\displaystyle\bigwedge_{i = 1}^n
\exists z.
\left(
\begin{array}{rl}
& z = b_i + \displaystyle\sum_{j=1}^m a_{i,j} \cdot x_j\\[4ex]
\wedge &
\left(
\begin{array}{rl}
& (z \le 0 ~\wedge~ y_i = 0)\\[1ex]
\vee & (z > 0 ~\wedge~ y_i = z)
\end{array}
\right)
\end{array}
\right)\,.
\]
Then, statement (\ref{eq:nntoformula}) holds by the very definitions.

Now assume
$\NN = (\Layer^{(1)},\ldots,\Layer^{(\nlayers)},\Layer^{(\nlayers+1)})$ with $\nlayers \ge 1$.
Let $\NN_1 = (\Layer^{(1)},\ldots,\Layer^{(\nlayers)})$
and $\NN_2 = (\Layer^{(\nlayers+1)})$
and assume we already have formulas
${\phi}_{\NN_1}(\varx_1,\ldots,x_m,\varz_1,\ldots,\varz_k)$
and ${\phi}_{\NN_2}(\varz_1,\ldots,\varz_k,\vary_1,\ldots,\vary_n)$
as required, where $k = \outdim{\NN_1} = \indim{\NN_2}$.
Then, we set
\[
{\phi}_{\NN}(\varx_1,\ldots,x_m,\vary_1,\ldots,\vary_n) =
\exists \varz_1,\ldots,\varz_k.
\left(
\begin{array}{rl}
& {\phi}_{\NN_1}(\varx_1,\ldots,x_m,\varz_1,\ldots,\varz_k)\\[0.5ex]
\wedge & {\phi}_{\NN_2}(\varz_1,\ldots,\varz_k,\vary_1,\ldots,\vary_n)
\end{array}
\right)\,.
\]
Indeed, for all $\nrealx_1,\ldots,\nrealx_m,\nrealy_1,\ldots,\nrealy_n \in \Reals$ we have 
\[
\begin{array}{rl}
& \nnfunction{\NN}(\nrealx_1,\ldots,\nrealx_m) = (\nrealy_1,\ldots,\nrealy_n)\\[1ex]
\textup{ iff } & (\nnfunction{\NN_2} \fcomp \nnfunction{\NN_1})(\nrealx_1,\ldots,\nrealx_m) = (\nrealy_1,\ldots,\nrealy_n)\\[1ex]
\textup{ iff } & \textup{there are }\nrealz_1,\ldots,\nrealz_k \in \Reals\textup{:}
\left(
\begin{array}{rl}
& \nnfunction{\NN_1}(\nrealx_1,\ldots,\nrealx_m) = (\nrealz_1,\ldots,\nrealz_k)\\[1ex]
\textup{and } \!\!\!\!\!&
\nnfunction{\NN_2}(\nrealz_1,\ldots,\nrealz_k) = (\nrealy_1,\ldots,\nrealy_n)
\end{array}
\right)
\\[3.5ex]
\textup{ iff } & \textup{there are }\nrealz_1,\ldots,\nrealz_k \in \Reals\textup{:}
\left(
\begin{array}{rl}
& \models \phi_{\NN_1}(\nrealx_1,\ldots,\nrealx_m,\nrealz_1,\ldots,\nrealz_k)\\[1ex]
\textup{and } \!\!\!\!\!&
\models \phi_{\NN_2}(\nrealz_1,\ldots,\nrealz_k,\nrealy_1,\ldots,\nrealy_n)
\end{array}
\right)\\[3.5ex]
\textup{ iff } & \models \phi_\NN(\nrealx_1,\ldots,\nrealx_m,\nrealy_1,\ldots,\nrealy_n)\,.
\end{array}
\]
Now, to translate a given \NNSL formula
$\phi$ into the LRA formula $\nnltrans{\phi}$ as required,
we replace, in $\phi$, every occurrence of
an atomic subformula $\NN(\varx_1,\ldots,\varx_m) = (\vary_1,\ldots,\vary_n)$
by $\phi_\NN(\varx_1,\ldots,\varx_m,\vary_1,\ldots,\vary_n)$.
\end{proof}

\begin{myexercise}{}
For the \NNSL sentence $\phi$ for the maximum function, and the neural network $\NN$ from Example~\ref{exmp:nn}(a), determine $\nnltrans{\phi}$ according to Proposition~\ref{prop:nntoformula}.
\end{myexercise}

As a corollary from Proposition~\ref{prop:nntoformula} and
Theorem~\ref{thm:declra} (decidability of $\SAT{\LRA}$),
we obtain the following result:


\begin{mytheorem}[label=thm:decnnl]{}
The problem $\SAT{\NNLrelu}$ is decidable.
\end{mytheorem}

We still need to show Theorem~\ref{thm:declra}, i.e., decidability of LRA,
which deserves its own section.

\section{Proof of Decidability of LRA}

We use automata-theoretic techniques, which are versatile
tools for deciding arithmetic theories and can often be easily extended
to cover even richer theories \cite{Haase2020}.
The automata-theoretic approach to deciding arithmetic
theories goes back to Büchi \cite{Buechi1966}.
Given an LRA formula $\phi$, the idea is to construct a Büchi automaton
$\BA_\phi$ such that $\phi$ is satisfiable iff the language of $\BA_\phi$
is nonempty.

In the description below, we adopt
several constructions and some notation
from the paper \cite{Saelzer2022}, where
Sälzer et al.\ provide translations of neural networks into Büchi automata.

\paragraph{Encoding Real Numbers as Words.}

The main idea is to encode a real number as a word over the alphabet
$\Sigma = \{0,1,\comma,+,-\}$, and a $\tsize$-tuple of real numbers
as a word over $\Sigma^\tsize$. We call a word $w \in \Sigma^\omega$
\emph{well-formed} if it is of the form
\[
w = \binsign \binint_{n-1} \ldots \binint_{0} \commaord \binfract_{1} \binfract_{2} \ldots \in \{+,-\}\{0,1\}^\ast\{\comma\}\{0,1\}^\omega
\]
with $n \ge 0$, $s \in \{+,-\}$, and $\binint_{n-1}, \ldots, \binint_{0}, \binfract_{1}, \binfract_{2}, \ldots \in \{0,1\}$.
Then, $w$ encodes the real number in binary
\[
\dec{w} = (-1)^{\sign{\binsign}} \cdot \left(\sum_{i=0}^{n-1} \binint_i \cdot 2^i + \sum_{i=1}^\infty \binfract_i \cdot \displaystyle\frac{1}{2^i}\right)
\]
where $\sign{+} = 0$ and $\sign{-} = 1$.
Note that $w$ determines a unique value $\dec{w}$.
On the other hand, given $\nrealx \in \Reals$,
there may be several well-formed words $w \in \Sigma^\omega$ such that
$\dec{w} = \nrealx$. For example, we have
$\dec{+0\commaord000\ldots} = \dec{-000\commaord000\ldots} = 0$ and
$\dec{+0110\commaord1000\ldots} = \dec{+110\commaord0111\ldots} = 6.5$.

Let $\tsize \ge 0$. A word
\[
w=
\begin{bmatrix}
\binint_{1,1}\\\vdots\\ \binint_{\tsize,1}
\end{bmatrix}
\begin{bmatrix}
\binint_{1,2}\\\vdots\\ \binint_{\tsize,2}
\end{bmatrix}
\begin{bmatrix}
\binint_{1,3}\\\vdots\\ \binint_{\tsize,3}
\end{bmatrix}
\ldots \in (\Sigma^\tsize)^\omega
\]
may be seen as the $\tsize$-tuple
$(\binint_{1,1}\binint_{1,2}\binint_{1,3} \ldots,
~\ldots~, \binint_{\tsize,1}\binint_{\tsize,2}\binint_{\tsize,3}\ldots)$
of words over $\Sigma$.
Thus, a language $L \subseteq (\Sigma^\tsize)^\omega$
can, equivalently, be considered as a relation $L \subseteq (\Sigma^\omega)^\tsize$
and the latter is the view that we mostly adopt in the following.

Now, a tuple $(w_1,\ldots,w_\tsize) \in (\Sigma^\tsize)^\omega$ of well-formed words
encodes the tuple of real numbers $(\dec{w_1},\ldots,\dec{w_\tsize}) \in \Reals^\tsize$.
However, the algorithmic manipulations we perform on automata to
simulate arithmetic operations require that the comma $\comma$ in the binary representations be aligned.
Therefore, we introduce the set of \emph{($\tsize$-ary) well-formed words}, denoted $\WF{\tsize}$.
A word from $(\Sigma^\tsize)^\omega$ is \emph{well-formed} if it is of the form
\[
\begin{bmatrix}
\binsign_1\\\vdots\\\binsign_\tsize
\end{bmatrix}
\begin{bmatrix}
\binint_{1,{n-1}}\\\vdots\\ \binint_{\tsize,{n-1}}
\end{bmatrix}
\ldots
\begin{bmatrix}
\binint_{1,{0}}\\\vdots\\ \binint_{\tsize,{0}}
\end{bmatrix}
\begin{bmatrix}
\comma\\ \vdots\\ \comma
\end{bmatrix}
\begin{bmatrix}
\binfract_{1,{1}}\\\vdots\\ \binfract_{\tsize,{1}}
\end{bmatrix}
\begin{bmatrix}
\binfract_{1,{2}}\\\vdots\\ \binfract_{\tsize,{2}}
\end{bmatrix}
\ldots
\]
such that $\binsign_i \in \{+,-\}$, $\binint_{i,j} \in \{0,1\}$, and $\binfract_{i,j} \in \{0,1\}$ for all $i$ and $j$. In particular, all $\comma$ are aligned in the same column.
If $\tsize = 0$, we have a unique infinite word over a singleton alphabet,
which we define to be well-formed.
Recall that we may consider $\WF{k} \subseteq (\Sigma^\omega)^k$.
For $\wtuple \in (\Sigma^\omega)^k$, we let $w_i$ refer to the $i$-th component of
$\wtuple$, i.e., $\wtuple = (w_1,\ldots,w_k)$.

\begin{figure}
\centering
\includegraphics[scale=0.5]{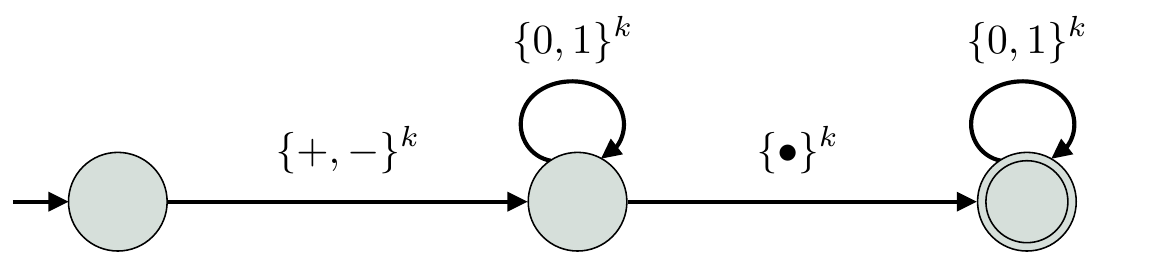}
\caption{The Büchi automaton $\wfBA{\tsize}$\label{fig:wfba}}
\end{figure}

\begin{myproposition}[label=prop:bawf]{}
For $\tsize \ge 0$, we can effectively construct
a Büchi automaton $\wfBA{\tsize}$
over $\Sigma^\tsize$ such that
$L(\wfBA{\tsize}) = \WF{\tsize}$.
\end{myproposition}

\begin{proof}
The automaton $\wfBA{\tsize}$ is given in Figure~\ref{fig:wfba}.
It checks that each ``row'' is contained in
$\{+,-\}\{0,1\}^\ast\{\comma\}\{0,1\}^\omega$
and that the $\comma$-symbols are aligned in the same column.
\end{proof}

We present two further useful automata constructions.

\begin{myproposition}{Büchi Automata Projection}
Let $k \ge 1$ and $i \in \{0,1,\ldots,\tsize\}$. Let $\BA$ be a Büchi automaton
over $\Sigma^k$. We can effectively construct a Büchi automaton $\BAproj{\BA}{i}$
over $\Sigma^i$ such that
\[L(\BAproj{\BA}{i}) = \{(w_1,\ldots,w_i) \mid (w_1,\ldots,w_k) \in L(\BA)\}\,.\]
\end{myproposition}

\begin{proof}
Let $\BA = (\States,\Trans,\init,\Final)$ be a Büchi automaton over $\Sigma^k$.
The Büchi automaton $\BAproj{\BA}{i}$ has the same structure (i.e., the same states, initial states, transitions, final states). The only thing that changes are the transition \emph{labels}, which instead of $k$ symbols,
$[\letter_1,\ldots,\letter_\tsize]$, only contain the first $i$ symbols $[\letter_1,\ldots,\letter_i]$.
That is, we set $\BAproj{\BA}{i} = (\States,\Trans',\init,\Final)$ with
\[
\Trans' = \{ (\statex,[\letter_1,\ldots,\letter_i],\statey)   \mid (\statex,[\letter_1,\ldots,\letter_\tsize],\statey) \in \Trans\}\,.
\]
Note that, when $i=0$, $\BAproj{\BA}{i}$ is a Büchi automaton over a single-letter alphabet.
The correctness proof is now straightforward.
\end{proof}

Applying projection to well-formed words does not necessarily preserve
closure under removing leading zeros.
Therefore, we will make use of another useful closure property.

\newcommand{\BAclosure}[1]{\mathit{cl}(#1)}

\begin{myproposition}[label=prop:baclosure]{}
Let $\tsize \ge 1$ and let $\BA$ be a Büchi automaton over $\Sigma^\tsize$
such that $L(\BA) \subseteq \WF{\tsize}$. We can construct
a Büchi automaton $\BAclosure{\BA}$ over $\Sigma^\tsize$ such that
$L(\BAclosure{\BA})$ is the least set satisfying the following:
\begin{boxitemize}
\item $L(\BA) \subseteq L(\BAclosure{\BA})$ and
\item for all words $\wtuple \in L(\BAclosure{\BA})$ of the form
$\wtuple = (\binsign_1,\ldots,\binsign_\tsize)(0,0,\ldots,0)\wtuple'$
(i.e., $\binsign_1,\ldots,\binsign_\tsize \in \{+,-\}$),
we have $(\binsign_1,\ldots,\binsign_\tsize)\wtuple'  \in L(\BAclosure{\BA})$.
\end{boxitemize}
\end{myproposition}

\begin{myexercise}{}
Prove Proposition~\ref{prop:baclosure}.
\end{myexercise}

\paragraph{From LRA to Büchi automata.}

We are now ready to describe how Büchi automata can be used to
decide whether a given LRA sentence is true. We start with
some constructions for $k$-ary relations of real numbers
that will serve as building blocks in the translation.
They are due to \cite{Saelzer2022}.

\newcommand{\widthleft}{8.5em}
\newcommand{\widthiff}{1.2em}
\newcommand{\widthright}{13em}

\begin{myproposition}[label=prop:baconstructions]{Büchi Automata Constructions}
Let $k \ge 1$, $i,j,i_1,i_2 \in \{1,\ldots,\tsize\}$, and $a, b \in \Rationals$. We can effectively construct
automata $\BAeq{\tsize}{i}{j}$, $\BAle{\tsize}{i}{j}$, $\BAadd{\tsize}{i}{i_1}{i_2}$, $\BAmult{\tsize}{i}{a}{j}$, and $\BAconst{\tsize}{i}{b}$ over $\Sigma^k$ such that
\begin{align*}
L(\BAeq{\tsize}{i}{j}) &= \{\wtuple \in \WF{\tsize} \mid \dec{w_i} = \dec{w_j}\}\\[1ex]
L(\BAle{\tsize}{i}{j}) &= \{\wtuple \in \WF{\tsize} \mid \dec{w_i} \le \dec{w_j}\}\\[1ex]
L(\BAadd{\tsize}{i}{i_1}{i_2}) &= \{\wtuple \in \WF{\tsize} \mid \dec{w_i} = \dec{w_{i_1}} + \dec{w_{i_2}}\}\\[1ex]
L(\BAmult{\tsize}{i}{a}{j}) &= \{\wtuple \in \WF{\tsize} \mid \dec{w_i} = a \cdot \dec{w_j}\}\\[1ex]
L(\BAconst{\tsize}{i}{b}) &= \{\wtuple \in \WF{\tsize} \mid \dec{w_i} = b\}
\end{align*}
\end{myproposition}

\begin{proof}
We start with equality and addition. The other automata will build on them.
We only consider constructions with pairwise distinct indices.
The remaining cases follow similar patterns.

\paragraph{Equality.} The equality Büchi automaton $\BAeq{2}{1}{2}$
is depicted in Figure~\ref{fig:BAeq}. Note that
two binary representations of real numbers can still be equal
when they have different signs. The right-hand side of the automaton
also takes care of real numbers that have a suffix of the form $1^\omega$.
The general case $\BAeq{\tsize}{i}{j}$ allows for arbitrary values
in the components different from $i$ and $j$ (while checking that the word be well-formed).

\paragraph{Addition.}

Consider first the Büchi automaton $\mathcal{B}^{k}_{i=\mathsf{add}(i_1,i_2)}$
as illustrated in Figure~\ref{fig:BAadd} for the special case $\mathcal{B}^3_{3=\mathsf{add}(1,2)}$
(the cases with different signs are similar, since $x_3 = x_1 - x_2$ iff $x_1=x_2 + x_3$ and so on).
It performs bitwise addition of the first two numbers, while checking, each time,
whether a carry bit has to be produced.
Whereas this procedure yields, for all possible input strings, at least one valid result,
$L(\mathcal{B}^{k}_{i=\mathsf{add}(i_1,i_2)})$ does not contain \emph{all}
$\wtuple \in \WF{\tsize}$ such that $\dec{w_{i}} = \dec{w_{i_1}} + \dec{w_{i_2}}$.
For example, $L(\mathcal{B}^{3}_{3=\mathsf{add}(1,2)})$ does not contain the word
\[
\begin{bmatrix}
+\\+\\+
\end{bmatrix}
\begin{bmatrix}
0\\0\\1
\end{bmatrix}
\begin{bmatrix}
0\\0\\0
\end{bmatrix}
\begin{bmatrix}
\comma\\\comma\\\comma
\end{bmatrix}
\begin{bmatrix}
1\\1\\0
\end{bmatrix}
\begin{bmatrix}
1\\1\\0
\end{bmatrix}
\begin{bmatrix}
1\\1\\0
\end{bmatrix}
\begin{bmatrix}
1\\1\\0
\end{bmatrix}
\ldots
\]
but instead
\[
\begin{bmatrix}
+\\+\\+
\end{bmatrix}
\begin{bmatrix}
0\\0\\0
\end{bmatrix}
\begin{bmatrix}
0\\0\\1
\end{bmatrix}
\begin{bmatrix}
\comma\\\comma\\\comma
\end{bmatrix}
\begin{bmatrix}
1\\1\\1
\end{bmatrix}
\begin{bmatrix}
1\\1\\1
\end{bmatrix}
\begin{bmatrix}
1\\1\\1
\end{bmatrix}
\begin{bmatrix}
1\\1\\1
\end{bmatrix}
\ldots
\]
We include all admissible triples using an additional tape and the equality automaton
defining
\[\BAadd{\tsize}{i}{i_1}{i_2} \df \BAclosure{\BAproj{\BAcap{\mathcal{B}^{\tsize+1}_{\tsize+1=\mathsf{add}(i_1,i_2)}}{\BAeq{\tsize+1}{i}{\tsize+1}}}{\tsize}}\,.\]


\begin{figure}
\centering
\includegraphics[scale=0.5]{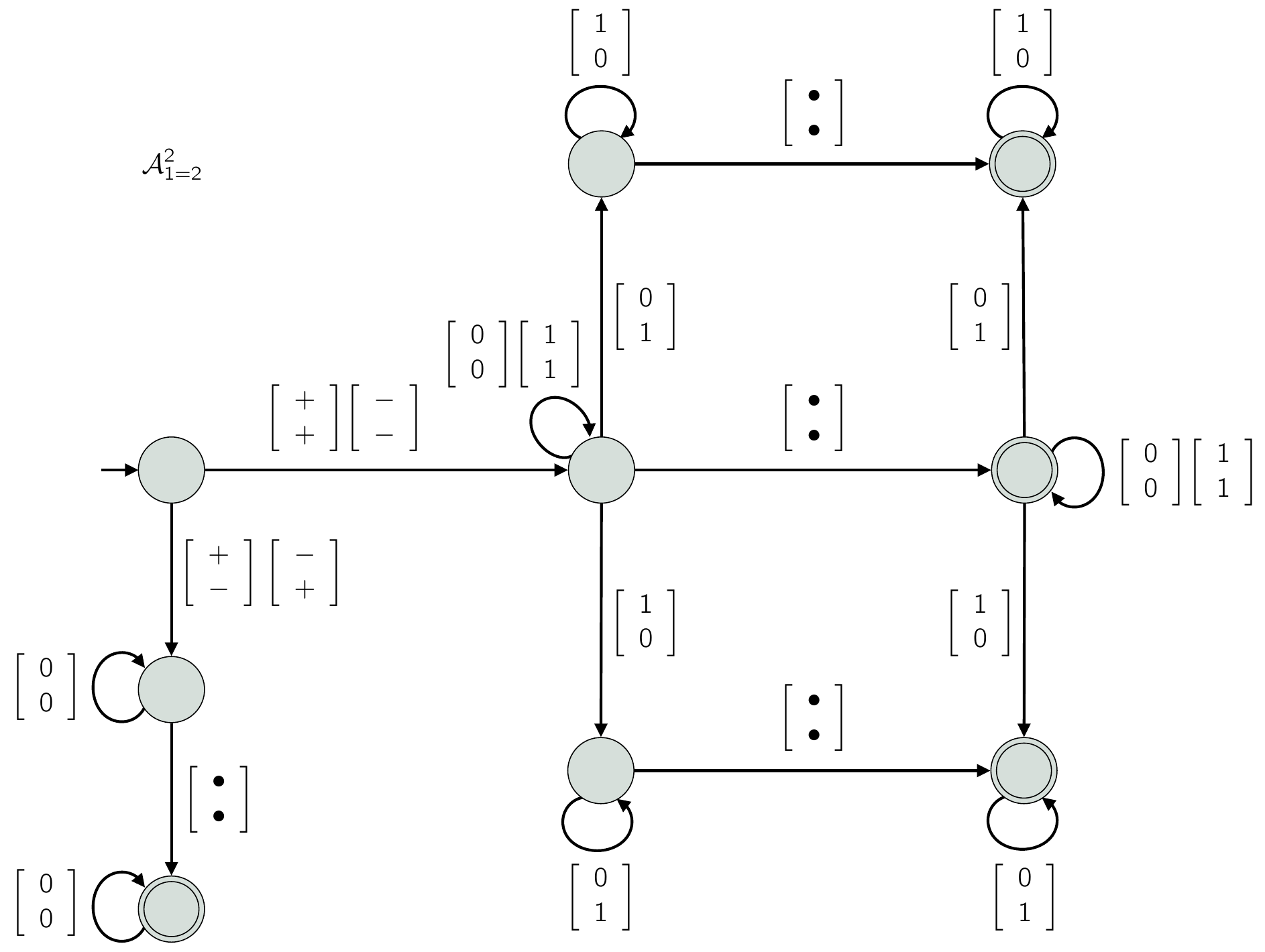}
\caption{The Büchi automaton $\BAeq{2}{1}{2}$\label{fig:BAeq}}
\end{figure}

\begin{figure}
\centering
\includegraphics[scale=0.5]{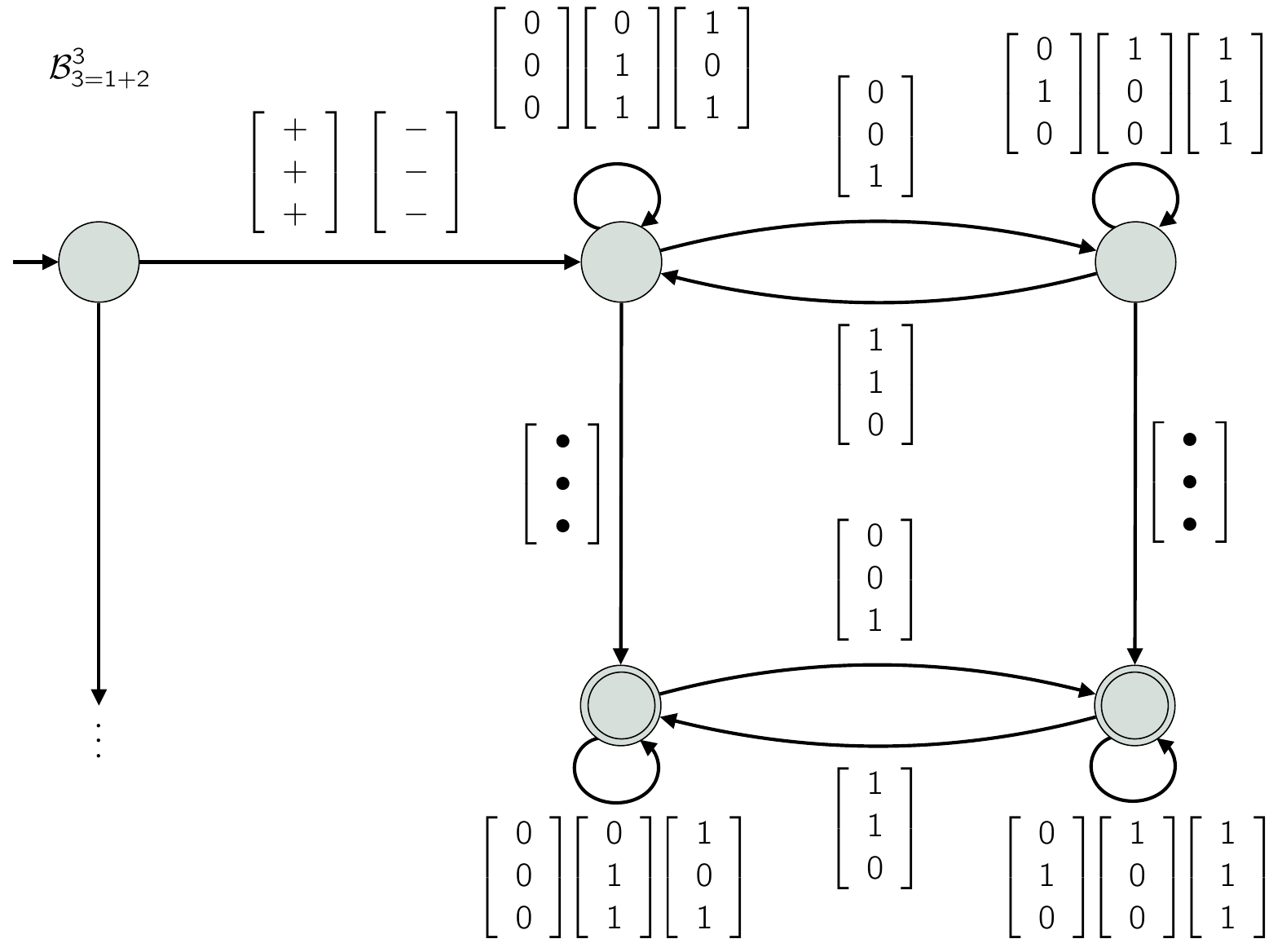}
\caption{The intermediate Büchi automaton $\mathcal{B}^3_{3=\mathsf{add}(1,2)}$\label{fig:BAadd}}
\end{figure}

%
%

\paragraph{Multiplication.}

The cases $a \in \{0,1\}$ are easy. Suppose $a=2$.
We create an additional tape, $k+1$, to ``copy'' the value of tape $j$ using the Büchi automaton for equality.
Tape $i$ should then contain the sum of tapes $j$ and $k+1$. We finally erase tape $k+1$. All this is realized by
\[\BAmult{\tsize}{i}{2}{j} \df \BAclosure{\BAproj{\BAcap{\BAeq{\tsize+1}{k+1}{j}}{\BAadd{\tsize+1}{i}{j}{k+1}}}{\tsize}}\,.\]
Now suppose $a$ is an integer such that $a \ge 3$ (negative constants are handled similarly). Let $\binint_{n-1} \ldots \binint_{0} \in \{0,1\}^\ast$ be the
binary representation of $a$ of minimal length, i.e., $a = \dec{+\binint_{n-1} \ldots \binint_{0}\commaord 000\ldots}$. Furthermore, let $K = \{i_1,\ldots,i_d\} \subseteq \{0,\ldots,n-1\}$ be the set of
indices $\ell$ such that $\binint_{\ell} = 1$. In particular, $n-1 \in K$. Note that, for all $x \in \Reals$, we have
\[a \cdot x = \sum_{\ell=0}^{n-1} \binint_{\ell} \cdot 2^\ell \cdot x = \sum_{\ell \in K} \binint_{\ell} \cdot 2^\ell \cdot x\,.\]
Thus, we can set
\[
\BAmult{\tsize}{i}{a}{j}
\df
\BAclosure{\BAproj{\BAeq{\tsize+n}{k+1}{j} \cap \bigcap_{\ell=1}^{n-1} \BAmult{\tsize+n}{k+1+\ell}{2}{k+\ell} \cap \BA^{\tsize+n}_{i = \mathsf{add}(\tsize+1+i_1,\ldots,\tsize+1+i_d)}}{\tsize}}\,.
\]
Note that we use a Büchi automaton performing an addition of possibly more than two elements.
We obtain the corresponding Büchi automaton by induction, letting, for $n \ge 3$,
\[\BA^{\tsize}_{i = \mathsf{add}(i_1,\ldots,i_n)}
\df
\BAclosure{\BAproj{\BA^{\tsize+1}_{k+1 = \mathsf{add}(i_1,\ldots,i_{n-1})} \cap \BAadd{\tsize+1}{i}{i_n}{\tsize+1}}{\tsize}}
\]
Finally, suppose that $a = \frac{m}{n}$ for integers $m \in \Integers$ and $n \in \posNaturals$.
For any real numbers $x,y \in \Reals$, we have $x = \frac{m}{n} \cdot y$ iff $n \cdot x = m \cdot y$.
Thus, we can set
\[\BAmult{\tsize}{i}{a}{j}
\df
\BAclosure{\BAproj{\BAmult{\tsize+2}{k+1}{n}{i} \cap \BAmult{\tsize+2}{k+2}{m}{j} \cap \BAeq{\tsize+2}{k+1}{k+2}}{\tsize}}\,.
\]
The remaining constructions are left as an exercise.
\end{proof}

\begin{myexercise}{}
Determine the Büchi automata $\BAle{\tsize}{i}{j}$ and $\BAconst{\tsize}{i}{b}$ (for $b \in \Rationals$).
\end{myexercise}

\newcommand{\termBA}{\BA_{\mathsf{term}}}

\paragraph{From LRA to Büchi Automata.}

We now have all the ingredients to transform a given formula into a Büchi automaton that we can then test for nonemptiness.
For simplicity, we suppose we are given a sentence, without free variables.
Given an LRA sentence, we first transform it into a logically equivalent formula in Prenex normal form, i.e., into
a sentence of the form \[\Psi = \quant_1 x_1 \ldots \quant_m x_m. \Phi(x_1,\ldots,x_m)\] where $\quant_1, \ldots, \quant_m \in \{\exists,\neg\exists\}$, the $x_1,\ldots,x_m$ are pairwise distinct, and $\Phi$ is quantifier-free.
For background on the Prenex normal form, we refer the reader to \cite{EbbinghausFT1994}.
Let $t_1,\ldots,t_{n}$ be all \emph{distinct occurrences} of terms in $\Phi$, including\footnote{Strictly speaking, a variable is not a term.} $x_1,\ldots,x_m$. That is, $m \le n$ and $t_i = x_i$ for all $i \in \{1,\ldots,m\}$.

With $\wtuple = (w_1,\ldots,w_m,w_{m+1},\ldots,w_{n}) \in \WF{n}$,
we associate an arbitrary interpretation function $\Int_\wtuple$ such that
$\Int_\wtuple(x_i) = \dec{w_i}$ for all $i \in \{1,\ldots,m\}$.

\begin{mylemma}{}
We can construct a Büchi automaton
$\termBA$ over $\Sigma^{n}$
such that
\[L(\termBA) =
\{\wtuple \in \WF{n} \mid
\termsem{t_i}{\Int_\wtuple} = \dec{w_i} \textup{ for all } i \in \{m+1,\ldots,n\}\}\,.
\]
\end{mylemma}

\begin{proof}
We set
\[
\termBA \df \bigcap_{i=m+1}^n \mathcal{B}_i^n
\]
where
\[
\mathcal{B}_i^n = 
\begin{cases}
\BAadd{n}{i}{j}{k} & \textup{if } t_i = t_j + t_k\\
\BAmult{n}{i}{a}{j} & \textup{if } t_i = a \cdot x_j\\
\BAconst{n}{i}{b} & \textup{if } t_i = b
\end{cases}\]
with $a,b \in \Rationals$.
Correctness follows from Proposition~\ref{prop:baconstructions} by induction.
\end{proof}

\begin{myexample}{}
Consider the sentence
\[ \Psi = \neg\exists x_1.\exists x_2.\underbrace{(x_1 \le x_2 \wedge \neg ((x_1 \le 0.5 \cdot x_1 + 0.5 \cdot x_2) \;\wedge\;
(0.5 \cdot x_1 + 0.5 \cdot x_2 \le x_2)))}_{\displaystyle\Phi(x_1,x_2)}\]
in Prenex normal form. The terms occurring in $\Phi$ are $x_1,x_2,0.5 \cdot x_1,0.5 \cdot x_2,0.5 \cdot x_1 + 0.5 \cdot x_2$,
and we have
\[
\begin{array}{r}
x_1\\x_2\\0.5 \cdot x_1\\0.5 \cdot x_2\\0.5 \cdot x_1 + 0.5 \cdot x_2
\end{array}
\begin{bmatrix}
+\\+\\+\\+\\+
\end{bmatrix}
\begin{bmatrix}
0\\1\\0\\0\\0
\end{bmatrix}
\begin{bmatrix}
1\\0\\0\\1\\1
\end{bmatrix}
\begin{bmatrix}
\comma\\\comma\\\comma\\\comma\\\comma
\end{bmatrix}
\begin{bmatrix}
0\\0\\1\\0\\1
\end{bmatrix}
\begin{bmatrix}
0\\0\\0\\0\\0
\end{bmatrix}
\begin{bmatrix}
0\\0\\0\\0\\0
\end{bmatrix}
\begin{bmatrix}
0\\0\\0\\0\\0
\end{bmatrix}
\ldots
\in L(\termBA)\,.
\]
\end{myexample}

\newcommand{\BAint}{\mathcal{B}}

Using $\termBA$, we now proceed by induction to transform any subformula $\phi(x_1,\ldots,x_m)$ of
$\Phi$ (including $\Phi$) into a Büchi automaton $\BAint_\phi$ over
$\Sigma^{n}$ such that, for all $\wtuple \in L(\termBA)$,
\begin{align}
\label{eq:BAinduction}
\wtuple \in L(\BAint_\phi) \textup{ iff } \Int_\wtuple \models \phi(x_1,\ldots,x_m)\,.
\end{align}

\paragraph{Term Comparison.}
Suppose $\phi(x_1,\ldots,x_m) = (t_i \le t_j)$.
We set \[\BAint_{t_i \le t_j} = \BAle{n}{i}{j}\,.\]
Let $\wtuple=(w_1,\ldots,w_{n}) \in L(\termBA)$. We will show (\ref{eq:BAinduction}).

Suppose $\wtuple \in L(\BAint_\phi)$.
Since $\wtuple \in L(\termBA)$, we have $\termsem{t_i}{\Int_\wtuple} = \dec{w_i}$ and $\termsem{t_j}{\Int_\wtuple} = \dec{w_j}$.
By $\wtuple \in L(\BAle{n}{i}{j})$ and Proposition~\ref{prop:baconstructions}, we have $\dec{w_i} \le \dec{w_j}$.
Thus, $\termsem{t_i}{\Int_\wtuple} \le \termsem{t_j}{\Int_\wtuple}$, which implies $\Int_\wtuple \models t_i \le t_j$.

Conversely, suppose $\Int_\wtuple \models t_i \le t_j$.
Then, $\Int_\wtuple(t_i) \le \Int_\wtuple(t_j)$.
Since $\wtuple \in L(\termBA)$, we also have $\dec{w_i} \le \dec{w_j}$,
which implies $\wtuple \in L(\BAle{n}{i}{j})$. We conclude $\wtuple \in L(\BAint_{t_i \le t_j})$.

\paragraph{Disjunction.}
Suppose $\phi(x_1,\ldots,x_m) = \phi_1 \vee \phi_2$.
We set \[\BAint_{\phi_1 \vee \phi_2} = \BAcup{\BAint_{\phi_1}}{\BAint_{\phi_2}}\,.\]
To show (\ref{eq:BAinduction}), let $\wtuple \in L(\termBA)$. We have $\wtuple \in L(\BAint_{\phi_1 \vee \phi_2})$ iff
$\wtuple \in L(\BAint_{\phi_1}) \cup L(\BAint_{\phi_2})$ iff (by induction hypothesis)
$\Int_\wtuple \models \phi_1 \vee \phi_2$.

\paragraph{Negation.}
Suppose $\phi(x_1,\ldots,x_m) = \neg\psi$.
We set \[\BAint_{\neg\psi} = \BAcompl{\BAint_{\psi}}\,.\]
Let us show (\ref{eq:BAinduction}). For $\wtuple \in L(\termBA)$, we have $\wtuple \in L(\BAint_{\neg\psi})$
iff $\wtuple \not\in L(\BAint_{\psi})$ iff (by induction hypothesis)
$\Int_\wtuple \not\models \psi$ iff
$\Int_\wtuple \models \neg\psi$.

\bigskip

We thus hold a Büchi automaton $\BAint_\Phi$ over $\Sigma^{n}$ such that,
for all $\wtuple \in L(\termBA)$,
\[\wtuple \in L(\BAint_\Phi) \textup{ iff } \Int_\wtuple \models \Phi(x_1,\ldots,x_m)\,.\]

Consider the Büchi automaton $\BA_\Phi = \BAclosure{\BAproj{\BAcap{\BAint_\Phi}{\termBA}}{m}}$.

\begin{mylemma}[label=lemma:closure]{}
We have
\[L(\BA_\Phi) = \{\wtuple \in \WF{m} \mid
\Int_\wtuple \models \Phi(x_1,\ldots,x_m)\}\,.\]
\end{mylemma}

\begin{proof}
Let $\wtuple \in L(\BA_\Phi)$.
There is $\wtuple' = (w_1',\ldots,w_n') \in L(\BAcap{\BAint_\Phi}{\termBA})$
such that, for all $i \in \{1,\ldots,m\}$,
the word $w_i'$ equals $w_i$ modulo (possibly) extra leading zeros.
In particular, $\dec{w_i'} = \dec{w_i}$ for all $i \in \{1,\ldots,m\}$.
Due to $\wtuple' \in L(\BAint_\Phi)$, we have $\Int_{\wtuple'} \models \Phi$ and, therefore, $\Int_{\wtuple} \models \Phi$.

Conversely, assume $\wtuple = (w_1,\ldots,w_m) \in \WF{m}$ such that $\Int_\wtuple \models \Phi(x_1,\ldots,x_m)$.
Let $\wtuple' = (w_1',\ldots,w_n') \in \WF{n}$ such that,
\begin{itemize}
\item for all $i \in \{1,\ldots,m\}$, $w_i'$ equals $w_i$ but possibly has extra leading zeros,
\item for all $i \in \{m+1,\ldots,n\}$, $\dec{w_i'} = \Int_\wtuple(t_i)$.
\end{itemize}
Then, we have $\wtuple' \in L(\termBA)$ and, since $\Int_{\wtuple'} \models \Phi$,
$\wtuple' \in L(\BAint_\Phi)$. Altogether, we have $(w_1',\ldots,w_m') \in \BAproj{\BAcap{\BAint_\Phi}{\termBA}}{m}$
and, finally, $\wtuple = (w_1,\ldots,w_m)\in L(\BA_\Phi)$.
\end{proof}

Towards a Büchi automaton for $\Psi = \quant_1 x_1 \ldots \quant_m x_m. \Phi(x_1,\ldots,x_m)$
where $\quant_i \in \{\exists,\neg\exists\}$,
we add quantifiers (and so eliminate free variables) by induction:
Let $\Psi_m = \Phi(x_1,\ldots,x_m)$ and, for $i = m-1,\ldots,0$,
let
\[\Psi_i = \quant_{i+1} x_{i+1} \ldots \quant_{m} x_{m}. \Phi(x_1,\ldots,x_m)\,.\]
For $i = m,\ldots,0$ (the number of free variables yet to be eliminated),
we build a Büchi automaton $\BA_{\Psi_i}$ over $\Sigma^{i}$ such that
\begin{align}
\label{statement:APsi}
L(\BA_{\Psi_i}) = \{\wtuple \in \WF{i} \mid
\Int_\wtuple \models \Psi_i(x_1,\ldots,x_{i})\}\,.
\end{align}
Note that $\BA_{\Psi_{m}} = \BA_{\Phi}$. Now, let $i = m-1,\ldots,0$.
Towards $\BA_{\Psi_i}$, we will have to eliminate free variable~$x_{i+1}$.

\paragraph{Existential Quantification.}
If $\quant_{i+1} = \exists$, then we set \[\BA_{\Psi_i} = \BAclosure{\BAproj{\BA_{\Psi_{i+1}}}{i}}\,.\]
We show (\ref{statement:APsi}).
We certainly have $\BA_{\Psi_i} \subseteq \WF{i}$.
Let $\wtuple = (w_1,\ldots,w_i) \in \WF{i}$.

Suppose $\wtuple \in L(\BA_{\Psi_i})$.
Then, there is $\wtuple' \in L(\BA_{\Psi_{i+1}}) \subseteq \WF{i+1}$ such that, for all $j \in \{1,\ldots,i\}$,
$w_j'$ equals $w_j$ modulo extra leading zeros. By induction hypothesis,
we have that $\Int_{\wtuple'} \models \Psi_{i+1}(x_1,\ldots,x_{i+1})$.
But this implies $\Int_{\wtuple} \models \exists x_{i+1}.\Psi_{i+1}(x_1,\ldots,x_{i+1})$.

Conversely, suppose $\Int_{\wtuple} \models \Psi_i = \exists x_{i+1}.\Psi_{i+1}(x_1,\ldots,x_{i+1})$.
There is a real number $\nrealx \in \Reals$ such that
$\Int_{\wtuple}[x_{i+1} \mapsto \nrealx] \models \Psi_{i+1}(x_1,\ldots,x_{i+1})$.
Let $\wtuple' = (w_1',\ldots,w_{i+1}') \in \WF{i+1}$ such that $\dec{w_{i+1}'} = \nrealx$ and,
for all $j \in \{1,\ldots,i\}$, $w_j'$ equals $w_j$ but possibly has extra leading zeros.
We have $\Int_{\wtuple'} \models \Psi_{i+1}(x_1,\ldots,x_{i+1})$.
By induction hypothesis, $\wtuple' \in L(\BA_{\Psi_{i+1}})$.
We conclude $\wtuple \in L(\BA_{\Psi_i})$.

\paragraph{Negated Existential Quantification.}
If $\quant_{i+1} = \neg\exists$, then we set \[\BA_{\Psi_i} = \BAcompl{\BAclosure{\BAproj{\BA_{\Psi_{i+1}}}{i}}} \cap \wfBA{i}\,.\]
That is, we first apply projection and closure as in the previous case, then complement the automaton,
and finally intersect it with $\wfBA{i}$. By the preceding discussion, we have
\[
\begin{array}{rl}
& \wtuple \in L(\BA_{\Psi_i})\\[1ex]
\textup{ iff } & \wtuple \in \WF{i} \textup{ and } \Int_{\wtuple} \not\models \exists x_{i+1}.\Psi_{i+1}(x_1,\ldots,x_{i+1})\\[1ex]
\textup{ iff } & \wtuple \in \WF{i} \textup{ and } \Int_{\wtuple} \models
\underbrace{\neg\exists x_{i+1}.\Psi_{i+1}(x_1,\ldots,x_{i+1})}_{\displaystyle =\Psi_i}\\[1ex]
\,.
\end{array}
\]
To wrap up, we obtain the Büchi automaton $\BA_{\Psi_0}$ over $\Sigma^0$ (a singleton alphabet)
such that $L(\BA_{\Psi_0}) \neq \emptyset$ iff $\models \Psi_0$ (recall that $\Psi_0 = \Psi$).
Nonemptiness of the language of a Büchi automaton
is a decidable problem due to Theorem~\ref{thm:BAemptiness}.

\newcommand{\isPowerOfTwo}[1]{\mathsf{isPowerOfTwo}(#1)}

\begin{myexercise}{}
Consider the extension NNL$^+$ of \NNSL given by the following grammar:
\[
\begin{array}{rcl}
t & ::= & \consta \cdot \varx ~\mid~ \constb ~\mid~ t + t\\[1ex]
\phi & ::= & \isPowerOfTwo{x} ~\mid~ t \le t ~\mid~ \neg \phi ~\mid~ \phi \vee \phi ~\mid~ \exists \varx.\phi ~\mid\\
& & \NN(\varx_1,\ldots,\varx_m) = (\vary_1,\ldots,\vary_n)
\end{array}
\]
where $\NN$ is a neural network with input dimension $m \in \posNaturals$ and output dimension $n \in \posNaturals$, $\varx,\varx_1\ldots,\varx_m,\vary_1,\ldots,\vary_n \in \Var$, and $\consta, \constb \in \Rationals$.

\medskip

The semantics of the new atomic formula
$\isPowerOfTwo{x}$ is given as follows (for an interpretation function $\Int$):
\[
\Int \models \isPowerOfTwo{x} \text{ if there is $n \in \Naturals$ such that } \Int(x) = 2^n
\]
Show that the problem below is decidable:
\begin{boxdescription}
\item[Input:] An NNL$^+$ specification $\phi[\NN_1,\ldots,\NN_k]$ such that $\NN_1,\ldots,\NN_k$ are all ReLU neural networks.
\item[Question:] Does $\models \phi[\NN_1,\ldots,\NN_k]$ hold?
\end{boxdescription}

It is enough to describe the modifications wrt.\ the decidability proof for NNL.
\end{myexercise}

\newcommand{\FOTh}{\textup{FO}}

\paragraph{Historical Notes and Remarks on Complexity.}

LRA corresponds to the first-order theory of real numbers with addition, denoted by $\FOTh(\Reals,+,<)$.
Tarski showed that even $\FOTh(\Reals,+,\,\cdot\,,<)$, i.e., non-linear real arithmetic (including multiplication),
is decidable \cite{Tarski1948}. While his algorithm was nonelementary, $\FOTh(\Reals,+,\,\cdot\,,<)$ was shown
to be solvable in doubly exponential time \cite{Collins74}.
The automata-based decidability proof for $\FOTh(\Reals,+,\mathord{<})$ presented in this section
can be easily extended to $\FOTh(\Reals,+,<,\Integers)$, which
has an additional unary predicate for the integers $\Integers$. Another well-known decidable theory is $\FOTh(\Naturals,+,<)$, also known as Presburger arithmetic. On the other hand, $\FOTh(\Naturals,+,\,\cdot\,,<)$, i.e., Peano arithmetic,
is undecidable due to Gödel \cite{Goedel1931}.

The automata-theoretic approach to deciding $\FOTh(\Reals,+,<,\Integers)$ is due to Büchi \cite{Buechi1966}.
Note that projection and negation together may cause an exponential blow-up in the automata
size, which a priori does not allow us to infer an elementary bound on the automata size and computation time.
Boigelot, Jodogne, and Wolper showed that the models of every $\FOTh(\Reals,+,<,\Integers)$-formula
are recognized by a \emph{weak} deterministic Büchi automaton, in which every
strongly connected component contains either only final states or only non-final states \cite{BoigelotJW01}.
Löding showed that minimization of these automata can be reduced, in linear time, to the
minimization of DFAs \cite{Loeding01}. Moreover, a doubly and triply exponential upper bound on the size of
minimal weak deterministic Büchi automata for formulas from $\FOTh(\Reals,+,<)$ and $\FOTh(\Reals,+,<,\Integers)$
were shown by Klaedtke \cite{Klaedtke10} and, respectively, Eisinger \cite{Eisinger08}.
For Presburger arithmetic and finite automata, a triply exponential upper bound is due to
Klaedtke \cite{Klaedtke08}. These (finite) automata can even be computed
in triply exponential time, as was shown by Durand-Gasselin and Habermehl \cite{Durand-GasselinH12}.
These results suggest that automata-based decision procedures may still enjoy a good complexity
and run well in practice. Various automata-based libraries for deciding
arithmetic theories are available \cite{BeckerDEK07,BoigelotJW01,LASH}.

\NNSL specifications can often do without quantifier alternation, i.e., they
belong to the existential fragment, which has a favorable complexity.
We will address this in the next section. We will also discuss theories that correspond
to deciding \NNSL sentences with activation functions such as $\sigmoid$ and $\tanh$
\cite{IsacZBK23}.

\section{An Efficiently Solvable Fragment of NNL}

Our goal is now to identify a fragment of $\NNLrelu$ that comes with an efficiently solvable satisfiability problem.
Observe that many specifications that we have seen previously have
a relatively simple structure in the sense that they can do without quantifier alternation.
For example, consider Exercise~\ref{ex:inj_sur}.
While $\phi_1$ has one quantifier alternation, $\phi_2$ is the negation of an existential formula:
$$\phi_2[\NN] = \neg \exists x_1,x_2,x_1',x_2',y_1,y_2.
\left(
\begin{array}{c}
\NN(x_1,x_2) = (y_1,y_2) ~\wedge~ \NN(x_1',x_2') = (y_1,y_2)\\[0.8ex]
\wedge\\[1ex]
\neg(x_1 = x_1') ~\vee~ \neg(x_2 = x_2')\\[0.8ex]
\end{array}
\right)$$
We will indeed show that satisfiability for existential $\NNLrelu$ formulas
is NP-complete (and, therefore, satisfiability for universal $\NNLrelu$
formulas coNP-complete).

Let us first define the corresponding fragment:

\begin{mydefinition}{Existential \NNSL ($\existsNNL$)}
Formulas from $\existsNNL$ are given by the following grammar:
\[
\begin{array}{rcl}
t & ::= & \consta \cdot \varx ~\mid~ \constb ~\mid~ t + t\\[1ex]
\phi & ::= & t \le t ~\mid~ t < t ~\mid~ \phi \vee \phi ~\mid~ \phi \wedge \phi ~\mid~ \exists x.\phi ~\mid\\[1ex]
& & \NN(\varx_1,\ldots,\varx_m) = (\vary_1,\ldots,\vary_n) ~\mid~
\neg \left(\NN(\varx_1,\ldots,\varx_m) = (\vary_1,\ldots,\vary_n)\right)
\end{array}
\]
where $\NN$ is a neural network with input dimension $m \in \posNaturals$ and output dimension $n \in \posNaturals$, $\varx,\varx_1,\ldots,\varx_m,\vary_1,\ldots,\vary_n \in \Var$, and $\consta, \constb \in \Rationals$.
\end{mydefinition}


Note that, as formulas are in negation normal form (negation is pushed inwards),
we include atomic formulas of the form $t_1 < t_2$ and
$\neg \left(\NN(\varx_1,\ldots,\varx_m) = (\vary_1,\ldots,\vary_n)\right)$,
which are otherwise no longer expressible.
We also use the abbreviations
$t_1 = t_2 ~\equiv~ t_1 \le t_2 ~\wedge~ t_2 \le t_1$ and
$t_1 \neq t_2 ~\equiv~ t_1 < t_2 ~\vee~ t_2 < t_1$.
Note that existential quantifiers are not strictly necessary when
considering satisfiability. However, we include them to be able to
preserve the number of free variables when translating forth and back between
various logics.

We will see that we can translate $\existsNNLrelu$ into existential LRA,
which is defined analogously:

\begin{mydefinition}{Existential Linear Real Arithmetic ($\existsLRA$)}
Formulas from $\existsLRA$ are generated by the following grammar:
\[
\begin{array}{rcl}
t & ::= & \consta \cdot \varx ~\mid~ \constb ~\mid~ t + t\\[1ex]
\phi & ::= & t \le t ~\mid~ t < t ~\mid~ \phi \vee \phi ~\mid~ \phi \wedge \phi  ~\mid~ \exists x.\phi
\end{array}
\]
where $\varx \in \Var$ and $\consta, \constb \in \Rationals$.
\end{mydefinition}

We use the abbreviations $t_1 = t_2$ and $t_1 \neq t_2$ as in $\existsNNL$.


\begin{mytheorem}[label=thm:decexistslra]{Complexity of $\existsLRA$}
The problem $\SAT{\existsLRA}$ is NP-complete.
\end{mytheorem}

The decision procedure can resort to SMT solvers
combining, for example, DPLL(T), Simplex (linear programming)
and Tseitin's transformation. See \cite{albarghouthi-book} or \cite{KroeningS16} for
a detailed exposition.

Actually, we still need to say what the size of a formula is. It is the length of the formula
when constants are written in binary encoding: The length of a rational number $m/n$ (the fraction being simplified), with $m \in \Integers$ and $n \in \posNaturals$, is defined as $1 + \lceil \log(|m| + 1) + 1 \rceil + \lceil \log(n + 1) + 1 \rceil$.
Similarly, concerning $\existsNNLrelu$, we define the size of a matrix/vector as the sum of the sizes of all its coefficients.


%

Our goal is to show the following result:

\begin{mytheorem}[label=thm:decexistsnnl]{}
The problem $\SAT{\existsNNLrelu}$ is NP-complete.
\end{mytheorem}

The lower bound follows from the obvious fact that $\existsLRA \polyred \existsNNLrelu$.
However, we will show it for a smaller fragment of $\existsNNL$.

The upper bound stated in Theorem~\ref{thm:decexistsnnl} is due to the following reduction:

\begin{mylemma}{}
We have $\existsNNLrelu \polyred \existsLRA$.
\end{mylemma}

\begin{proof}
We proceed like in the general case of Lemma~\ref{prop:nntoformula} and translate
$\NN$, say with input dimension $m$ and output dimension $n$,
into $\existsLRA$ formulas $\phi_\NN(\varx_1,\ldots,\varx_m,\vary_1,\ldots,\vary_n)$
and $\phi_\NN'(\varx_1,\ldots,\varx_m,\vary_1,\ldots,\vary_n)$
such that, for all $\nrealx_1,\ldots,\nrealx_m,\nrealy_1,\ldots,\nrealy_n \in \Reals$ the following hold:
\begin{itemize}
\item $\nnfunction{\NN}(\nrealx_1,\ldots,\nrealx_m) = (\nrealy_1,\ldots,\nrealy_n)
\textup{ iff } \models \phi_\NN(\nrealx_1,\ldots,\nrealx_m,\nrealy_1,\ldots,\nrealy_n)$
\item $\nnfunction{\NN}(\nrealx_1,\ldots,\nrealx_m) \neq (\nrealy_1,\ldots,\nrealy_n)
\textup{ iff } \models \phi_\NN'(\nrealx_1,\ldots,\nrealx_m,\nrealy_1,\ldots,\nrealy_n)$
\end{itemize}
We define $\phi_\NN(\varx_1,\ldots,\varx_m,\vary_1,\ldots,\vary_n)$ exactly
like in the proof of Lemma~\ref{prop:nntoformula}. Note that this gives indeed
rise to an $\existsLRA$ formula.
Moreover, we set
\[
{\phi}_{\NN}'(\varx_1,\ldots,x_m,\vary_1,\ldots,\vary_n) =
\exists \varz_1,\ldots,\varz_n.
\left(
\begin{array}{rl}
& {\phi}_{\NN}(\varx_1,\ldots,x_m,\varz_1,\ldots,\varz_n)\\[0.5ex]
\wedge & \displaystyle\bigvee_{i=1}^n y_i \neq z_i
\end{array}
\right)\,,
\]
which is an $\existsLRA$ formula, too.
The overall translation produces a formula of polynomial size.
\end{proof}

We will now present a stronger lower bound, due to \cite{SaelzerL21,SaelzerL22,KatzBDJK17}, for the \emph{reachability}
fragment of $\existsNNLrelu$:
\begin{mydefinition}{Reachability Formulas}
A formula from $\basicexistsNNLrelu$ is an $\existsNNL$ formula of the form
\[
\NN(x_1,\ldots,x_m) = (y_1,\ldots,y_n)
~\wedge~ \phi_1(x_1,\ldots,x_m)
~\wedge~ \phi_2(y_1,\ldots,y_n)
\]

where $\phi_1$ and $\phi_2$ are quantifier-free formulas generated by the grammar
\[
\begin{array}{rcl}
t & ::= & \consta \cdot \varx ~\mid~ \constb ~\mid~ t + t\\[1ex]
\phi & ::= & t \le t ~\mid~ \phi \wedge \phi~
\end{array}
\]
with $\varx \in \Var$ and $\consta, \constb \in \Rationals$.
\end{mydefinition}

In particular, all variables are implicitly existentially quantified.

\begin{mytheorem}[label=thm:decbasicnnl]{Reachability is NP-hard \cite{SaelzerL21,SaelzerL22}}
The problem $\SAT{\basicexistsNNLrelu[\ReLU]}$ is NP-hard.
\end{mytheorem}

\begin{proof}
We follow \cite{SaelzerL21,SaelzerL22} and provide a polynomial-time reduction from 3SAT (satisfiability of 3CNF formulas).
A 3CNF formula is a conjunction of clauses. Every clause is the disjunction of three literals.
A literal is a propositional variable $X_i$ or its negation $\neg X_i$.
An example formula over the variables $X_1,X_2,X_3,X_4$ with three clauses is
\[
\underbrace{(X_1 \vee X_2 \vee X_3)}_{\displaystyle C_1} ~\wedge~ \underbrace{(\neg X_1 \vee X_2 \vee \neg X_3)}_{\displaystyle C_2} ~\wedge~ \underbrace{(\neg X_2 \vee X_3 \vee X_4)}_{\displaystyle C_3}\,.
\]

The satisfiability problem for 3CNF formulas is NP-complete:
Given a 3CNF formula $\Phi = C_1 \wedge \ldots \wedge C_k$ over variables $X_1,\ldots,X_m$,
is there a valuation $\val: \{X_1,\ldots,X_m\} \to \{0,1\}$ such that
$\val(\Phi) = 1$ (with the canonical extension of $\val$ to formulas)?

\begin{figure}[t]
\centering
\includegraphics[scale=0.62]{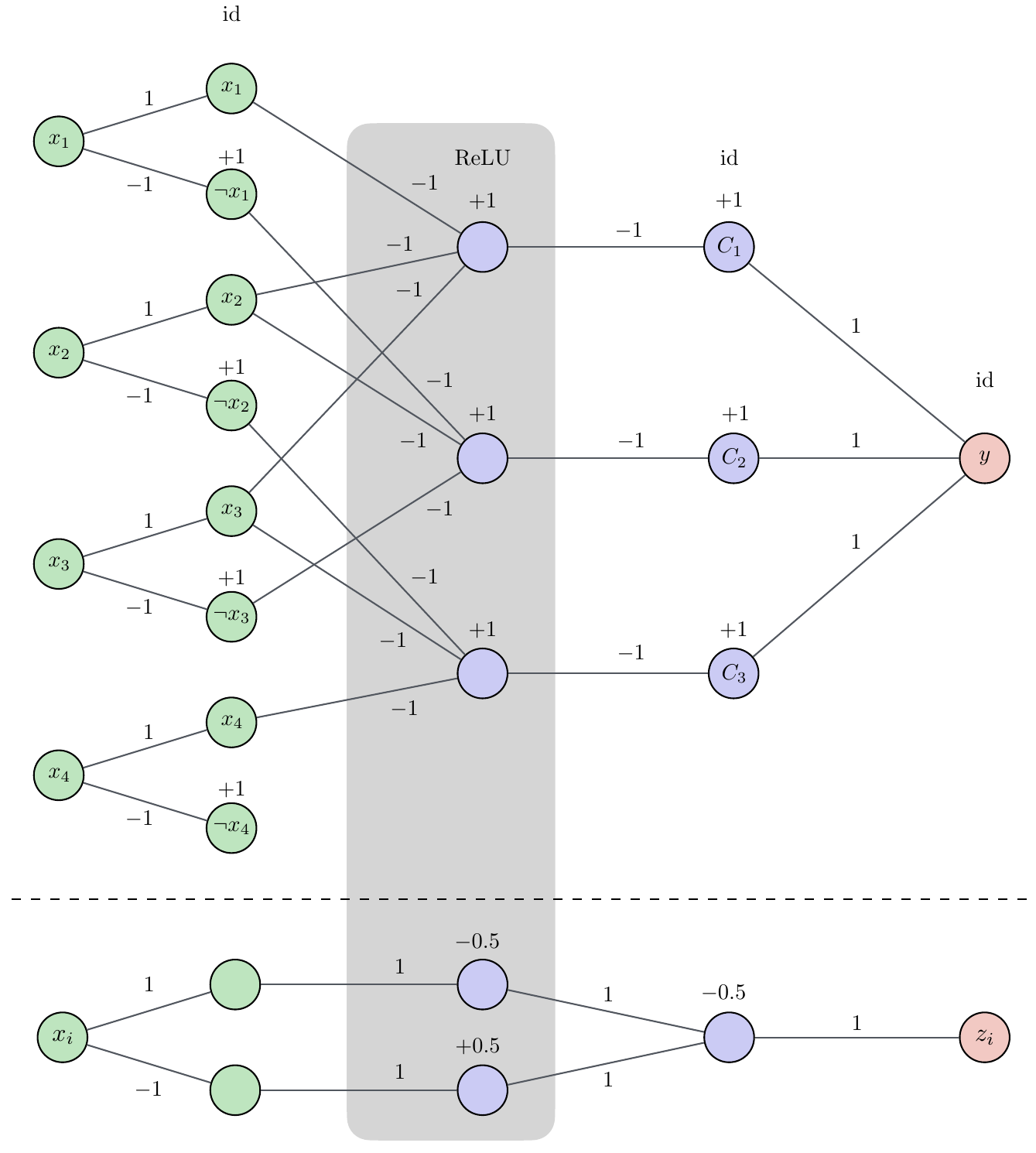}
\caption{Neural network for $(X_1 \vee X_2 \vee X_3) ~\wedge~ (\neg X_1 \vee X_2 \vee \neg X_3) ~\wedge~ (\neg X_2 \vee X_3 \vee X_4)$\label{fig:nn3cnf}}
\end{figure}

We will first build a neural network $\NN$ such that
\[\frestr{\nnfunction{\NN}}{\{0,1\}^m}:
\begin{cases}
\{0,1\}^m \to \Reals\\[1ex]
\vect{r} = (\nrealx_1,\ldots,\nrealx_m) \mapsto \val_\vect{r}(C_1) +  \ldots + \val_\vect{r}(C_k)
\end{cases}\]
where $\val_\vect{r}(X_i) = r_i$ for all $i \in \{1,\ldots,m\}$.
The neural network $\NN$ is illustrated in the upper part of Figure~\ref{fig:nn3cnf}.
Correctness follows because, for all $r_1,r_2,r_3 \in \{0,1\}$, we have
\[
\begin{array}{rl}
& 1 - \ReLU(1 - r_1 - r_2 - r_3)\\[1ex]
= & 1 - \max(0, 1 - r_1 - r_2 - r_3)\\[1ex]
= & \begin{cases}
0 & \textup{ if } r_1 = r_2 = r_3 = 0\\[1ex]
1 & \textup{ otherwise.}
\end{cases}
\end{array}
\]
It follows that
\[
\Phi \textup{ is satisfiable ~~iff~~ }
\left(
\begin{array}{rl}
& \NN(x_1,\ldots,x_m) = y\\[1ex]
\wedge & \displaystyle\bigwedge_{i=1}^m (x_i = 0 ~\vee~ x_i = 1)\\[3ex]
\wedge & y = k
\end{array}
\right) \text{ is satisfiable}.
\]
However, the latter is not a $\basicexistsNNLrelu$ formula (there cannot be a suitable $\basicexistsNNLrelu$ formula, as $\{0,1\}^m$ is not a hyperrectangle).
So we add some more neurons to $\NN$ that allow us to enforce $x_i \in \{0,1\}$ in the sentence
(cf.\ lower part of Figure~\ref{fig:nn3cnf}).
We then obtain a neural network $\NN'$ with $\indim{\NN'} = m$ and $\outdim{\NN'} = m + 1$.

Note that, for all $r \in \Reals$, the following holds:
\[
\begin{array}{rl}
& \mathit{bool}(r)\\[1ex]
\df & \ReLU(r - 0.5) + \ReLU(0.5 - r) - 0.5\\[1ex]
= &
\begin{cases}
r - 1 & \textup{if } r \ge 0.5\\
-r & \textup{if } r < 0.5
\end{cases}
\end{array}
\]
In particular, we have $\mathit{bool}(r) = 0$ iff $r \in \{0,1\}$.
Therefore,
\[
\Phi \textup{ is satisfiable ~~iff~~ }
\left(
\begin{array}{rl}
& \NN'(x_1,\ldots,x_m) = (y,z_1,\ldots,z_m)\\[1ex]
\wedge & \displaystyle\bigwedge_{i=1}^m z_i = 0\\[3ex]
\wedge & y = k
\end{array}
\right)
\text{ is satisfiable}.
\]
As the latter is a $\basicexistsNNLrelu$ formula whose size (in particular $\NN'$) is polynomial, we are done.
\end{proof}

\begin{figure}[t]
\centering
\includegraphics[scale=0.5]{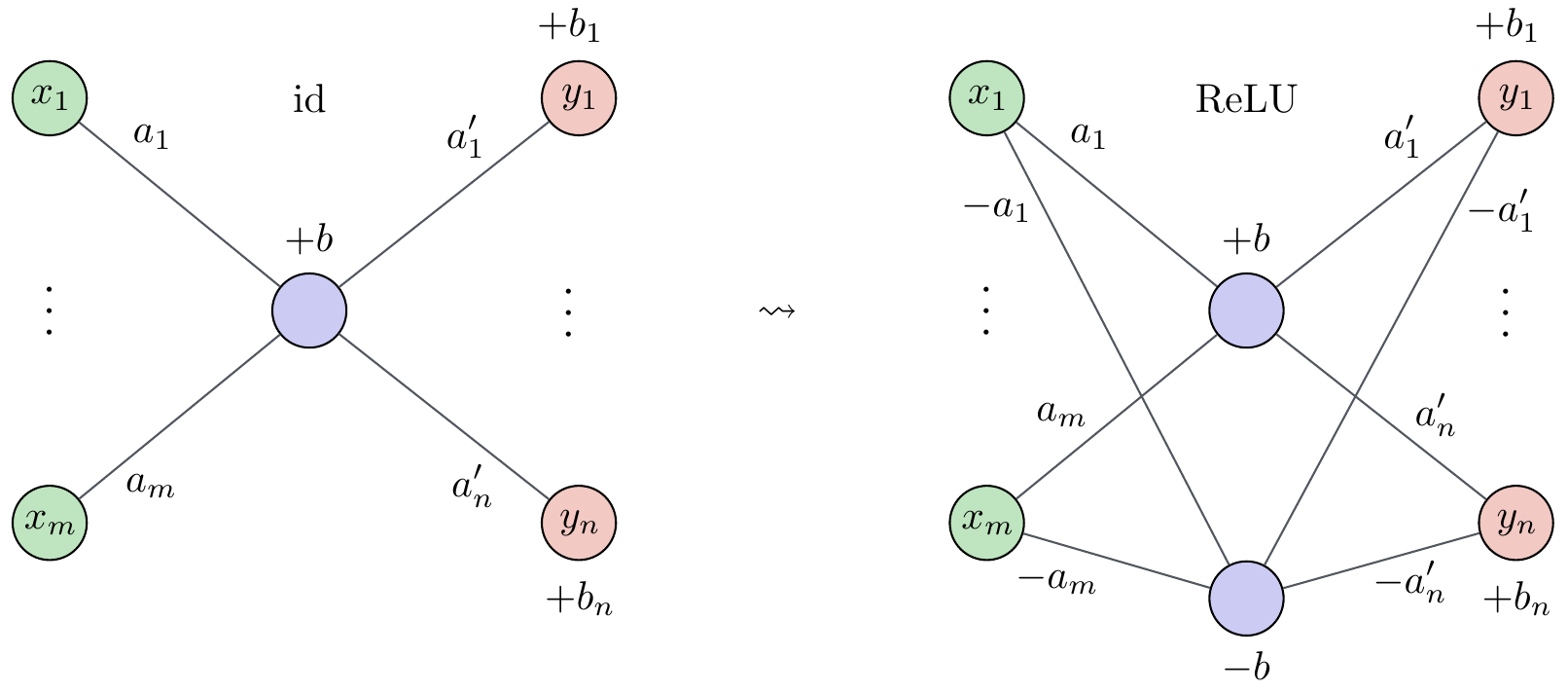}
\caption{Replacing an id neuron with ReLU neurons
\label{fig:id_relu_neuron}}
\end{figure}

\begin{mytheorem}{NP-hardness for ReLU layers \cite{SaelzerL21,SaelzerL22}}
NP-hardness from Theorem~\ref{thm:decbasicnnl}
also holds for ReLU neural networks such that
all but the last layers are ReLU layers.
\end{mytheorem}

\begin{proof}
We replace a neuron with activation function $\idactivation$
by a gadget employing activation function $\ReLU$. This
is illustrated in Figure~\ref{fig:id_relu_neuron}.
Indeed, we have
\begin{align*}
&\phantom{=}~ \displaystyle\Big(b + \sum_{i=1}^m a_ix_i\Big) \cdot a_j'\\[1ex]
&= \ReLU\Big(b + \sum_{i=1}^m a_ix_i\Big) \cdot a_j' +
\ReLU\Big({-} \big(b + \sum_{i=1}^m a_ix_i \bigr)\Big) \cdot (-a_j')\\[1ex]
&= \ReLU\Big(b + \sum_{i=1}^m a_ix_i\Big) \cdot a_j' +
\ReLU\Big({-}b + \sum_{i=1}^m {-}a_ix_i\Big) \cdot (-a_j')
\end{align*}
This concludes the proof.
\end{proof}

NP-hardness results for even stronger restrictions on the neural networks
can be found in \cite{SaelzerL21,SaelzerL22,Wurm23}.

Observe that, when the only activation function allowed is the identity function $\idactivation$,
satisfiability of $\basicexistsNNLrelu[\emptyset]$
can be reduced to a linear-programming problem.
Therefore, we obtain the following result:

\begin{mytheorem}{}
The problem $\SAT{\basicexistsNNLrelu[\emptyset]}$ is solvable in polynomial time.
\end{mytheorem}

Recall that many NNL specifications are negations of $\existsNNLrelu$ sentences.
As a corollary of Theorem~\ref{thm:decexistsnnl}, we can cover this case, too:

\begin{mytheorem}[label=thm:coNPforallNNLrelu]{}
The following decision problem is coNP-complete.
\begin{boxdescription}
\item[Input:] A sentence $\phi \in \existsNNLrelu$.
\item[Question:] Do we have $\models \neg\phi$?
\end{boxdescription}
\end{mytheorem}

\begin{myexercise}{}
In Theorem~\ref{thm:coNPforallNNLrelu}, why do we have to restrict inputs to sentences?
\end{myexercise}


\section{Beyond ReLU Neural Networks}

So far, we mainly considered ReLU neural networks.
But how about decidability for general neural networks?
In this section, we establish equivalence between verification for a particular set of activation functions
and LRA extended by the exponential function~\cite{IsacZBK23}.

\begin{figure}[t]
\centering
\begin{tikzpicture}
  \begin{axis}[
    domain=-3:3,           
    samples=100,           
    axis lines=middle,      
    xlabel={$x$},          
    ylabel={$g(x)$},       
    ymin=-1,
    legend pos=north west, 
    legend entries={ReLU, NLReLU, ~sigmoid ($\sigma$)~, tanh},
    legend style={font=\footnotesize, xshift=-30pt},
  ]
  
  \addplot[blue, thick] {relu(x)};

  \addplot[orange, thick] {ln(1 + relu(x))};
  
  \addplot[red, thick] {sigmoid(x)};
  
  \addplot[teal, thick] {tanh(x)};  
  
  \end{axis}
  \node[anchor=west] at (6.9,5.7) {$\textup{ReLU}(x) = \max(0, x)$}; 
    \node[anchor=west] at (6,3.7) {$\textup{NLReLU}(x) = \ln(1 + \textup{ReLU}(x))$}; 
  \node[anchor=west] at (-3.2,2.1) {$\sigma(x) = \displaystyle{\frac{1}{1 + e^{-x}}}$}; 
  \node[anchor=west] at (-3.9,0.1) {$\textup{tanh}(x) = \displaystyle{\frac{e^x - e^{-x}}{e^x + e^{-x}}}$}; 

\end{tikzpicture}
\caption{(Local) activation functions given by $g: \Reals \to \Reals$ \label{fig:extactivations}}
\end{figure}

\begin{mydefinition}{First-Order Formulas over the Real Exponential Field}
Formulas from $\REF$ (first-order formulas over the \emph{real exponential field}) are defined as follows:
\[
\begin{array}{rcl}
t & ::= & \consta ~\mid~  \varx ~\mid~ t + t ~\mid~ t \cdot t ~\mid~ e^t\\[1ex]
\phi & ::= & t \le t ~\mid~ \neg \phi ~\mid~ \phi \vee \phi ~\mid~ \exists \varx.\phi
\end{array}
\]
where $\varx \in \Var$ and $\consta \in \Rationals$.
\end{mydefinition}

The semantics of the new terms is given as follows (for an interpretation function $\Int$):
\begin{itemize}
\item $\termsem{e^t}{\Int} = e^{\Int(t)}$
\item $\termsem{t_1 \cdot t_2}{\Int} = \termsem{t_1}{\Int} \cdot \termsem{t_2}{\Int}$
\end{itemize}

Let $\NNLplus$ be a shorthand for $\Formrestr{\NNSL}{\ReLU,\NLReLU,\sigmoid,\tanh}$.
We recall these activation functions in Figure~\ref{fig:extactivations}.
The decidability status of $\SAT{\REF}$, also called \emph{Tarski's exponential function problem}, is unknown.
We show a result, due to \cite{IsacZBK23}, stating that the problem is effectively equivalent to $\SAT{\NNLplus}$.

\begin{mytheorem}[label=thm:decbasicrelunnl]{Expressive Equivalence of $\NNLplus$ and \REF \cite{IsacZBK23}}
We have $\NNLplus \le \REF$ and $\REF \le \NNLplus$.
\end{mytheorem}

\begin{proof}
The statement follows from the two propositions below.
\end{proof}

\begin{myproposition}{}
We have $\NNLplus \le \REF$.
\end{myproposition}

\begin{proof}
It remains to consider neural networks $\NN = (\Layer)$ with
$\Layer=(\wmatrix{W}, \bias{b}, \activation)$ and $\activation = \sigmoid$ or $\activation = \tanh$ or
$\activation = \NLReLU$.
Let $m = \indim{\Layer}$ and $n = \outdim{\Layer}$.

We construct an \REF formula
$\phi_\NN(\varx_1,\ldots,\varx_m,\vary_1,\ldots,\vary_n)$ such that,
for all real numbers $\nrealx_1,\ldots,\nrealx_m,\nrealy_1,\ldots,\nrealy_n \in \Reals$,
\[\nnfunction{\NN}(\nrealx_1,\ldots,\nrealx_m) = (\nrealy_1,\ldots,\nrealy_n)
\textup{ iff } \models \phi_\NN(\nrealx_1,\ldots,\nrealx_m,\nrealy_1,\ldots,\nrealy_n)\,.
\]

We set
\[
{\phi}_\NN(x_1,\ldots,x_m,y_1,\ldots,y_n) ~=~
\displaystyle\bigwedge_{i = 1}^n
\exists z.
\left(
\begin{array}{rl}
& z = b_i + \displaystyle\sum_{j=1}^m a_{i,j} \cdot x_j\\[4ex]
\wedge & \alpha_f(z, y_i)
\end{array}
\right)\,.
\]
where
\[
\alpha_f(z, y_i) =
\begin{cases}
y_i \cdot (1 + e^z) = e^z & \text{if } \activation = \sigmoid\\[1ex]
y_i \cdot (e^{2z} + 1) = (e^{2z} - 1) & \text{if } \activation = \tanh\\[1ex]
\left(
\begin{array}{rl}
& (z \le 0 ~\wedge~ y_i = 0)\\
\vee & (z > 0 ~\wedge~ e^{y_i} = z + 1)
\end{array}
\right) & \text{if } \activation = \NLReLU\\
\end{cases}
\]
The rest of the induction follows exactly the lines of the proof
of Proposition~\ref{prop:nntoformula}.
\end{proof}


\newcommand{\nlrelu}{\eta}

Now, we translate \REF formulas into $\NNLplus$ formulas.
In the following, let $\nlrelu = \NLReLU$.

\begin{myproposition}[label=prop:reftonnlplus]{}
We have $\REF \polyred \NNLplus$.
\end{myproposition}

Towards the proof, we establish a series of intermediary results:

\begin{mylemma}{}
For every function $f: \Reals \to \Reals$ from $\{\ReLU, \nlrelu, \sigmoid, \tanh, \sigmoid^{-1}, \tanh^{-1}\}$,
there is an $\NNLplus$ formula $\phi_f(x,y)$ such that,
for all interpretation functions $\Int$, we have
\[
\Int \models \phi_f(x,y) \textup{ ~iff~ } f(\Int(x)) = \Int(y)\,.
\]
\end{mylemma}

\begin{proof}
For $f \in \{\ReLU, \nlrelu, \sigmoid, \tanh\}$,
let $\NN_f = (\Layer)$ where $\Layer = ((1), (0), \activation)$.
We have $\nnfunction{\NN_f} = f$. Therefore, we can set
$\phi_f(x,y) = (\NN_f(x) = y)$.
Moreover, we let $\phi_{\sigmoid^{-1}}(x,y) = (\NN_\sigmoid(y) = x)$
and $\phi_{\tanh^{-1}}(x,y) = (\NN_{\tanh}(y) = x)$.
\end{proof}

Henceforth, we will write $f(x) = y$ as a shorthand for $\phi_f(x,y)$.

\begin{mylemma}{}
There is an $\NNLplus$ formula $\phi_{\ln}(x,y)$ such that, for all interpretation functions
$\Int$, we have
\[
\Int \models \phi_{\ln}(x,y) \textup{ ~iff~ } {\ln}(\Int(x)) = \Int(y)\,.
\]
\end{mylemma}

\begin{proof}
We will express $\ln$ in terms of $\sigmoid^{-1}$, $\tanh^{-1}$, and $\nlrelu$:
\begin{itemize}
\item For all $r \in \Reals$ such that $r \ge 1$:
\begin{align*}
\nlrelu(r - 1) &= \ln(r)
\end{align*}
\item For all $r \in \Reals$ such that $0 < r < 1$:
\begin{align*}
\sigmoid^{-1}(r) &= \ln(r) - \ln(1 - r)\\[1ex]
\tanh^{-1}(r) &= \displaystyle\frac{1}{2}(\ln(1+r) - \ln(1 - r))\\[1ex]
\sigmoid^{-1}(r) - 2 \tanh^{-1}(r) + \nlrelu(r) &=
\ln(r) - \ln(1 - r) - \ln(1+r) + \ln(1 - r) + \ln(1 + r)\\[1ex]
&= \ln(r)
\end{align*}
\end{itemize}
Thus, we can set
\[\phi_{\ln}(x,y) =
\left(
\begin{array}{rl}
& x > 0\\[1ex]
\wedge & (x \ge 1 ~\Rightarrow~ \exists z.(z = x-1 \wedge \nlrelu(z) = y))\\[1ex]
\wedge & (x < 1 ~\Rightarrow~
\exists z_1,z_2,z_3.
\left(
\begin{array}{rl}
& y = z_1 - 2 \cdot z_2 + z_3\\[0.5ex]
\wedge & \sigmoid^{-1}(x) = z_1\\[0.5ex]
\wedge & \tanh^{-1}(x) = z_2\\[0.5ex]
\wedge & \nlrelu(x) = z_3
\end{array}
\right)
\end{array}
\right)\,.
\]
This concludes the proof.
\end{proof}

\begin{mylemma}{}
There is an $\NNLplus$ formula $\phi_{\textup{exp}}(x,y)$ such that, for all interpretation functions
$\Int$, we have
\[
\Int \models \phi_{\textup{exp}}(x,y) \textup{ ~iff~ } e^{\Int(x)} = \Int(y)\,.
\]
\end{mylemma}

\begin{proof}
We set $\phi_{\textup{exp}}(x,y) = \phi_{\ln}(y,x)$ (obtained by exchanging $x$ and $y$ in $\phi_{\ln}(x,y)$).
\end{proof}

Henceforth, we will write $e^x = y$ as a shorthand for $\phi_{\textup{exp}}(x,y)$.

\begin{mylemma}[label=lem:multxyz]{}
There is an $\NNLplus$ formula $\phi_{\mathsf{mult}}(x,y,z)$ such that, for all interpretation functions
$\Int$, we have
\[
\Int \models \phi_{\mathsf{mult}}(x,y,z) \textup{ ~iff~ } \Int(x) \cdot \Int(y) = \Int(z)\,.
\]
\end{mylemma}

\begin{proof}
Recall that, for all $r,s \in \Reals$ with $r,s > 0$, we have $r \cdot s = e^{\ln(r) + \ln(s)}$.
We can set
\[\phi_{\mathsf{mult}}(x,y,z) =
\left(
\begin{array}{rl}
& (x = 0 \vee y = 0) ~\Rightarrow~ z = 0\\[1ex]
\wedge & (x > 0 \wedge y > 0) ~\Rightarrow~ \exists z_1,z_2,z_3.
\left(
\begin{array}{rl}
& \ln(x) = z_1\\
\wedge & \ln(y) = z_2\\
\wedge & z_1 + z_2 = z_3\\
\wedge & e^{z_3} = z
\end{array}
\right)\\[6ex]
\wedge & (x < 0 \wedge y > 0) ~\Rightarrow~ \exists z_1,z_2,z_3,x',z'.
\left(
\begin{array}{rl}
& x' = -x\\
& \ln(x') = z_1\\
\wedge & \ln(y) = z_2\\
\wedge & z_1 + z_2 = z_3\\
\wedge & e^{z_3} = z'\\
\wedge & z = -z'
\end{array}
\right)\\[8ex]
\wedge & \cdots
\end{array}
\right)
\]
Completing the formula is left as an exercise.
\end{proof}

\begin{myexercise}{}
Fill the dots in the definition of $\phi_{\mathsf{mult}}(x,y,z)$ in the proof of Lemma~\ref{lem:multxyz}.
\end{myexercise}

Henceforth, we will write $x \cdot y = z$ as a shorthand for $\phi_{\mathsf{mult}}(x,y,z)$.

We are now ready to prove Proposition~\ref{prop:reftonnlplus} stating that $\REF \le \NNLplus$:

\begin{proof}[Proof of Proposition~\ref{prop:reftonnlplus}]
Consider a formula $\phi = (t_1 \le t_2)$ and let $\mathfrak{T}$ be the set of
subterms occurring in $t_1$ or $t_2$.
We replace $\phi$ with
\[\exists (z_t)_{t \in \mathfrak{T}}.
(z_{t_1} \le z_{t_2} ~\wedge~ \psi_{t_1} ~\wedge~ \psi_{t_2})\]
where $\psi_{t_1}$ and $\psi_{t_2}$ are defined inductively by
\[
\begin{array}{lcl}
\psi_{a} &=& (z_{a} = a)\\
\psi_{x} &=& (z_{x} = x)\\
\psi_{t + t'} &=& (z_{t} + z_{t'} = z_{t + t'} ~\wedge~ \psi_t ~\wedge~ \psi_{t'})\\
\psi_{t \,\cdot\, t'} &=& (z_{t} \cdot z_{t'} = z_{t \,\cdot\, t'} ~\wedge~ \psi_t ~\wedge~ \psi_{t'})\\
\psi_{e^t} &=& (e^{z_{t}} = z_{e^t} ~\wedge~ \psi_t)
\end{array}
\]
Doing this with all inequalities $t_1 \le t_2$, we obtain the desired $\NNLplus$ formula.
\end{proof}

\chapter{Recurrent Neural Networks}

Feed-forward neural networks process one single input vector and produce an output vector.
In this chapter, we look at neural networks that process \emph{sequences}. More precisely,
they may translate sequences of input vectors into sequences of output vectors,
or serve as sequence \emph{classifiers}.

\newcommand{\RNN}{\mathcal{R}}
\newcommand{\statematrix}{{\bf A}}
\newcommand{\inputmatrix}{{\bf I}}
\newcommand{\outputmatrix}{{\bf O}}
\newcommand{\rnnbias}{{\bf b}}
\newcommand{\rnninit}{{\bf h}^{(0)}}
\newcommand{\rnnstate}{{\bf h}}
\newcommand{\rnninput}{{\bf x}}
\newcommand{\statedim}[1]{\mathit{state}(#1)}
\newcommand{\inLayer}{\Layer^{\mathsf{in}}}
\newcommand{\outLayer}{\Layer^{\mathsf{out}}}

\newcommand{\rnntrans}[1]{\delta_{#1}}
\newcommand{\extrnntrans}[1]{\hat\delta_{#1}}
\newcommand{\rnnoutdim}{o}
\newcommand{\rnninvect}{\vect{x}}
\newcommand{\rnnoutvect}{\vect{y}}
\newcommand{\rtime}{\uptau}
\newcommand{\inlength}{\ell}
\newcommand{\inseq}{w}
\newcommand{\outseq}{w'}
\newcommand{\onehot}[1]{\mathit{enc}_{#1}}
\newcommand{\onehotdec}[1]{\mathit{dec}_{#1}}
\newcommand{\rnnLang}[2]{L_{#1}(#2)}
\newcommand{\thr}{\uptheta}
\newcommand{\rnnLangthr}[4]{L_{#1}^{{#3}\, #4}(#2)}
\newcommand{\Langthr}[3]{L^{{#2}\, #3}(#1)}
\newcommand{\initvect}{\boldsymbol{\lambda}}
\newcommand{\finalvect}{\boldsymbol{\gamma}}
\newcommand{\initweight}{\lambda}
\newcommand{\finalweight}{\gamma}

\newcommand{\probmatrix}{{\bf P}}

\newcommand{\rnntransducer}[3]{\nnfunction{#1}_{#2,#3}}

\newcommand{\laststate}[1]{\nnfunction{#1}^{\mathsf{lstate}}}
\newcommand{\laststatetrans}[3]{\nnfunction{#1}_{#2,#3}^{\mathsf{lstate}}}
\newcommand{\lastoutput}[1]{\nnfunction{#1}^{\mathsf{lout}}}
\newcommand{\lastoutputtrans}[3]{\nnfunction{#1}_{#2,#3}^{\mathsf{lout}}}

\newcommand{\rnnstos}[1]{\nnfunction{#1}}
\newcommand{\rnnstol}[1]{\llangle #1 \rrangle}
\newcommand{\alphrnnstos}[3]{\nnfunction{#1}_{#2,#3}}
\newcommand{\alphrnnstol}[3]{\llangle #1 \rrangle_{#2,#3}}

\section{Definition and Semantics}

We consider a simple class of recurrent neural networks, also known as Elman neural networks \cite{Elman90},
rather than more advanced architectures such as long short-term memory (LSTM) \cite{HochreiterS97}.

\begin{mydefinition}{Recurrent Neural Network}
A \emph{recurrent neural network (RNN)} is a triple
$\RNN=(\inLayer, \outLayer, \rnninit)$
where, for some $m,n,\rnnoutdim \in \posNaturals$,
\begin{boxitemize}
\item $\inLayer$ is a feed-forward layer with $\nndim{\inLayer} = (n+m, n)$,
\item $\outLayer$ is a feed-forward layer with $\nndim{\outLayer} = (n, \rnnoutdim)$,
\item $\rnninit \in \Rationals^n$ is the \emph{initial (hidden) state vector}.
\end{boxitemize}
\end{mydefinition}

\begin{figure}[t]
\centering
\includegraphics[scale=0.46]{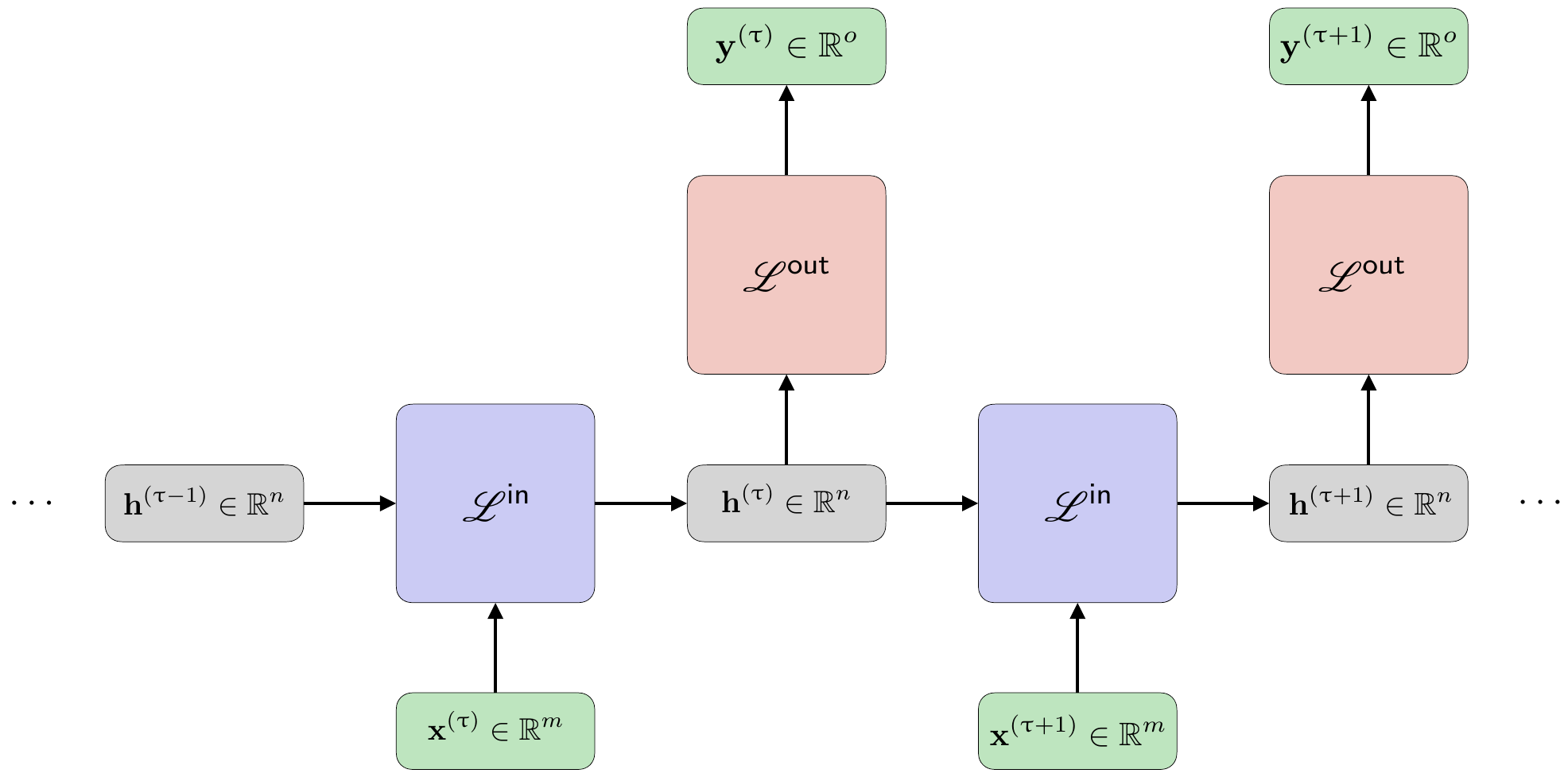}
\caption{Sequence processing by an RNN through time
\label{fig:rnn}}
\end{figure}

If $\inLayer$ uses activation function $f$ and $\outLayer$ uses activation function $g$,
we call $\RNN$ an $(f,g)$-RNN. We call
\begin{itemize}
\item $\statedim{\RNN} \df n$ the \emph{state dimension} of $\RNN$ (i.e., the dimension of each hidden state),
\item $\indim{\RNN} \df m$ the \emph{input dimension} of $\RNN$ (i.e., the dimension of each input symbol),
\item $\outdim{\RNN} \df \rnnoutdim$ the \emph{output dimension} of $\RNN$ (i.e., the dimension of each output symbol).
\end{itemize}
However, states are considered to be hidden, and the type of the input-output relation is solely described by $\nndim{\RNN} = (m,o)$.
When we see $\RNN$ as a \emph{sequence-to-sequence} transducer\footnote{We use the terms \emph{sequence} and \emph{string} interchangeably. However, we rather avoid the term \emph{word} in this context. In NLP, words often refer to what we call letters (or tokens).}, we consider the mapping
\[
\rnnstos{\RNN}:
\begin{cases}
(\Reals^m)^+ \to (\Reals^{\rnnoutdim})^+\\
\vect{x}^{(1)} \ldots \vect{x}^{(\inlength)} \mapsto \rnnoutvect^{(1)} \ldots \rnnoutvect^{(\inlength)}
\end{cases}
\]
where,
for all discrete time points $\rtime \in \{1,\ldots,\ell\}$,
\begin{align*}
\rnnstate^{(\rtime)} &= \nnfunction{\inLayer}(\vectconc{\rnnstate^{(\rtime-1)}}{\rnninput^{(\rtime)}})\\[0.5ex]
\rnnoutvect^{(\rtime)} &= \nnfunction{\outLayer}(\rnnstate^{(\rtime)})
\end{align*}
Recall that $\vectconc{\rnnstate^{(\rtime-1)}}{\rnninput^{(\rtime)}}$ is
the vertical concatenation of $\rnnstate^{(\rtime-1)}$ and $\rnninput^{(\rtime)}$.
The computation is illustrated in Figure~\ref{fig:rnn}.
That is, along the way, $\RNN$ also computes a sequence of \emph{hidden states}
$\rnnstate^{(1)} \ldots \rnnstate^{(\inlength)} \in (\Reals^n)^+$.
Note that $\rnnstos{\RNN}$ is length preserving,
i.e., an input sequence containing $\ell$ vectors is always mapped
to an output sequence containing $\ell$ vectors.

It can actually be useful to view $\RNN$ as an (infinite) automaton coming with
a state-transition function $\delta_\RNN: \Reals^n \times \Reals^m \to \Reals^n$
defined by $\delta_\RNN(\rnnstate, \rnninvect) =
\nnfunction{\inLayer}(\vectconc{\rnnstate}{\rnninvect})$, which is canonically extended to sequences.
When considering $\RNN$ as a \emph{sequence-to-vector} transducer,
we would be interested in the mapping
\[
\rnnstol{\RNN}:
\begin{cases}
(\Reals^m)^+ \to \Reals^{\rnnoutdim}\\
w \mapsto
\nnfunction{\outLayer}(\delta_\RNN(\rnninit, w))\,.
\end{cases}
\]




\paragraph{RNNs as String Transducers.}

Though input and output symbols are vectors over real numbers,
they allow us to deal with symbols from a finite alphabet as well.
This is important in natural language processing, where words
are considered as single symbols and are
encoded, for example via one-hot encoding, as vectors.

\begin{mydefinition}{One-Hot Encoding}
Let $\Sigma$ be a finite alphabet with $|\Sigma| = m$.
A \emph{one-hot encoding} $\onehot{\Sigma}$ of $\Sigma$ is an injective mapping
$\Sigma \to \{0,1\}^m \subseteq \Reals^m$ such that, for all $\alpha \in \Sigma$,
the vector $\vect{x} = \onehot{\Sigma}(\alpha)$ satisfies
$\sum_{i=1}^m x_i = 1$. For example, for $\Sigma = \{\alpha,\beta\}$,
we may set $\onehot{\Sigma}(\alpha) = (1,0)^\top$ and
$\onehot{\Sigma}(\beta) = (0,1)^\top$.

\medskip

The corresponding \emph{one-hot decoding} is the mapping
$\onehotdec{\Sigma}: \Reals^m \to \Sigma$ defined by
$\onehotdec{\Sigma}(\vect{x}) = \alpha$ where $\alpha$
is such that $\min(\argmax(\onehot{\Sigma}(\alpha))) = \min(\argmax(\vect{x}))$.
For example, we have $\onehotdec{\Sigma}((0.7, 0.5)^\top) =
\onehotdec{\Sigma}((0.5, 0.5)^\top) = \alpha$, and
$\onehotdec{\Sigma}((0.2, 0.5)^\top) = \beta$.
\end{mydefinition}

\begin{myremark}{Binary Alphabets}
If $\Sigma$ is a binary alphabet such as $\{\alpha_0,\alpha_1\}$,
we may also choose the dimension to be $m = 1$ and define
$\onehot{\Sigma}(\alpha_i) = i$ as well as
$\onehotdec{\Sigma}(x) = \alpha_1$ iff
$x > \thr$ (or $x \ge \thr$) for some given threshold $\thr$.
This is particularly common in the output layer
in combination with activation
function $\sigmoid$ and threshold $\thr = 0.5$.
\end{myremark}

The mappings $\onehot{\Sigma}$ and $\onehotdec{\Sigma}$ are extended to sequences
as expected, i.e., $\onehot{\Sigma}(\alpha_1 \ldots \alpha_\ell)
= \onehot{\Sigma}(\alpha_1) \ldots \onehot{\Sigma}(\alpha_\ell)$
and $\onehotdec{\Sigma}(\vect{x}^{(1)} \ldots \vect{x}^{(\ell)})
= \onehotdec{\Sigma}(\vect{x}^{(1)}) \ldots \onehotdec{\Sigma}(\vect{x}^{(\ell)})$.
In the following, we assume that all finite alphabets
come with their one-hot encoding/decoding.

Let $\Sigma$ and $\Gamma$ be finite alphabets with
associated one-hot encodings $\onehot{\Sigma}$ and $\onehot{\Gamma}$,
and decodings $\onehotdec{\Sigma}$ and $\onehotdec{\Gamma}$.
Furthermore, assume $|\Sigma| = \indim{\RNN}$ and
$|\Gamma| = \outdim{\RNN}$.
Then, RNN $\RNN$ defines the mappings
\begin{align*}
\alphrnnstos{\RNN}{\Sigma}{\Gamma}&:
\begin{cases}
\Sigma^+ \to \Gamma^+\\
w \mapsto \onehotdec{\Gamma}(\rnnstos{\RNN}(\onehot{\Sigma}(w)))
\end{cases} &
\alphrnnstol{\RNN}{\Sigma}{\Gamma}&:
\begin{cases}
\Sigma^+ \to \Gamma\\
w \mapsto \onehotdec{\Gamma}(\rnnstol{\RNN}(\onehot{\Sigma}(w)))
\end{cases}
\end{align*}

\begin{myremark}[label=rem:altsemrnn]{Alternative Semantics}
If the activation function of $\outLayer$ is $\softmax$,
we may consider the mapping
\[
\alphrnnstol{\RNN}{\Sigma}{\Gamma}^{\mathsf{prob}}:
\begin{cases}
\Sigma^+ \to \mathit{Distr}({\Gamma})\\
w \mapsto \rnnstol{\RNN}(\onehot{\Sigma}(w))
\end{cases}
\]
where $\mathit{Distr}({\Gamma})$ is the set of probability distributions
over $\Gamma$. The idea is that, for 
the vector $\vect{y} = \alphrnnstol{\RNN}{\Sigma}{\Gamma}^{\mathsf{prob}}(w)$,
the probability of selecting $\beta \in \Gamma$ as the
next letter is $y_{\min(\argmax(\onehot{\Gamma}(\beta)))}$. This view has applications in
character-level language modeling or text generation.

\medskip

In the context of reactive systems, which
are supposed to run forever, it is particularly interesting
to consider \emph{infinitary} versions
of $\rnnstos{\RNN}$ and $\rnnstol{\RNN}$, which are of the following form and
whose definitions are as expected:
\begin{align*}
\rnnstos{\RNN}^\omega&: (\Reals^m)^\omega \to (\Reals^{\rnnoutdim})^\omega\\[1ex]
\alphrnnstos{\RNN}{\Sigma}{\Gamma}^\omega&: \Sigma^\omega \to \Gamma^\omega
\end{align*}
\end{myremark}

\paragraph{RNNs as String Classifiers.}
Let $\RNN$ be an RNN and $m = \indim{\RNN}$.
If $\outdim{\outLayer} = 1$, then $\RNN$ can be viewed as
a classifier of strings over $\Reals^m$: For ${\bowtie} \in \{\ge, >, =\}$ and
$\thr \in \Reals$, we define
\[
\rnnLangthr{}{\RNN}{\bowtie}{\thr} = \{w \in (\Reals^m)^+ \mid \rnnstol{\RNN}(w) \bowtie \thr\}
\]
to be the \emph{language} of $\RNN$, i.e., the set of strings that
have a sufficient ``score``, or \emph{acceptance probability},
and are therefore classified by $\RNN$ as positive.\footnote{If the activation function of
$\outLayer$ is the sigmoid function $\sigmoid$, we can also interpret the score as \emph{acceptance probability}.}

Suppose we are given a finite alphabet $\Sigma$.
The RNN $\RNN$ is called an RNN \emph{over} $\Sigma$ if $m=|\Sigma|$.
Like above, $\RNN$ can then take one-hot encoded letters from $\Sigma$ as inputs.
If, again, $\outdim{\outLayer} = 1$,
we can interpret $\RNN$ as a classifier of strings over $\Sigma$.
For ${\bowtie} \in \{\ge, >, =\}$ and $\thr \in \Reals$, we let
\[
\rnnLangthr{\Sigma}{\RNN}{\bowtie}{\thr} = \{w \in \Sigma^+ \mid \rnnstol{\RNN}(\onehot{\Sigma}(w)) \bowtie \thr\}
\]
be the \emph{language} of $\RNN$ over $\Sigma$.

\begin{myexample}{Example Applications of RNNs}
Among the classical practical applications of RNNs, let us mention price prediction, text translation, or text classification.
Several concrete use cases are given in the blog article \cite{karpathy2015}.
Consider, for example, an email spam classifier that processes a text, i.e., a sequence over a finite alphabet $\Sigma$ (where each letter represents a single word), and outputs a probability of being classified as spam. In other words, we are interested in filtering emails contained in a language of the form $\rnnLangthr{\Sigma}{\RNN}{\ge}{\thr}$. Here, it is up to the user to decide on the threshold $\theta$ to achieve a reasonable trade-off between sensitivity (true positive rate) and specificity (true negative rate).
\end{myexample}

\section{Undecidability of the Emptiness Problem}

\newcommand{\prob}{p}

In this section, we demonstrate that, unfortunately, determining the emptiness of the language of a given RNN, when treated as a string classifier, is undecidable.

\begin{mytheorem}[label=thm:emptinessrnn]{Undecidability of RNN Language (Non)emptiness}
For all ${\bowtie} \in \{\ge, >, =\}$, the following decision problem is undecidable:
\begin{boxdescription}
\item[Input:] A finite alphabet $\Sigma$ and a $(\ReLU,\sigmoid)$-RNN $\RNN$ over $\Sigma$ with $\outdim{\RNN} = 1$.
\item[Question:] Do we have $\rnnLangthr{\Sigma}{\RNN}{\bowtie}{\frac{1}{2}} \neq \emptyset$\,?
\end{boxdescription}
\end{mytheorem}

As this fundamental decision problem is already undecidable for
RNNs, this will also be the case for the majority of non-trivial verification tasks.
Thus, it is important to identify suitable abstractions
or restrictions. There has been an ongoing and fruitful effort
to establish positive verification results specifically designed for RNNs
\cite{AkintundeKLP19,JacobyBK20,TranCY0HP23,RyouCBSDV21,KhmelnitskyNRXBBFHLY23}.

The rest of this section is dedicated to the proof of Theorem~\ref{thm:emptinessrnn}.
We first show that $(\ReLU,\sigmoid)$-RNNs can simulate
all \emph{probabilistic finite automata (PFAs)}. PFAs were introduced and studied
by Rabin~\cite{Rabin63}. Their (non)emptiness problem is
undecidable and they are strictly more expressive than finite automata. 

\begin{mydefinition}{Probabilistic Finite Automaton (PFA)}
Let $\Sigma$ be a finite alphabet. A \emph{probabilistic finite automaton (PFA)}
over $\Sigma$ is a tuple $\BA = ((\probmatrix^{\alpha})_{\alpha \in \Sigma},\initvect,\finalvect)$
where, for some $n \in \posNaturals$,
\begin{boxitemize}
\item $\probmatrix^\alpha = (\prob^\alpha_{j,i}) \in ([0,1] \cap \Rationals)^{n \times n}$
such that, for all $i \in \{1,\ldots,n\}$,
$\sum_{j=1}^n p_{j,i}^\alpha = 1$,
\item $\initvect \in \{0,1\}^n$ is the \emph{initial state vector} (a column vector)
such that $\sum_{i=1}^n \initweight_i = 1$,
\item $\finalvect \in \{0,1\}^n$ is the \emph{final state vector} (a row vector).
\end{boxitemize}
\end{mydefinition}

Note that $\probmatrix^\alpha$ is a \emph{left stochastic matrix}.
We call it the \emph{transition-probability matrix} of $\alpha \in \Sigma$.
The intuition is that $\BA$ has an (implicit) set of states
$Q=\{q_1,\ldots,q_n\}$, and $p_{j,i}^\alpha$ is the probability that,
when reading $\alpha$ in state $q_i$, we go to state $q_j$.%
\footnote{We choose $p_{j,i}^\alpha$ rather than the more standard
$p_{i,j}^\alpha$, as it reflects the way computations are represented
in the layers of RNNs, where the transition matrix on the left is multiplied with the current state
on the right-hand side.}
Thus, $\BA$ has an equivalent representation as a \emph{state-transition graph}
$(Q,\init,\Delta,F)$,
which is basically a finite automaton with transition probabilities.
The initial state is $\init = q_i$ for $\initweight_i = 1$, and
the set of final states is $F = \{q_i \mid i \in \{1,\ldots,n\}$ with $\finalweight_i = 1\}$.
Moreover, we have a set of transitions $\Delta \subseteq Q \times \Sigma \times (0,1] \times Q$
containing, for all $\alpha \in \Sigma$ and
all $i,j \in \{1,\ldots,n\}$ such that $\prob^\alpha_{j,i} > 0$, the transition
$(q_i,\alpha,\prob^\alpha_{j,i},q_j)$.
We will switch between these representations at discretion.

PFA $\BA$ defines the mapping
\[
\delta_\BA:
\begin{cases}
\Reals^n \times \Sigma \to \Reals^n\\
(\vect{x}, \alpha) \mapsto \probmatrix^\alpha \cdot \vect{x}\,.
\end{cases}
\]
Suppose that $\delta_\BA$ gets $(\vect{x},\alpha)$ as arguments
and that $\vect{x}$ represents a probability distribution
over $\{q_1,\ldots,q_n\}$,
i.e., $x_i$ is the probability of being in state $q_i$.
Then, letting $\vect{x}' = \delta_\BA(\vect{x},\alpha)$,
$x_i'$ can naturally be interpreted as the probability of
being in state $q_i$ after reading $\alpha$.
Using the extension of $\delta_\BA$ to strings,
we obtain the mapping
\[
\nnfunction{\BA}:
\begin{cases}
\Sigma^\ast \to [0,1]\\
w \mapsto \finalvect \cdot \delta_\BA(\initvect,w)\,.
\end{cases}
\]
Thus, starting from the initial state, $\nnfunction{\BA}(w)$ is the probability of reaching a final state
after reading $w$.
Finally, for ${\bowtie} \in \{\ge, >, =\}$ and $\thr \in [0,1]$, we define the language
\[\Langthr{\BA}{\bowtie}{\thr} = \{w \in \Sigma^+ \mid \nnfunction{\BA}(w) \bowtie \thr\}\,.\]
We restrict to nonempty strings here, aligning with the definition of RNN
languages.

\begin{myremark}{Reactive vs.\ Generative Probabilistic Automata}
PFAs due to Rabin are also called \emph{reactive},
as they yield a probability distribution when
providing an input. On the other hand, \emph{generative}
probabilistic automata (also called Segala automata \cite{Segala95})
generate the next letter according
to a probability distribution over $\Sigma \times Q$.
In formal terms, we may consider that reactive PFAs
(as we study here) are equipped with a transition function
$\delta: Q \times \Sigma \to \mathit{Distr}(Q)$,
whereas generative PFAs have a transition function of type
$\delta: Q \to \mathit{Distr}(\Sigma \times Q)$.
\end{myremark}

\begin{myexample}[label=ex:pfa]{Probabilistic Finite Automaton}
Let $\Sigma = \{\alpha, \beta\}$.
Consider the PFA $\BA = ((\probmatrix^{\alpha}, \probmatrix^{\beta}),\initvect,\finalvect)$
over $\Sigma$ given by
\[
\initvect =
\begin{pmatrix}
1\vspace{1.5ex}\\
0
\end{pmatrix}
\hspace{2em}
\probmatrix^{\alpha} =
\begin{pmatrix}
\frac{1}{3} & \frac{1}{3}\vspace{1.5ex}\\
\frac{2}{3} & \frac{2}{3}
\end{pmatrix}
\hspace{2em}
\probmatrix^{\beta} =
\begin{pmatrix}
0 & 1\vspace{1.5ex}\\
1 & 0
\end{pmatrix}
\hspace{2em}
\finalvect =
\begin{pmatrix}
1 & 0
\end{pmatrix}
\]
Alternatively, $\BA$ can be represented by its state-transition diagram:\bigskip
\begin{center}
\includegraphics[scale=0.5]{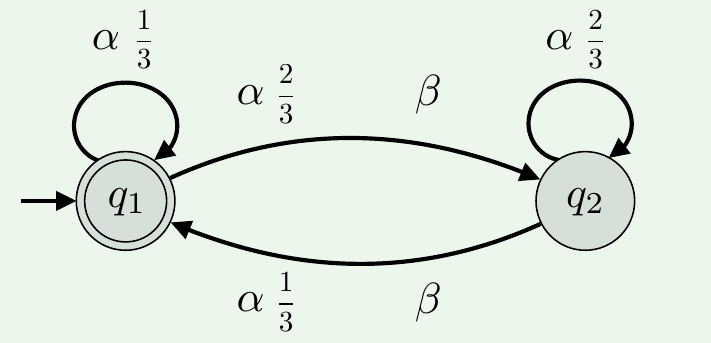}
\end{center}

For $k \in \Naturals$, we have
\[
\nnfunction{\BA}(\alpha\beta^k) =
\begin{pmatrix}
1 & 0
\end{pmatrix}
\cdot
\begin{pmatrix}
0 & 1\vspace{1.5ex}\\
1 & 0
\end{pmatrix}^k
\cdot
\begin{pmatrix}
\frac{1}{3} & \frac{1}{3}\vspace{1.5ex}\\
\frac{2}{3} & \frac{2}{3}
\end{pmatrix}
\cdot
\begin{pmatrix}
1\vspace{1.5ex}\\
0
\end{pmatrix}
=
\begin{cases}
\frac{1}{3} & \text{if~} k \text{ is even}\\[3ex]
\frac{2}{3} & \text{if~} k \text{ is odd}
\end{cases}
\]
It is easy to see that every string ending in $\alpha$ leads to the final state
with probability~$\frac{1}{3}$. Thus, no such string is in $\Langthr{\BA}{\ge}{\frac{1}{2}}$.
Altogether, we obtain
\[\Langthr{\BA}{\ge}{\frac{1}{2}} = \{\beta^k \mid k \in \posNaturals \text{ is even}\} \cup \{w\alpha \beta^k \mid w \in \{\alpha,\beta\}^\ast \text{ and } k \text{ is odd}\}\,.\]
Note that the image of $\nnfunction{\BA}$ is finite since we have
$\nnfunction{\BA}(w) \in \{0,\frac{1}{3},\frac{2}{3},1\}$ for all $w \in \Sigma^+$.
In general, however, this is not the case.
\end{myexample}


PFAs enjoy some essential closure properties:

\begin{mylemma}[label=thm:pfaclosure]{Closure Properties of PFAs}
Let $\Sigma$ be a finite alphabet, $p \in [0,1]$ be rational,
and $\mathcal{A}$ and $\mathcal{B}$ be PFAs over $\Sigma$.
We can effectively construct PFAs $\mathcal{C}_1,\mathcal{C}_2,\mathcal{C}_3$
over $\Sigma$ such that, for all $w \in \Sigma^+$,
\begin{boxitemize}
\item $\nnfunction{\mathcal{C}_1}(w) = 1 - \nnfunction{\mathcal{A}}(w)$,
\item $\nnfunction{\mathcal{C}_2}(w) = p \cdot \nnfunction{\mathcal{A}}(w) + (1 - p) \cdot \nnfunction{\mathcal{B}}(w)$, and
\item $\nnfunction{\mathcal{C}_3}(w) = \nnfunction{\mathcal{A}}(w) \cdot \nnfunction{\mathcal{B}}(w)$.
\end{boxitemize}
\end{mylemma}

\begin{myexercise}{}
Prove Lemma~\ref{thm:pfaclosure}.
\end{myexercise}

The next theorem states that RNNs with a particular set of activation functions are at
least as expressive as PFAs. It constitutes the first step towards showing that
emptiness of RNNs is undecidable.

\begin{mytheorem}[label=thm:pfatornn]{RNNs Recognize All PFA Languages}
Let $\Sigma$ be a finite alphabet and $\BA$ be a PFA over $\Sigma$.
Moreover, let ${\bowtie} \in \{\ge, >, =\}$ and $\thr \in [0,1]$ be rational.
We can effectively construct a $(\ReLU,\sigmoid)$-RNN $\RNN$
over $\Sigma$ such that $\outdim{\RNN} = 1$ and
\[\rnnLangthr{\Sigma}{\RNN}{\bowtie}{\frac{1}{2}}
= \Langthr{\BA}{\bowtie}{\thr}\,.\]
\end{mytheorem}

Before we prove the theorem formally, we illustrate the construction
using an example.

\begin{figure}[t]
\centering
  \begin{subfigure}[b]{0.39\textwidth}
    \centering
    \includegraphics[width=\linewidth]{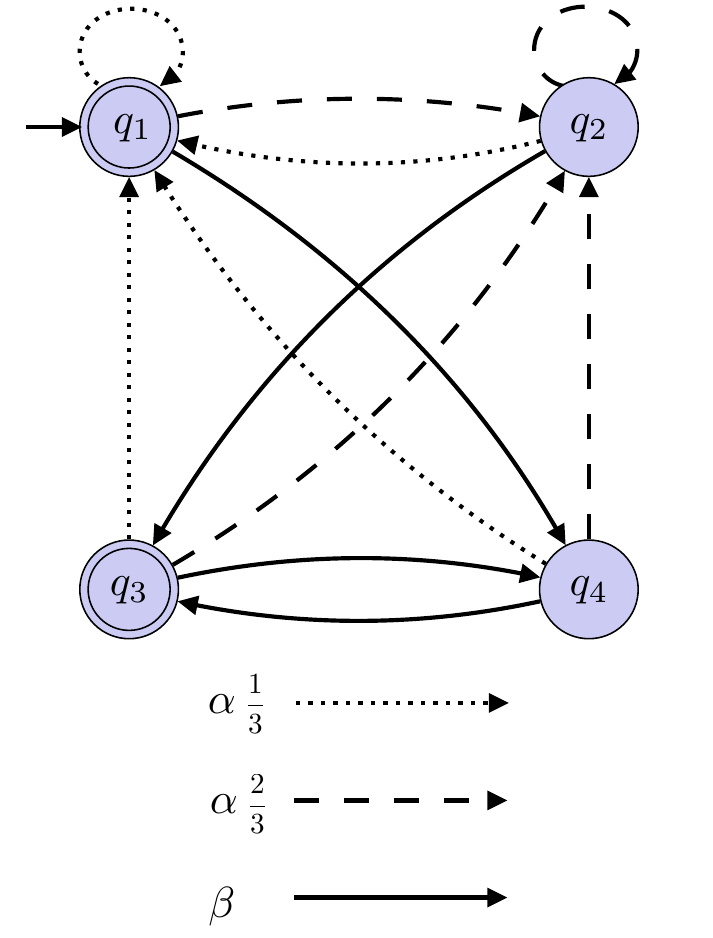}
    \caption{PFA $\BA'$ where every state has a unique ``incoming letter''\label{fig:extpfa}}
  \end{subfigure}~~~~~~~%
  \begin{subfigure}[b]{0.42\textwidth}
    \centering
    \includegraphics[width=\linewidth]{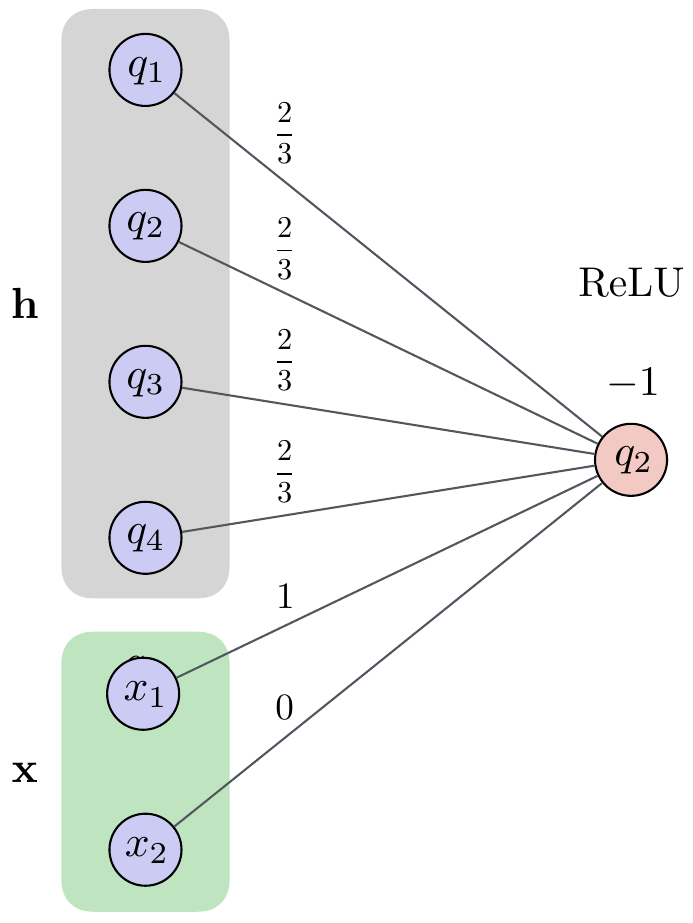}
    \caption{A neuron for target $\alpha$-state $q_2$\label{fig:pfaneuron}}
  \end{subfigure}
  \caption{Simulating a PFA with an RNN}
\end{figure}

\begin{myexample}{From PFAs to RNNs}
An obvious first idea is to choose
$\inLayer$ such that $\indim{\inLayer} = |\Sigma|$ (with
one-hot encoded letters as inputs) and
$\statedim{\inLayer} = |Q|$ as state dimension.
The corresponding matrix $\vect{A}^\sfin$ would
then have dimension $|Q| \times (|Q| + |\Sigma|)$.
However, in $\BA$, we have $|\Sigma| \cdot |Q|^2$
transition probabilities, which thus may not fit into
$\vect{A}^\sfin$.

\medskip

The solution will be to ``augment'' the set of states so that
every state has incoming transitions of a unique letter type.
Consider, for example, the PFA $\BA$ from Example~\ref{ex:pfa}.
Recall that $\Sigma = \{\alpha, \beta\}$
and suppose $\onehot{\Sigma}(\alpha) = (1, 0)^\top$ and
$\onehot{\Sigma}(\beta) = (0, 1)^\top$.
The PFA $\BA'$ illustrated in Figure~\ref{fig:extpfa},
is equivalent to the PFA $\BA$ from Example~\ref{ex:pfa},
i.e., $\nnfunction{\BA}(w) = \nnfunction{\BA'}(w)$ for all $w \in \Sigma^\ast$.
However, every state of $\BA'$ now has only incoming transitions labeled
with a dedicated letter $\alpha$ or $\beta$ (which we accordingly call
an $\alpha$- or $\beta$-state).
Note that $\BA'$ is given by the following ingredients:
\[
\initvect =
\begin{pmatrix}
1\vspace{1ex}\\
0\vspace{1ex}\\
0\vspace{1ex}\\
0
\end{pmatrix}
\hspace{1.5em}
\probmatrix^{\alpha} =
\begin{pmatrix}
\frac{1}{3} & \frac{1}{3} & \frac{1}{3} & \frac{1}{3}\vspace{1ex}\\
\frac{2}{3} & \frac{2}{3} & \frac{2}{3} & \frac{2}{3}\vspace{1ex}\\
0 & 0 & 0 & 0\vspace{1ex}\\
0 & 0 & 0 & 0
\end{pmatrix}
\hspace{1.5em}
\probmatrix^{\beta} =
\begin{pmatrix}
0 & 0 & 0 & 0\vspace{1ex}\\
0 & 0 & 0 & 0\vspace{1ex}\\
0 & 1 & 0 & 1\vspace{1ex}\\
1 & 0 & 1 & 0
\end{pmatrix}
\hspace{1.5em}
\finalvect =
\begin{pmatrix}
1 & 0 & 1 & 0
\end{pmatrix}
\]
Thanks to the transformation, the transition-probability matrices $\probmatrix^{\alpha}$ and $\probmatrix^{\beta}$
no longer ``overlap'':
for all $i,j \in \{1,2,3,4\}$, at most one of $\prob^\alpha_{j,i}$ and $\prob^\beta_{j,i}$ is non-zero.
That is, the sum $\probmatrix^{\alpha}$ and $\probmatrix^{\beta}$ accommodates all probabilities occurring in $\BA'$
in a single matrix of dimension $\Reals^{4 \times 4}$.
The layer $\inLayer$ of the desired RNN is illustrated in Figure~\ref{fig:pfaneuron}
for a neuron belonging to the $\alpha$-state $q_2$.
If the input letter is $\alpha$, i.e., the input vector is $(1, 0)^\top$,
the output neuron indeed receives the updated probability of being in $q_2$
after processing the input. If, on the other hand, the input is $\beta$
and, respectively, $(0, 1)^\top$, then the result before applying $\ReLU$
will be $\le 0$ so that the overall result is $0$. As $q_2$ is an $\alpha$-state,
the probability of being in $q_2$ after processing $\beta$ is indeed $0$.
\end{myexample}

The study of the expressive power of RNNs and their extensions and variants
in terms of formal languages has a long tradition and is still
an active research field \cite{MerrillWGSSY20}.
In this context, a prominent result by Siegelmann and Sontag states that RNNs
can simulate all Turing machines \cite{SiegelmannS95}.
A general technique of transforming finite-state machines
into equivalent RNNs goes back to Minsky \cite{Minsky1954}.
It has recently been generalized to
deterministic \emph{generative} PFAs \cite{SveteC23}, which define
probability distributions over $\Sigma^\ast$ (cf.\ Remark~\ref{rem:altsemrnn}).
Here, we apply this technique to translate (reactive) PFAs into RNNs,
aiming to show undecidability of the
RNN emptiness problem.

Now, let us formally prove~Theorem~\ref{thm:pfatornn}.

\begin{proof}[Proof of Theorem~\ref{thm:pfatornn}]
Let $\Sigma$ be a finite alphabet and $m = |\Sigma|$. Let
$\BA = ((\probmatrix^{\alpha})_{\alpha \in \Sigma},\initvect,\finalvect)$
be a PFA over $\Sigma$,
with state-transition representation $(Q,\init,\Delta,F)$ where
$Q = \{q_1,\ldots,q_n\}$. Moreover, fix ${\bowtie} \in \{\ge, >, =\}$
and a rational number $\thr \in [0,1]$.

Without loss of generality, we assume that,
for every $i_1,i_2,j \in \{1,\ldots,n\}$, $\alpha,\beta \in \Sigma$,
and $\prob_1,\prob_2 \in (0,1]$ such that
$(q_{i_1},\alpha,\prob_1,q_j) \in \Delta$ and
$(q_{i_2},\beta,\prob_2,q_j) \in \Delta$,
we have $\alpha = \beta$.\footnote{If $\BA$ does not
satisfy this property, we can convert it as follows.
Choose some $\beta_0 \in \Sigma$, replace $Q$ by
$Q \times \Sigma$, and take $(\init,\beta_0)$ as initial state.
For all $(q_i,\alpha,\prob,q_j) \in \Delta$ and
$\beta \in \Sigma$, add the transition
$((q_i,\beta),\alpha,\prob,(q_j,\alpha))$.
The final states are $F \times \Sigma$. Thus, after reading a
nonempty prefix ending in $\alpha$, the second component records
that last input letter; the first component has exactly the
same distribution as in $\BA$. Hence the resulting PFA is
equivalent to $\BA$ on all words.}
We call state $q_j$ whose incoming transitions all carry letter
$\alpha$ an $\alpha$-\emph{state}.
Note that the transition probability matrices are not ``overlapping'',
i.e., for all $\alpha, \beta \in \Sigma$ and $i,j \in \{1,\ldots,n\}$
such that $\prob^\alpha_{j,i} > 0$ and $\prob^\beta_{j,i} > 0$,
we have $\alpha = \beta$.

%

We now formally construct the corresponding
$(\ReLU,\sigmoid)$-RNN $\RNN=(\inLayer, \outLayer, \rnninit)$
over $\Sigma$ such that $\outdim{\outLayer} = 1$ and $\rnnLangthr{\Sigma}{\RNN}{\bowtie}{\frac{1}{2}} = \Langthr{\BA}{\bowtie}{\thr}$.

\begin{itemize}
\item We let $\rnninit=\initvect$.

\item Define $\inLayer=(\wmatrix{W}^\sfin, \bias{b}^\sfin, \ReLU)$ as follows:
Let $\wmatrix{W}^\sfin = \matrixconc{\vect{P}}{\vect{C}}$ be the
concatenation of the matrices $\vect{P} \in [0,1]^{n \times n}$ and
$\vect{C} \in \{0,1\}^{n \times m}$ defined by
\[
\vect{P} = \sum_{\alpha \in \Sigma} \probmatrix^\alpha
\]
and, for all $j \in \{1,\ldots,n\}$ and $k \in \{1,\ldots,m\}$,
\[
c_{j,k} =
\begin{cases}
1 & \begin{array}[t]{@{}l@{}}
\text{if there is } \alpha \in \Sigma \text{ such that } q_j \textup{ is an } \alpha\textup{-state}\\[-0.5ex]
\text{and } \min(\argmax(\onehot{\Sigma}(\alpha))) = k
\end{array}\\[4ex]
0 & \text{otherwise.} 
\end{cases}
\]
Finally, for all $j \in \{1,\ldots,n\}$, we set $b_j^\sfin = -1$.

\item We define $\outLayer=(\wmatrix{W}^\sfout, \bias{b}^\sfout, \sigmoid)$ by $\wmatrix{W}^\sfout = \finalvect$ and $\bias{b}^\sfout = (-\thr)$.
\end{itemize}

\paragraph{Correctness Proof.}

Let us show that, for all hidden states $\rnnstate \in [0,1]^n$ and
$\alpha \in \Sigma$,
$\delta_{\RNN}(\rnnstate, \onehot{\Sigma}(\alpha)) =
\delta_{\BA}(\rnnstate, \alpha)$.
Let $\rnninvect = \onehot{\Sigma}(\alpha)$. We have
\begin{align*}
& \delta_{\RNN}(\rnnstate, \onehot{\Sigma}(\alpha)) = \delta_{\RNN}(\rnnstate, \rnninvect)
= \nnfunction{\inLayer}(\vertconc{\rnnstate}{\rnninvect})\\[1ex]
=\;& \ReLU (\wmatrix{W}^\sfin \cdot (\vertconc{\rnnstate}{\vect{\realx}}) + \vect{b}^\sfin)\\[1ex]
=\;& \ReLU ((\matrixconc{\vect{P}}{\vect{C}}) \cdot (\vertconc{\rnnstate}{\vect{\realx}}) + \vect{b}^\sfin)\\[1ex]
=\;& \ReLU (\vect{P} \cdot \rnnstate + \vect{C} \cdot \vect{x} + \vect{b}^\sfin)\\[1ex]
=\;& \ReLU ((\textstyle\sum_{\beta \in \Sigma} \probmatrix^\beta) \cdot \rnnstate + \vect{C} \cdot \vect{x} + \vect{b}^\sfin)
\end{align*}
Let $\vect{y} = (y_1,\ldots,y_n)^\top = \vect{C} \cdot \vect{x} + \vect{b}^\sfin$.
For all $j \in \{1,\ldots,n\}$, we have
\[
y_{j} =
\begin{cases}
0 & \text{if } q_j \textup{ is an } \alpha\textup{-state}\\
-1 & \text{otherwise.} 
\end{cases}
\]
Thus, letting $\vect{z} = (\sum_{\beta \in \Sigma} \probmatrix^\beta) \cdot \rnnstate + \vect{y} \in \Reals^n$, we get, for all $j \in \{1,\ldots,n\}$,
\[
\begin{cases}
z_{j}= \probmatrix^\alpha_{j} \cdot \rnnstate & \text{if } q_j \textup{ is an } \alpha\textup{-state}\\
z_{j} \le 0 & \textup{otherwise.}
\end{cases}
\]
where $\probmatrix^\alpha_{j}$ is the $j$-th row of $\probmatrix^\alpha$. We conclude 
\[
\delta_{\RNN}(\rnnstate, \onehot{\Sigma}(\alpha)) =
\ReLU (\vect{z}) =
\probmatrix^\alpha \cdot \rnnstate =
\delta_\BA(\rnnstate, \alpha)\,.
\]
With this, we obtain that, for all $w \in \Sigma^+$,
\begin{align*}
& \rnnstol{\RNN}(\onehot{\Sigma}(w)) \bowtie \textstyle\frac{1}{2}\\[1ex]
\text{iff}\;& \nnfunction{\outLayer}(\delta_\RNN(\rnninit,\onehot{\Sigma}(w))) \bowtie \textstyle\frac{1}{2}\\[1ex]
\text{iff}\;& \nnfunction{\outLayer}(\delta_\BA(\initvect,w)) \bowtie \textstyle\frac{1}{2}\\[1ex]
\text{iff}\;& \sigmoid(\finalvect \cdot \delta_\BA(\initvect,w) - \thr) \bowtie \textstyle\frac{1}{2}\\[1ex]
\text{iff}\;& \finalvect \cdot \delta_\BA(\initvect,w) - \thr \bowtie 0\\[1ex]
\text{iff}\;& \finalvect \cdot \delta_\BA(\initvect,w) \bowtie \thr\\[1ex]
\text{iff}\;& \nnfunction{\BA}(w) \bowtie \thr
\end{align*}
We have shown $\rnnLangthr{\Sigma}{\RNN}{\bowtie}{\frac{1}{2}} = \Langthr{\BA}{\bowtie}{\thr}$.
\end{proof}


To prove Theorem~\ref{thm:emptinessrnn}, i.e., undecidability of RNN language emptiness,
it remains to establish the corresponding facts for PFAs, depending on ${\bowtie} \in \{\ge, >, =\}$.

The following undecidability results for PFAs are due to \cite{Bertoni74} and \cite{Paz71}.
The proofs were later simplified and strengthened in \cite{GimbertO10}.
For a concise overview of what is decidable and undecidable in PFAs,
we refer to \cite{Fijalkow17}.

\begin{mytheorem}[label=thm:emptinesspfaeq]{Undecidability of PFA Language Emptiness \cite{Bertoni74,Paz71}}
The following three decision problems are undecidable:
\begin{boxdescription}
\item[Input:] A finite alphabet $\Sigma$ and a PFA $\BA$ over $\Sigma$.
\item[Question 1:] Do we have $\Langthr{\BA}{=}{\frac{1}{2}} \neq \emptyset$~ (i.e., $\nnfunction{\BA}(w) = \frac{1}{2}$ for some $w \in \Sigma^+$)?
\item[Question 2:] Do we have $\Langthr{\BA}{\ge}{\frac{1}{4}} \neq \emptyset$\,?
\item[Question 3:] Do we have $\Langthr{\BA}{>}{\frac{1}{8}} \neq \emptyset$\,?
\end{boxdescription}
\end{mytheorem}


\newcommand{\morph}{f}
\newcommand{\mletter}{\alpha}
\newcommand{\malph}{\Sigma}
\newcommand{\binenc}[1]{\overline{#1}}

\begin{proof}
We first consider Question 1. The proof is a reduction from the following modified Post's correspondence
problem (modified PCP)\footnote{The standard undecidable PCP does not have the restriction $\morph_i(\mletter) \in 1(0 + 1)^\ast$.
However, when we start with an unrestricted instance of the form $g_1,g_2: \malph \to \{0,1\}^+$, we can translate it into $\morph_1,\morph_2: \malph \to \{0,1\}^\ast$ defined by $\morph_i = f \circ g_i$ where $f: \{0,1\}^\ast \to \{0,1\}^\ast$
is the morphism given by $f(0) = 10$ and $f(1) = 11$. Then, $\morph_i(\alpha)$ starts with $1$ for all $\alpha \in \malph$. Moreover, we easily see that $g_1(w) = g_2(w)$ for some $w \in \malph^+$ iff $\morph_1(w) = \morph_2(w)$ for some $w \in \malph^+$}.
\begin{boxdescription}
\item[Input:] A finite alphabet $\malph$ and morphisms $\morph_1,\morph_2: \malph^\ast \to \{0,1\}^\ast$ such that $\morph_i(\mletter) \in 1(0 + 1)^\ast$ for all $i \in \{1,2\}$ and $\mletter \in \malph$.
\item[Question:] Is there $w \in \malph^+$ such that $\morph_1(w) = \morph_2(w)$?
\end{boxdescription}

Given an instance $\morph_1,\morph_2$ of the modified PCP,
we will effectively construct a PFA $\BA$ over $\malph$ such that, for all
$w \in \malph^+$, we have $\morph_1(w) = \morph_2(w)$ iff
$\nnfunction{\BA}(w) = \frac{1}{2}$, which implies the theorem.

Let $\binenc{\epsilon} = 0$ and, for $u_1 \ldots u_k \in \{0,1\}$ with $k \ge 1$,
\begin{align*}
\binenc{u_1 \ldots u_k} &= 0.u_{k}u_{k-1} \ldots u_1 \textup{ (in binary)}\\[1ex]
&= \frac{u_k}{2^1} + \frac{u_{k-1}}{2^2} + \ldots + \frac{u_1}{2^k}\,.
\end{align*}
Thanks to the modified PCP, we have, for all $w \in \malph^\ast$,
$\morph_1(w) = \morph_2(w)$ iff $\binenc{\morph_1(w)} = \binenc{\morph_2(w)}$.

Towards the PFA $\BA$, we will construct two PFAs $\BA_1$ and $\BA_2$ over $\malph$
such that, for all $i \in \{1,2\}$ and $w \in \malph^\ast$,
we get
\begin{align}
\nnfunction{\BA_i}(w) = \binenc{\morph_i(w)}\label{align:PFAone}\,.
\end{align}
We then combine $\BA_1$ and $\BA_2$ using Lemma~\ref{thm:pfaclosure}
and obtain a PFA $\BA$ over $\Sigma$ such that, for all $w \in \malph^\ast$,
\[
\nnfunction{\BA}(w) = \frac{1}{2}(\nnfunction{\BA_1}(w) + (1 - \nnfunction{\BA_2}(w)))\,.
\]
Then, we are done as $\nnfunction{\BA}(w) = \frac{1}{2}$ iff $\binenc{\morph_1(w)} = \binenc{\morph_2(w)}$ iff
$\morph_1(w) = \morph_2(w)$.

\newcommand{\bword}{\nu}
\newcommand{\calB}{\mathcal{B}}

\paragraph{PFA Evaluating Binary Numbers.}

The main building block in the construction of both $\BA_1$ and $\BA_2$ will be a PFA $\calB = ((\probmatrix^0, \probmatrix^1),\initvect,\finalvect)$ over $\{0,1\}$ such that, for all $\bword \in \{0,1\}^\ast$, $\nnfunction{\calB}(\bword) = \binenc{\bword}$.\footnote{Note that $\probmatrix^0$ and $\probmatrix^1$ should not be confused with powers of some matrix $\probmatrix$.}
Thus, $\calB$ ``evaluates'' $\bword$. It is given by
\[
\initvect =
\begin{pmatrix}
1\vspace{1.5ex}\\
0
\end{pmatrix}
\hspace{2em}
\probmatrix^0 =
\begin{pmatrix}
  1 & \frac{1}{2}\vspace{1.5ex}\\
  0 & \frac{1}{2}
\end{pmatrix}
\hspace{2em}
\probmatrix^1 =
\begin{pmatrix}
\frac{1}{2} & 0\vspace{1.5ex}\\
\frac{1}{2} & 1
\end{pmatrix}
\hspace{2em}
\finalvect =
\begin{pmatrix}
0 & 1
\end{pmatrix}
\]

We show $\nnfunction{\calB}(\bword) = \binenc{\bword}$, by induction on $k = |\bword|$. The claim clearly holds for $k = 0$,
i.e., $\bword = \epsilon$.
Moreover, for the case $k = 1$,
\begin{center}
\begin{tabular}{rl}
  & $\nnfunction{\calB}(0) = \prob^0_{2,1} = 0.0 = \binenc{0}$\vspace{1.5ex}\\
  and &  $\nnfunction{\calB}(1) = \prob^1_{2,1} = 0.1 = \binenc{1}$
\end{tabular}
\end{center}
Now suppose, for $\bword = u_1 \ldots u_k$ with $k \ge 1$, that $\nnfunction{\calB}(\bword) =
\binenc{\bword}$ holds. Moreover, assume
\[
\delta_\calB(\initvect,\bword) =
\begin{pmatrix}
  1-p\vspace{1.5ex}\\
  p
\end{pmatrix}
\]
for suitable $p \in [0,1]$. In particular,
\begin{align}
p = \nnfunction{\calB}(\bword) = \binenc{\bword}\,. \label{align:encbword}
\end{align}
Let
$u \in \{0,1\}$. We have
\begin{align*}
  \nnfunction{\calB}(u_1 \ldots u_ku) &=
  \begin{pmatrix}
	0 & 1
  \end{pmatrix} \cdot
  \probmatrix^u \cdot
\begin{pmatrix}
  1-p\vspace{1.5ex}\\
  p
\end{pmatrix}
=
\begin{cases}
  \dfrac{p}{2} & \textup{if } u=0\\[4ex]
  \dfrac{1}{2} + \dfrac{p}{2} & \textup{if } u=1
\end{cases}\vspace{4ex}\\
&= \dfrac{u}{2} + \dfrac{p}{2}
\stackrel{(\ref{align:encbword})}{=}  0.u ~+~ 0.0u_k \ldots u_1
 = 0.uu_k \ldots u_1 ~=~ \binenc{u_1 \ldots u_ku}\\
\end{align*}

\paragraph{Construction of $\BA$.}

Let $i \in \{1,2\}$. Building on the PFA $\calB = ((\probmatrix^0, \probmatrix^1),\initvect,\finalvect)$, we now construct
the PFA $\BA_i = ((\probmatrix^{\alpha})_{\alpha \in \Sigma},\initvect,\finalvect)$
over $\Sigma$ such that, for all strings $w \in \Sigma^\ast$, we have $\nnfunction{\BA_i}(w) = \binenc{\morph_i(w)}$.
For $\bword = u_1 \ldots u_k$ with $k \ge 1$, define 
$\probmatrix^\bword = \probmatrix^{u_k} \cdot \ldots \cdot \probmatrix^{u_1}$.
With this, given a letter $\alpha \in \Sigma$, we let
$\probmatrix^\alpha = \probmatrix^{\morph_i(\alpha)}$.
Indeed, for all $w = \alpha_1 \ldots \alpha_\ell \in \Sigma^\ast$, we obtain
\begin{align*}
\nnfunction{\BA_i}(w) &= \finalvect \cdot \probmatrix^{\alpha_\ell} \cdot \ldots \cdot \probmatrix^{\alpha_1} \cdot \initvect\\
&= \finalvect \cdot \probmatrix^{\morph_i(\alpha_\ell)} \cdot \ldots \cdot \probmatrix^{\morph_i(\alpha_1)} \cdot \initvect\\
&= \finalvect \cdot \probmatrix^{\morph_i(w)} \cdot \initvect = \binenc{\morph_i(w)}\,.
\end{align*}


This concludes the proof of undecidability of the first problem.

\paragraph{Questions 2 and 3.}

Question 1 can be reduced to Question 2 as follows:
From the given PFA $\BA$, we construct, using Lemma~\ref{thm:pfaclosure}, a PFA $\mathcal{B}$ over $\Sigma$
such that, for all $w \in \Sigma^\ast$, we have
\[\nnfunction{\mathcal{B}}(w) = \nnfunction{\BA}(w) \cdot (1 - \nnfunction{\BA}(w))\,.\]
Note that $\frac{1}{4}$ is the global maximum of the function $r \mapsto r \cdot (1 - r)$,
which is only reached for the argument $r = \frac{1}{2}$.
Thus, we have $\nnfunction{\mathcal{B}}(w) \ge \frac{1}{4}$ iff
$\nnfunction{\BA}(w) = \frac{1}{2}$.

Undecidability of Question 3 can be obtained by
analyzing the transition probabilities in the automata considered so far.
We refer the reader to \cite{GimbertO10}.
\end{proof}


We have now shown Theorem~\ref{thm:emptinessrnn},
using the effective constructions from Theorems~\ref{thm:pfatornn} and \ref{thm:emptinesspfaeq}
to reduce PFA emptiness problems for ${\bowtie} \in \{\ge, >, =\}$
to RNN emptiness. Specifically, for every PFA $\BA$ over $\Sigma$, we constructed
$(\ReLU,\sigmoid)$-RNNs $\RNN_1,\RNN_2,\RNN_3$
over $\Sigma$ whose output layers have output dimension $1$ and such that
the following hold:
\begin{align*}
\rnnLangthr{\Sigma}{\RNN_1}{=}{\frac{1}{2}} &= \Langthr{\BA}{=}{\frac{1}{2}}\\
\rnnLangthr{\Sigma}{\RNN_2}{\ge}{\frac{1}{2}} &= \Langthr{\BA}{\ge}{\frac{1}{4}}\\
\rnnLangthr{\Sigma}{\RNN_3}{>}{\frac{1}{2}} &= \Langthr{\BA}{>}{\frac{1}{8}}
\end{align*}

\begin{myexercise}{}
Explicitly determine a $(\ReLU,\sigmoid)$-RNN $\RNN$ over $\Sigma=\{a\}$ with $\outdim{\RNN} = 1$ such that
$\rnnLangthr{\Sigma}{\RNN}{\ge}{\frac{1}{2}} = \{(aa)^n \mid n \in \posNaturals\}$.
\end{myexercise}

\begin{myexercise}{}
For (local) activation functions $f,g: \Reals \to \Reals$, we define the decision problem
$\textsc{RNN-NonEmptiness}(f,g)$ as follows:
\smallskip
\begin{boxdescription}\itemsep=0.5ex
\item[Input:] A finite alphabet $\Sigma$ and an $(f,g)$-RNN $\RNN$ over $\Sigma$ with $\outdim{\RNN} = 1$.
\item[Question:] Do we have $\smash{\rnnLangthr{\Sigma}{\RNN}{\ge}{\frac{1}{2}} \neq \emptyset}$\,?
\end{boxdescription}
\bigskip
\begin{itemize}\itemsep=1ex
\item[(a)] Show that $\textsc{RNN-NonEmptiness}(\ReLU,\ReLU)$ is undecidable.
\item[(b)] Show that $\textsc{RNN-NonEmptiness}(\heaviside,\ReLU)$ is decidable.
\end{itemize}
\end{myexercise}

%
%

%
%

\chapter{Attention and Transformers}

\newcommand{\TF}{\mathcal{T}}
\newcommand{\dmodel}{n}
\newcommand{\dkeyvalue}{{n_{\mathsf{kv}}}}
\newcommand{\dkey}{{n_{\mathsf{key}}}}
\newcommand{\dvalue}{{n_{\mathsf{value}}}}
\newcommand{\emb}{\mathsf{emb}}
\newcommand{\posenc}{\mathsf{PE}}
\newcommand{\wordemb}{\mathsf{WE}}
\newcommand{\AttLayer}{\mathscr{A}}
\newcommand{\SelfLayer}{\mathscr{A}_{\mathsf{Self}}}
\newcommand{\MaskedLayer}{\mathscr{A}_{\mathsf{Masked}}}
\newcommand{\CrossLayer}{\mathscr{A}_{\mathsf{Cross}}}
\newcommand{\EncLayer}{\mathcal{E}}
\newcommand{\DecLayer}{\mathcal{D}}
\newcommand{\FFLayer}{\mathscr{F}}

\newcommand{\Key}{\vect{K}}
\newcommand{\Query}{\vect{Q}}
\newcommand{\Value}{\vect{V}}
\newcommand{\seqlength}{\ell}
\newcommand{\seqind}{i}
\newcommand{\nheads}{d}
\newcommand{\ntflayers}{\kappa}
\newcommand{\nenclayers}{k}
\newcommand{\AttHead}{\mathcal{H}}
\newcommand{\AttMatrix}{\vect{W}}
\newcommand{\FFMatrix}{\vect{F}}
\newcommand{\Norm}{\mathit{Norm}}

\newcommand{\Encoder}{\mathcal{E}}
\newcommand{\Decoder}{\mathcal{D}}
\newcommand{\selfsem}[1]{\nnfunction{#1}_{\mathsf{self}}}
\newcommand{\maskedsem}[1]{\nnfunction{#1}_{\mathsf{masked}}}
\newcommand{\crosssem}[1]{\nnfunction{#1}}

\newcommand{\AttFunct}{\mathit{Weights}}

Transformers have been introduced by Vaswani et al.\ as a powerful alternative to RNNs and their variants \cite{VaswaniSPUJGKP17}.
Like RNNs, transformers can be used as \emph{sequence-to-sequence} transducers, as they
process sequences of arbitrary length.
For language-recognition tasks, we will consider only the encoder (or decoder) part, as is the case in
popular language-model architectures like BERT and GPT.

Verification issues for transformers have only been addressed sparingly so far.
The main purpose of this chapter is to highlight a few interesting questions for future research.


\section{Attention}

An essential component of a transformer is \emph{attention}, specifically, an \emph{attention head}.
There is an analogy between attention heads and
channels of convolutional neural networks. Both allow the network to focus on different aspects and parts of an input.
Therefore, one usually has several attention heads per layer.

\begin{mydefinition}{Attention Head}
An attention head $\AttHead = (\Query, \Key, \Value)$ is given by three matrices
(i.e., linear transformations)
$\Query, \Key \in \Rationals^{\dkey \times \dmodel}$
and $\Value \in \Rationals^{\dvalue \times \dmodel}$
for some $\dmodel, \dkey, \dvalue \in \posNaturals$. They are respectively called
\emph{query}, \emph{key}, and \emph{value matrix}.
\end{mydefinition}

We let $\indim{\AttHead} = n$, $\outdim{\AttHead} = \dvalue$, and
$\nndim{\AttHead} = (n, \dvalue)$.
Later, we will see that $n$ can be understood as the dimension of the
\emph{hidden state} of a transformer.
Attention head $\AttHead$ defines a mapping
\[
\nnfunction{\AttHead}:
\begin{cases}
(\Reals^\dmodel)^+ \times \Reals^\dmodel \to \Reals^\dvalue\\
(\vect{z}^{(1)} \ldots \vect{z}^{(\seqlength)}, \vect{x}) \mapsto \vect{y}
\end{cases}
\]
which allows it to situate a vector (word/letter embedding or hidden state) inside a whole sequence.
Let $\vect{q} = \Query \cdot \vect{x}$.
Moreover, for $i \in \{1,\ldots,\ell\}$, let
$\vect{k}^{(i)} = \Key \cdot \vect{z}^{(i)}$ and $\vect{v}^{(i)} = \Value \cdot \vect{z}^{(i)}$.
The semantics $\vect{y} = \nnfunction{\AttHead}(\vect{z}^{(1)} \ldots \vect{z}^{(\seqlength)}, \vect{x})$ is then given by
\[
\vect{y} = \sum_{i=1}^{\seqlength} p_{i} \cdot \vect{v}^{(i)}
\]
where
$(p_{1}, \ldots, p_{\seqlength}) = \AttFunct(a_{1}, \ldots, a_{\seqlength})$
with
\[
a_{i} = \frac{1}{\sqrt{\dkey}} \cdot (\vect{q}^\top \cdot \vect{k}^{(i)})\,.
\]
The scaling parameter $\frac{1}{\sqrt{\dkey}}$ is optional.
The function $\AttFunct: \Reals^+ \to \Reals^+$ is a length-preserving
\emph{weight function}. Usually, one chooses
$\AttFunct = \softmax^\ast\colon \Reals^+ \to \Reals^+$, where
the latter is $\softmax$ adapted for sequences, i.e., for a variable number of input arguments,
instead of a fixed number of input values.

For theoretical considerations, one sometimes replaces $\softmax^\ast$ by
other weight functions, in particular:
\begin{itemize}
\item $\textup{min-argmax}^\ast \colon \Reals^+ \to \{0,1\}^+$,
also called \emph{leftmost-hard attention}, and

\item $\textup{avg-argmax}^\ast \colon \Reals^+ \to \Reals^+$,
called \emph{average-hard attention}.
\end{itemize}

Here, $\textup{min-argmax}^\ast(x_1,\ldots,x_\ell) = (b_1,\ldots,b_\ell)$
where $b_i = 1$ iff $i = \min(\argmax(x_1,\ldots,x_\ell))$.
For example, $\textup{min-argmax}^\ast(3,7,4,7) = (0,1,0,0)$. Moreover,
$\textup{avg-argmax}^\ast$ takes the average for all maximal elements, i.e.,
$\textup{avg-argmax}^\ast(x_1,\ldots,x_\ell) = (b_1,\ldots,b_\ell)$
where \[b_i = \frac{1}{|\argmax(x_1,\ldots,x_\ell)|}\] for all
$i \in \argmax(x_1,\ldots,x_\ell)$, and $b_i = 0$ for all other $i$.
For example, we have $\textup{avg-argmax}^\ast(3,7,4,7) = (0, 0.5, 0, 0.5)$.

\begin{figure}[t]
\centering
\includegraphics[scale=0.46]{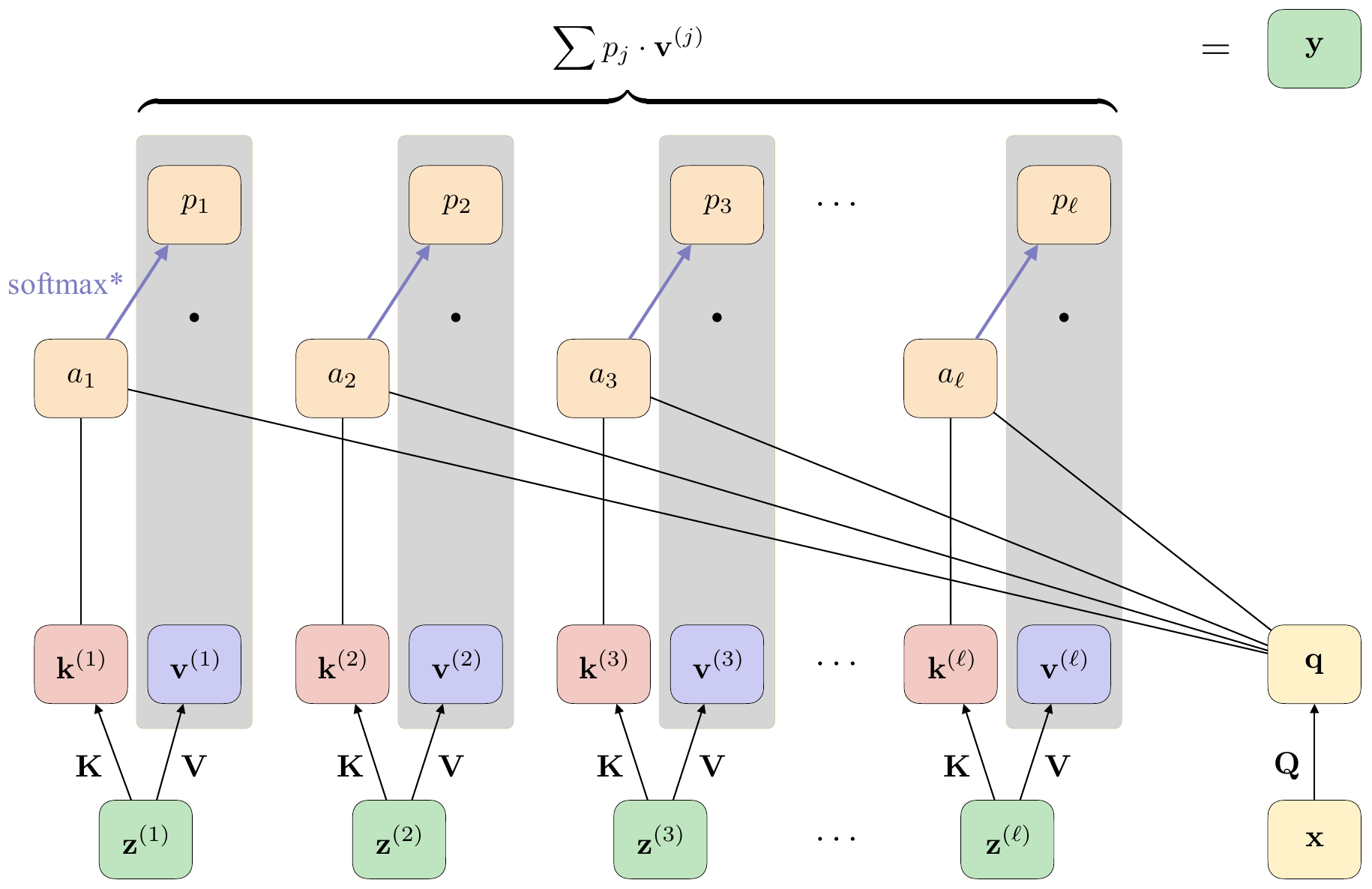}
\caption{Illustration of an attention head $\AttHead = (\Query, \Key, \Value)$
\label{fig:attentionhead}}
\end{figure}

The working principle of attention heads is illustrated in Figure~\ref{fig:attentionhead}.

In the following, we develop an attention head that will
later serve as a building block of transformers
(cf.\ Examples~\ref{ex:argmaxtransformer} and~\ref{ex:tfsorted}).

\begin{myexercise}[label=exercise:attentionhead]{Attention Head Computing the Maximum}
Define an attention head
$\AttHead_\mathsf{max} = (\Query, \Key, \Value)$, using $\textup{avg-argmax}^\ast$ as weight function
and with parameters $\dmodel = 3$ and $\dkey = \dvalue = 1$,
such that $\nnfunction{\AttHead_{\mathsf{max}}}: (\Reals^3)^+ \times \Reals^3 \to \Reals$
where, for all $z_1,\ldots,z_\ell, x \in \Reals$,
\[
\nnfunction{\AttHead_{\mathsf{max}}}((z_1,0,1)^\top \ldots (z_\ell,0,1)^\top, (x,0,1)^\top)
= \max\{z_1,\ldots,z_\ell\}\,.
\]
\end{myexercise}

\begin{mysolution}{}
Note that $\nndim{\AttHead_\mathsf{max}} = (3, 1)$.
The matrices can be chosen as follows:
\[
\Query =
\begin{pmatrix}
0 & 0 & 1
\end{pmatrix}
\hspace{4em}
\Key =
\Value =
\begin{pmatrix}
1 & 0 & 0
\end{pmatrix}
\]
This actually also works with $\textup{min-argmax}^\ast$ instead of
$\textup{avg-argmax}^\ast$.
An illustration of $\AttHead_\mathsf{max}$ can be found in
Figure~\ref{fig:tfargmax}.
\end{mysolution}

Several attention heads can be combined to form layers.

\begin{mydefinition}{Multi-Head Attention Layer}
A \emph{(multi-head) attention layer} $\AttLayer = (\AttHead^{(1)}, \ldots, \AttHead^{(\nheads)}, \AttMatrix)$ has $\nheads \ge 1$ attention heads $\AttHead^{(i)} = (\Query^{(i)}, \Key^{(i)}, \Value^{(i)})$ and one additional linear transformation in terms of a matrix $\AttMatrix$. We require that all attention heads share the same dimensions $\dmodel, \dkey, \dvalue \in \posNaturals$
and that $\AttMatrix \in \Rationals^{\dmodel \times (\nheads \,\cdot\, \dvalue)}$.
\end{mydefinition}

\begin{figure}[t]
\centering
\includegraphics[scale=0.4]{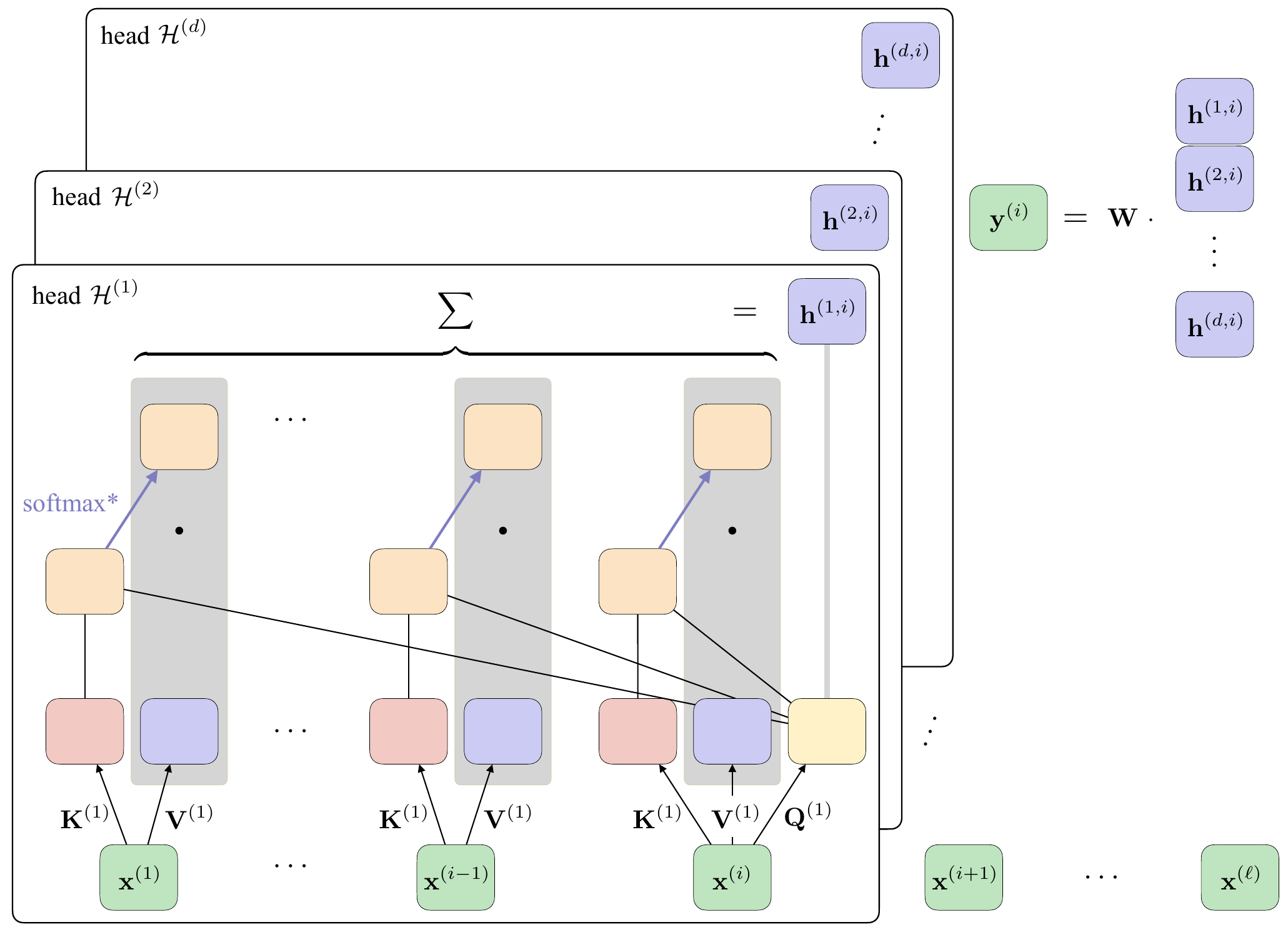}
\caption{A multi-head attention layer with masked self-attention semantics
\label{fig:maskedattention}}
\end{figure}

We let $\indim{\AttLayer} = \outdim{\AttLayer} = n$ and $\nndim{\AttLayer} = (n, n)$.
We can assign to $\AttLayer$ three different semantics:
\begin{boxitemize}
\item The \emph{(cross-)attention semantics} is given by
\[
\nnfunction{\AttLayer}:
\begin{cases}
(\Reals^\dmodel)^+ \times (\Reals^\dmodel)^+ \to (\Reals^\dmodel)^+\\
(w, \vect{x}^{(1)} \ldots \vect{x}^{(\inlength)}) \mapsto \vect{y}^{(1)} \ldots \vect{y}^{(\inlength)}
\end{cases}
\]
where
\[
\vect{y}^{(\seqind)} =
\AttMatrix \cdot (\AttHead^{(1)}(w, \vect{x}^{(\seqind)}) \vertconcwo \ldots \vertconcwo \AttHead^{(\nheads)}(w, \vect{x}^{(\seqind)}))\,.
\]
\item The \emph{self-attention semantics} is defined by
\[
\selfsem{\AttLayer}:
\begin{cases}
(\Reals^\dmodel)^+ \to (\Reals^\dmodel)^+\\
w \mapsto \crosssem{\AttLayer}(w,w)\,.
\end{cases}
\]
\item Finally, the \emph{masked self-attention semantics} is defined by
\[
\maskedsem{\AttLayer}:
\begin{cases}
(\Reals^\dmodel)^+ \to (\Reals^\dmodel)^+\\
\vect{x}^{(1)} \ldots \vect{x}^{(\inlength)} \mapsto \vect{y}^{(1)} \ldots \vect{y}^{(\inlength)}
\end{cases}
\]
where (letting $w_i=\vect{x}^{(1)} \ldots \vect{x}^{(\seqind)}$)
\[\vect{y}^{(\seqind)} =
\AttMatrix \cdot (\AttHead^{(1)}(w_i, \vect{x}^{(\seqind)}) \vertconcwo \ldots \vertconcwo \AttHead^{(\nheads)}(w_i, \vect{x}^{(\seqind)}))\,.
\]
\end{boxitemize}
Note that $\selfsem{\AttLayer}$ and $\maskedsem{\AttLayer}$ are length-preserving, and
$\crosssem{\AttLayer}$ is length-preserving in the second argument.
Masked self-attention is illustrated in Figure~\ref{fig:maskedattention}.

\begin{myexample}[label=ex:attentionlayer]{Attention Layer}
We continue Exercise~\ref{exercise:attentionhead}.
Let $\AttHead_{\mathsf{max}}$ be the attention head developed there.
We obtain a (single-head) attention layer $\AttLayer_{\mathsf{max}}$ when we add
the matrix $\AttMatrix_{\mathsf{max}} = (0, 1, 0)^\top \in \Rationals^{3 \times 1}$,
which writes the result delivered by $\AttHead_{\mathsf{max}}$
into the second component of the three-dimensional zero-vector.
For an illustration, consider Figure~\ref{fig:tfupper}.
\end{myexample}

\section{The Transformer Architecture}

\paragraph{Encoder Layer.}

An \emph{encoder layer} is of the form $\EncLayer = (\AttLayer, \NN)$. It has two components, a multi-head attention layer $\AttLayer$ (with self-attention semantics) and a feed-forward neural network $\NN$ with $\nndim{\AttLayer} = \nndim{\NN} = (n, n)$ for some $n \in \posNaturals$.\footnote{Typically (but not mandatorily), $\NN$ is a two-layer neural network with $\ReLU$ and $\idactivation$ as activation functions, respectively.}
Abusing notation, $\nnfunction{\NN}$ can be extended to a mapping $(\Reals^\dmodel)^+ \to (\Reals^{\dmodel})^+$ letting $\nnfunction{\NN}(\vect{x}^{(1)} \ldots \vect{x}^{(\seqlength)})
= \nnfunction{\NN}(\vect{x}^{(1)}) \ldots \nnfunction{\NN}(\vect{x}^{(\seqlength)})$.
We can now define $\nnfunction{\EncLayer}: (\Reals^\dmodel)^+ \to (\Reals^\dmodel)^+$ by
\begin{align*}
\nnfunction{\EncLayer}(w) &= \Norm(\widehat{w} + \nnfunction{\NN}(\widehat{w}))\\[1ex]
\text{where~ } \widehat{w} &= \Norm(w + \selfsem{\AttLayer}(w))\,.
\end{align*}
Here, $\Norm: (\Reals^\dmodel)^+ \to (\Reals^\dmodel)^+$
is the length-preserving \emph{layer norm}\footnote{The layer norm usually includes learnable parameters, which we omit in the definition of $\EncLayer$ for simplicity.} and addition is position-wise (thus length preserving, too).
Accordingly, we define $\nndim{\EncLayer} = (n, n)$.

\begin{figure}[t]
\centering
\includegraphics[scale=0.46]{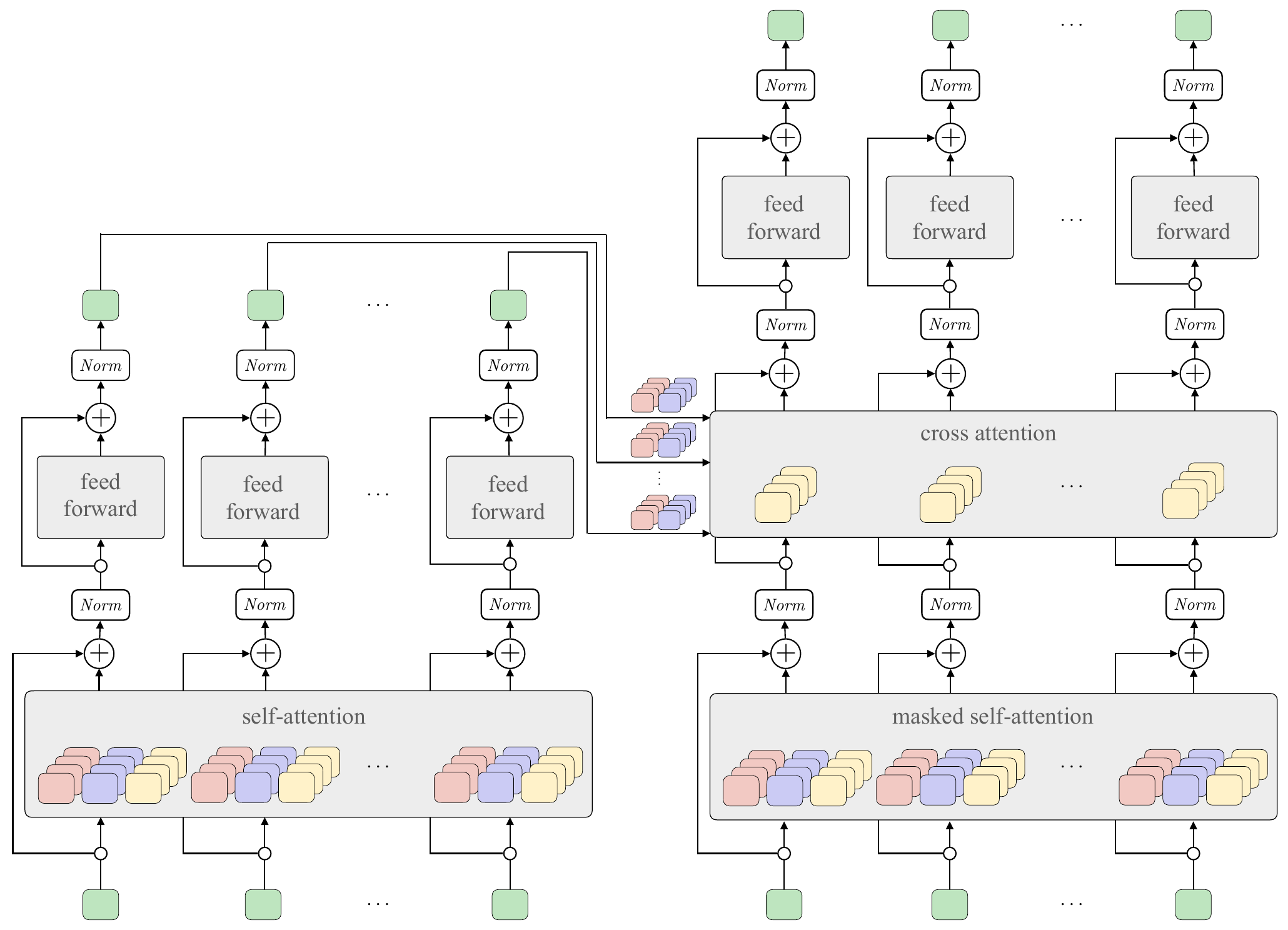}
\caption{The interplay between an encoder layer (left) and a decoder layer (right)
\label{fig:encdec}}
\end{figure}

The structure of an encoder layer is illustrated on the left-hand side of Figure~\ref{fig:encdec}.

\begin{myremark}[label=rem:normalization]{Normalization and Addition}
Optionally, the layer norm(s) may be chosen to be the identity function.
Moreover, addition (also called residual connection) may be omitted.
However, both greatly help in the training process of transformers.
Moreover, in some applications one may consider adopting the masked
self-attention semantics.
That is, alternative semantics for encoder layer $\EncLayer$
can be given as
\begin{align*}
\nnfunction{\EncLayer} &= \nnfunction{\NN} \circ \selfsem{\AttLayer}\\[0.5ex]
\textup{or }\nnfunction{\EncLayer} &= \nnfunction{\NN} \circ \maskedsem{\AttLayer}\,.
\end{align*}
\end{myremark}

\newcommand{\encword}{w_\mathsf{enc}}
\newcommand{\win}{w_\mathsf{in}}
\newcommand{\wout}{w_\mathsf{out}}

\paragraph{Decoder Layer.}

A decoder layer is similar to an encoder layer, but since part of its
input comes from an encoder layer, its semantics is described in a different way.
A \emph{decoder layer} (of dimension $n$) is of the form $\DecLayer = (\AttLayer^{(1)}, \AttLayer^{(2)}, \NN)$.
It features two multi-head attention layers $\AttLayer^{(1)}$ and $\AttLayer^{(2)}$ and, like the encoder layer,
a feed-forward neural network $\NN$ such that $\nndim{\AttLayer^{(1)}} = \nndim{\AttLayer^{(2)}} = \nndim{\NN}=(n, n)$.
Its semantics is a mapping $\nnfunction{\DecLayer}: (\Reals^\dmodel)^+ \times (\Reals^\dmodel)^+ \to (\Reals^\dmodel)^+$
defined by
\begin{align*}
\nnfunction{\DecLayer}(\encword, w) &= \Norm(w_2 + \nnfunction{\NN}(w_2))\\[1ex]
\text{where~ } w_2 &= \Norm(w_1 + \crosssem{\AttLayer^{(2)}}(\encword, w_1))\\[1ex]
w_1 &= \Norm(w + \maskedsem{\AttLayer^{(1)}}(w))\,.
\end{align*}
Thus, $\AttLayer^{(1)}$ is actually a masked multi-head attention layer,
and $\AttLayer^{(2)}$ is a cross multi-head attention layer.
We define $\nndim{\DecLayer} = (n, n)$.
Note that $\nnfunction{\DecLayer}$ is length-preserving in its second argument.
The decoder layer is illustrated on the right-hand side of Figure~\ref{fig:encdec}.

Note that Remark~\ref{rem:normalization} applies here as well, i.e.,
normalization and residual connections are optional. However,
in NLP tasks, applying the masked-self attention semantics in decoder layers
allows one to feed complete input and output sequences during training
while avoiding that the decoder can ``look into the future''.
In fact, its decisions should be based solely on what it has read/produced so far.

\newcommand{\SOS}{\mathsf{SOS}}
\newcommand{\EOS}{\mathsf{EOS}}
\newcommand{\Prob}{\mathbb{P}}
\newcommand{\NNin}{\NN_\mathsf{in}}
\newcommand{\NNout}{\NN_\mathsf{out}}

\paragraph{Transformer.}

Transformers were initially introduced for machine translation.
For that case, we assume ordered finite alphabets $\Sigma$ and $\Gamma$
(of words or tokens). We assume that $\Gamma$ contains a \emph{start-of-sequence}
symbol $\SOS$ and an \emph{end-of-sequence} symbol $\EOS$.
They indicate when the translation of the output sentence
will start and end, respectively.
The semantics of a transformer $\TF$ over $\Sigma$ and $\Gamma$ will define a
(partial and not necessarily length-preserving) mapping
\[
\alphrnnstos{\TF}{\Sigma}{\Gamma}: \Sigma^+ \to \Gamma^\ast\,.
\]
We start with a length-preserving encoding of the input sequence, realized by an \emph{embedding}
$\emb_\Sigma: \Sigma^+ \to (\Rationals^\dmodel)^+$. It is obtained from a \emph{word embedding}\footnote{The one-hot encoding $\onehot{\Sigma}$ is a special case of a word embedding.}
$\wordemb_\Sigma: \Sigma \to \Rationals^\dmodel$
and a \emph{positional encoding} $\posenc: \posNaturals \to \Rationals^\dmodel$
and defined, for $w = \alpha_1 \ldots \alpha_\seqlength \in \Sigma^+$ by
$\emb_\Sigma(w) = \vect{x}^{(1)} \ldots \vect{x}^{(\seqlength)}$ where
$\vect{x}^{(\seqind)} = \wordemb_\Sigma(\alpha_\seqind) + \posenc(\seqind)$.
We include the mappings $\emb_\Sigma$ and $\emb_\Gamma$ in the
definition of a transformer, as they are in principle learnable.

A transformer consists of a stack of encoder layers and a stack of
decoder layers.\footnote{RNNs and LSTMs can also be presented in that way,
especially when we consider text generation. We will present transformers
in their full form, but then focus on encoders and their capability as
language recognizers.}

\begin{mydefinition}{Transformer}
A \emph{(machine-translation) transformer} with hidden-state dimension $n \in \posNaturals$
over $\Sigma$ and $\Gamma$ is a tuple
\[
\TF = (\emb_\Sigma,\emb_\Gamma, (\EncLayer^{(1)},\ldots,\EncLayer^{(\ntflayers)}), (\DecLayer^{(1)},\ldots,\DecLayer^{(\ntflayers)}), \NNout)
\]
such that
\begin{boxitemize}
\item $\emb_\Sigma: \Sigma^+ \to (\Rationals^\dmodel)^+$ and $\emb_\Gamma: \Gamma^+ \to (\Rationals^\dmodel)^+$ are embeddings,
\item $\EncLayer^{(1)},\ldots,\EncLayer^{(\ntflayers)}$ are encoder layers with $\nndim{\EncLayer^{(i)}} = (n,n)$,
\item $\DecLayer^{(1)},\ldots,\DecLayer^{(\ntflayers)}$ are decoder layers with $\nndim{\DecLayer^{(i)}} = (n,n)$,
\item $\NNout$ is a feed-forward neural network with a softmax activation function in its last layer and
$\nndim{\NNout} = (\dmodel, |\Gamma|)$.
\end{boxitemize}
\end{mydefinition}

\newcommand{\lastwin}{\textcolor{blue}{\win^{(\ntflayers)}}}

\begin{figure}[t]
\centering
\includegraphics[scale=0.5]{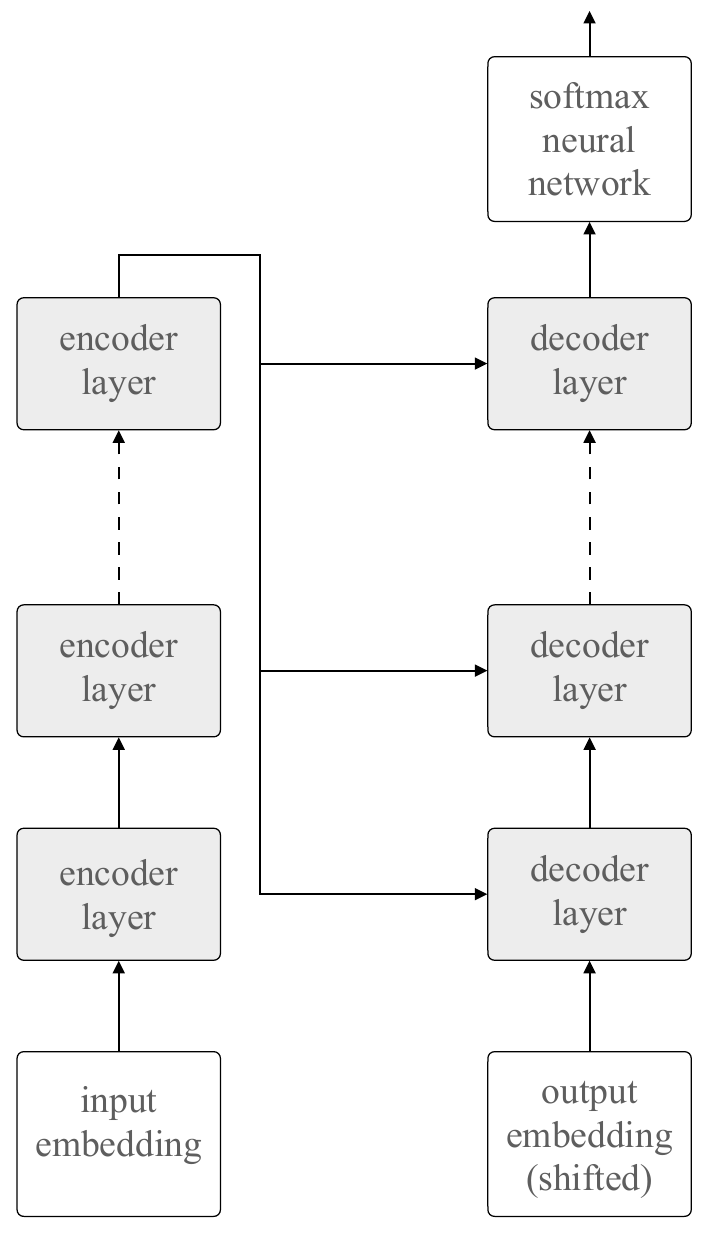}
\caption{The interplay between encoder layers (left) and decoder layers (right)
\label{fig:transformer}}
\end{figure}

Before defining $\alphrnnstos{\TF}{\Sigma}{\Gamma}$, we define an intermediate semantics
$\alphrnnstos{\TF}{\Sigma}{\Gamma}^{\mathsf{next}}: \Sigma^+ \times \Gamma^+ \to \Gamma$,
which, for a given input sequence $\win \in \Sigma^+$ and an
output sequence $\wout \in \Gamma^+$ generated so far, provides the next output letter
$\alphrnnstos{\TF}{\Sigma}{\Gamma}^{\mathsf{next}}(\win,\wout) \in \Gamma$
to be appended to $\wout$.
To determine $\alphrnnstos{\TF}{\Sigma}{\Gamma}^{\mathsf{next}}(\win,\wout)$, we compute
\begin{align*}
\lastwin &= \nnfunction{\EncLayer^{(\ntflayers)}}(\win^{(\ntflayers-1)})
& \wout^{(\ntflayers)} &= \nnfunction{\DecLayer^{(\ntflayers)}}(\lastwin, \wout^{(\ntflayers-1)}) \\[0.5ex]
& ~~\!\vdots & & ~~\!\vdots\\[1ex]
\win^{(2)} &= \nnfunction{\EncLayer^{(2)}}(\win^{(1)})
& \wout^{(2)}&= \nnfunction{\DecLayer^{(2)}}(\lastwin, \wout^{(1)})\\[1ex]
\win^{(1)} &= \nnfunction{\EncLayer^{(1)}}(\win^{(0)})
& \wout^{(1)}&= \nnfunction{\DecLayer^{(1)}}(\lastwin, \wout^{(0)})\\[1ex]
\win^{(0)} &= \emb_\Sigma(\win)
& \wout^{(0)} &= \emb_\Gamma(\wout) 
\end{align*}
This interplay between encoder and decoder layers is illustrated in
Figures~\ref{fig:encdec} and \ref{fig:transformer}.

With this, $\alphrnnstos{\TF}{\Sigma}{\Gamma}^{\mathsf{next}}(\win,\wout)$
is the $\min(\argmax(\NNout(\vect{x})))$-th letter from $\Gamma$ 
where $\vect{x}$ is the last vector in $\wout^{(\ntflayers)}$.
Now, $\SOS$ is a dummy symbol that allows the decoder to produce a first output.
Thus, $\alphrnnstos{\TF}{\Sigma}{\Gamma}^{\mathsf{next}}(\win,\SOS)$ generates the first letter
after ``reading'' the input string $\win$. Continuing this scheme, we let
\begin{align*}
\beta_1 &= \alphrnnstos{\TF}{\Sigma}{\Gamma}^{\mathsf{next}}(\win,\SOS)\\[1ex]
\beta_2 &= \alphrnnstos{\TF}{\Sigma}{\Gamma}^{\mathsf{next}}(\win,\SOS\,\beta_1)\\[1ex]
\beta_3 &= \alphrnnstos{\TF}{\Sigma}{\Gamma}^{\mathsf{next}}(\win,\SOS\,\beta_1\beta_2)\\[0.5ex]
& ~~\vdots
\end{align*}
Consider the smallest index $\ell \ge 0$ such that $\beta_{\ell+1} = \EOS$.
If $\ell$ does not exist, $\alphrnnstos{\TF}{\Sigma}{\Gamma}(\win)$ is undefined.
Otherwise, $\alphrnnstos{\TF}{\Sigma}{\Gamma}(\win) = \beta_1 \ldots \beta_{\ell}$.


%
%

\section{Encoder-Only Transformers}

Based on the general transformer architecture, we now extract
simple architectures, solely based on encoder layers,
that define simpler functions or serve
as language recognizers.

Henceforth, all encoder layers $\EncLayer = (\AttLayer, \NN)$
may or may not use \emph{masked} self-attention.
Also recall that residual connections and the normalization are optional
in every layer.

\begin{mydefinition}{Encoder-Only Transformer}
An \emph{encoder-only transformer} is a tuple
\[
\TF = (\NNin,\EncLayer^{(1)},\ldots,\EncLayer^{(\ntflayers)},\NNout)
\]
where, for some $m,n,o \in \posNaturals$,
\begin{boxitemize}
\item $\NNin$ and $\NNout$ are feed-forward neural networks
with $\nndim{\NNin} = (m,n)$ and $\nndim{\NNout} = (n,o)$, and
\item $\EncLayer^{(1)},\ldots,\EncLayer^{(\ntflayers)}$ are encoder layers with $\nndim{\EncLayer^{(i)}} = (n,n)$.
\end{boxitemize}
\end{mydefinition}

Similarly to RNNs, we let $\indim{\TF} \df m$, $\outdim{\TF} \df o$, $\nndim{\TF} = (m, o)$, and $\statedim{\TF} \df n$.
Analogously, we can now define a \emph{length-preserving} mapping
$\rnnstos{\TF}: (\Reals^m)^+ \to (\Reals^{\rnnoutdim})^+$
as the function composition
\[\rnnstos{\TF} \df \nnfunction{\NNout} \circ \nnfunction{\EncLayer^{(\ntflayers)}} \fcomp \ldots \fcomp \nnfunction{\EncLayer^{(1)}} \circ \nnfunction{\NNin}\]
where, again, $\nnfunction{\NNin}$ and $\nnfunction{\NNout}$ are straightforwardly extended to sequences.
We also define $\rnnstol{\TF}: (\Reals^m)^+ \to \Reals^{\rnnoutdim}$
such that $\rnnstol{\TF}(w)$ returns the last vector in the sequence
$\rnnstos{\TF}(w)$.

Just as for RNNs, we can view an encoder-only transformer as
a sequence classifier. Suppose $\nndim{\TF} = (m, 1)$. Then,
for ${\bowtie} \in \{\ge, >, =\}$ and $\thr \in \Reals$,
we can define the language
\[
\rnnLangthr{}{\TF}{\bowtie}{\thr} = \{w \in (\Reals^m)^+ \mid \rnnstol{\TF}(w) \bowtie \thr\}\,.
\]
Again, we can adjust this definition to cope with languages over a finite alphabet $\Sigma$
coming with a one-hot encoding $\onehot{\Sigma}$.
If $m = |\Sigma|$ and $\nndim{\TF} = (m, 1)$, then we let
\[
\rnnLangthr{\Sigma}{\TF}{\bowtie}{\thr} = \{w \in \Sigma^+ \mid \rnnstol{\TF}(\onehot{\Sigma}(w)) \bowtie \thr\}\,.
\]

\begin{figure}[ht]
\centering
\hspace{2cm}
\includegraphics[scale=0.46]{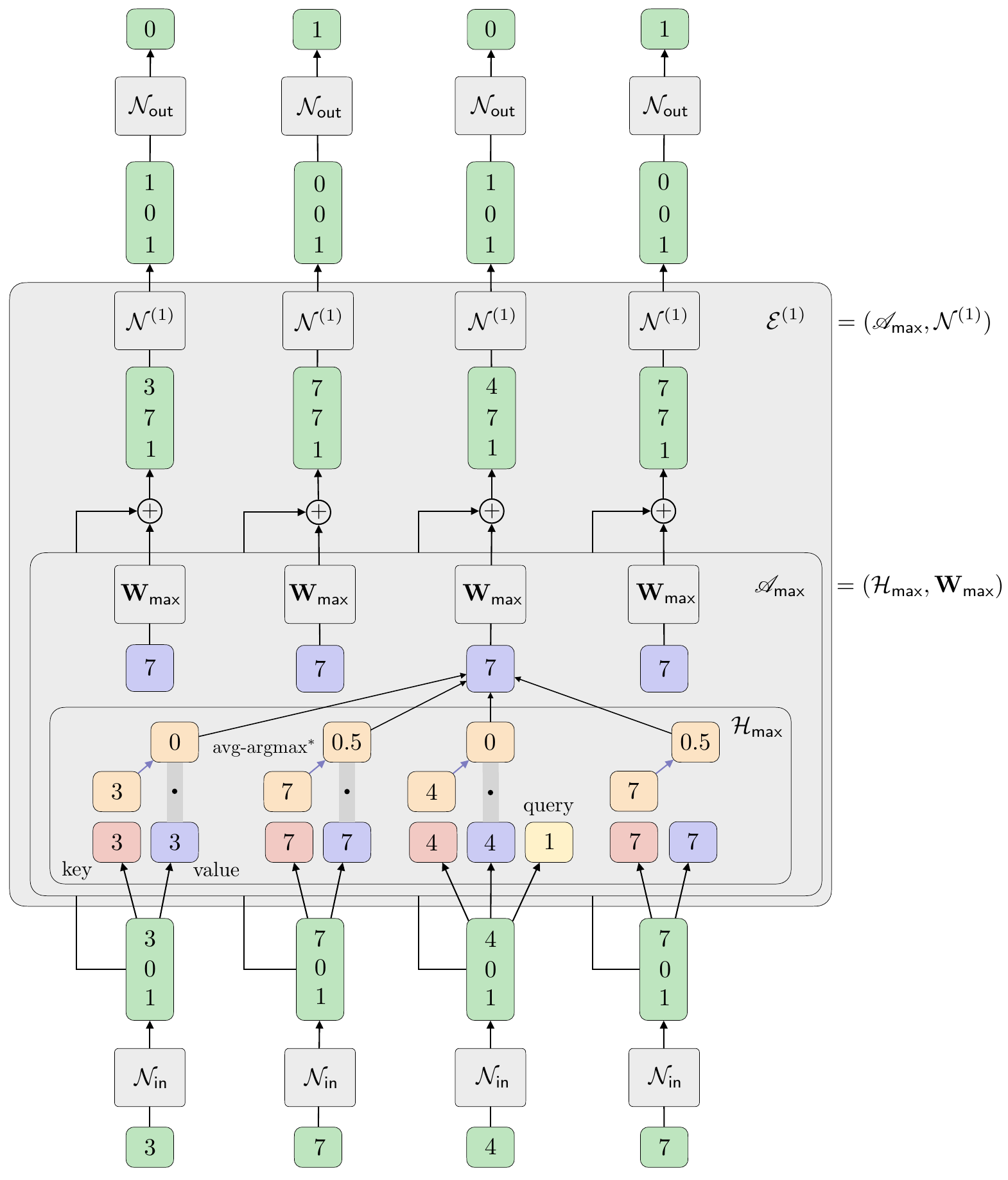}
\caption{Transformer implementing $\argmax^\ast$
\label{fig:tfargmax}}
\end{figure}

\begin{myexample}[label=ex:argmaxtransformer]{Encoder-Only Transformer For $\argmax^\ast$}
We continue Example~\ref{ex:attentionlayer}.
Consider the length-preserving
mapping $\argmax^\ast: \Reals^+ \to \{0,1\}^+$
over arbitrarily long sequences of real numbers.
We will define an encoder-only transformer
$\TF$ with $\nndim{\TF} = (1, 1)$ and $\statedim{\TF} = 3$ such that
$\rnnstos{\TF} = \argmax^\ast$.
It is illustrated in Figure~\ref{fig:tfargmax}.
\\[-1ex]

In a preprocessing step, we apply $\NNin$ such that
$\NNin(x) = (x, 0, 1)^\top$.
Next, we use self-attention in $\AttLayer_{\mathsf{max}}$.
Thus, $\AttHead$ outputs, at every position,
the maximum number occurring in the sequence.
Thanks to the residual connection, the output is added
to the input vectors. It remains to identify
the positions where the first two components are identical
(those are the positions originally carrying the maximum
number). This is taken care of by the neural network
$\NN^{(1)}$, which outputs $0$ iff the first two components
are equal. More precisely, for $\vect{x} = (x_1, x_2, x_3)^\top$,
\[
\nnfunction{\NN^{(1)}}(\vect{x}) =
\begin{cases}
(0, 0, 1)^\top & \text{if } x_1 = x_2\\
(1, 0, 1)^\top & \text{if } x_1 \neq x_2\\
\end{cases}
\]
Here, we leave the third component unchanged, as this will be
useful in the subsequent example.
The neural network $\NN^{(1)}$
makes use of $\heaviside$ as (local) activation function,
defined by
\[
\heaviside(x) =
\begin{cases}
0 & \text{if } x \le 0\\
1 & \text{if } x > 0\,.
\end{cases}
\]
However, we could also use a combination of $\ReLU$ and
$\sigmoid$, adjusting the interpretation of the output
accordingly.
Finally, $\NNout$ inverts the first component.
The specification of $\NNin$, $\NN^{(1)}$, and $\NNout$ is left
to the reader as an exercise.
\end{myexample}

\begin{figure}[t]
\centering
\hspace{2cm}
\includegraphics[scale=0.46]{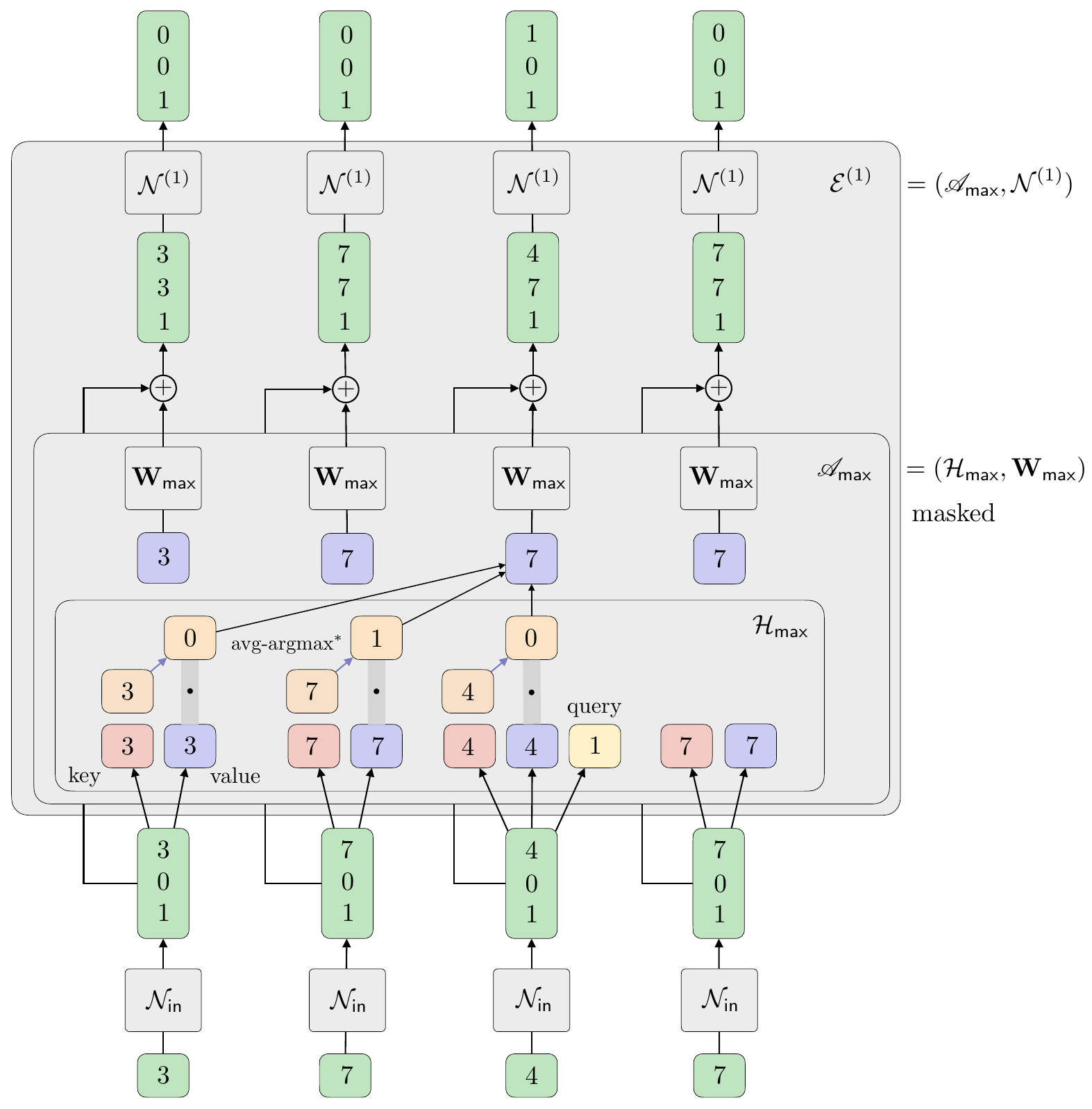}
\caption{Transformer recognizing sorted sequences (lower part)
\label{fig:tflower}}
\end{figure}

\begin{figure}[t]
\centering
\hspace{2cm}
\includegraphics[scale=0.46]{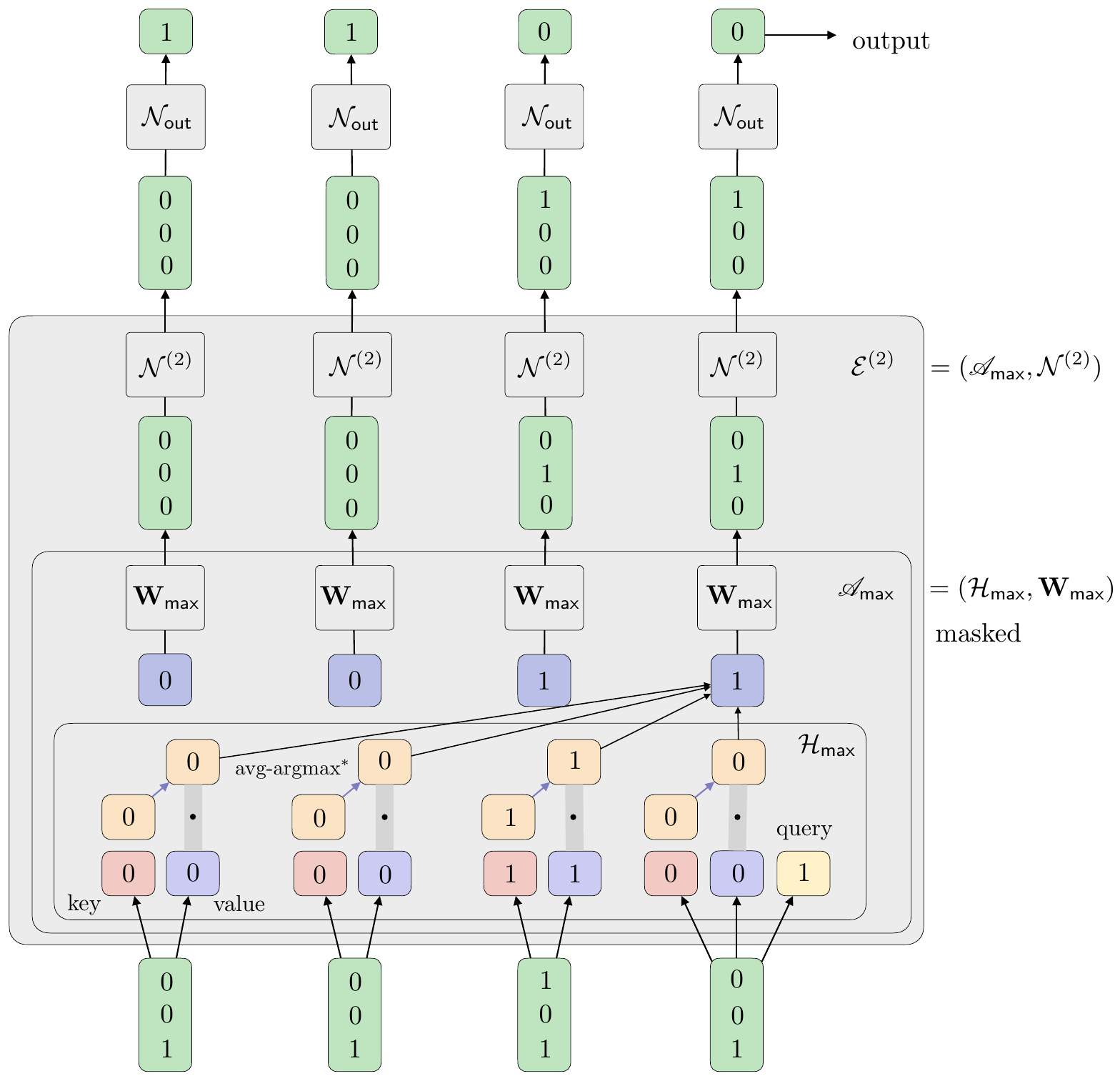}
\caption{Transformer recognizing sorted sequences (upper part)
\label{fig:tfupper}}
\end{figure}

\begin{myexample}[label=ex:tfsorted]{Encoder-Only Transformer Recognizing Sorted Sequences}
Again, we will build on Example~\ref{ex:attentionlayer}.
Our goal now is to model a function
$f: \Reals^+ \to \{0,1\}$ that returns 1 if the input sequence is
sorted, and 0 otherwise.
We will define an encoder-only transformer
$\TF$ with $\nndim{\TF} = (1, 1)$ and $\statedim{\TF} = 3$ such that,
for all $w \in \Reals^+$, the last element in the
sequence $\rnnstos{\TF}(w)$ is $f(w)$.
In other words,
$\rnnLangthr{}{\TF}{=}{1} = \rnnLangthr{}{\TF}{\ge}{0.5}$ is the set of sorted sequences over $\Reals$.
The transformer $\TF = (\NNin,\EncLayer^{(1)},\EncLayer^{(2)},\NNout)$
is illustrated in Figures~\ref{fig:tflower}
and~\ref{fig:tfupper}.\\[-1ex]

We use the very same components as in
Example~\ref{ex:argmaxtransformer}.
However, we now adopt the \emph{masked} self-attention semantics.
In that case, every position $i$ in the sequence can only ``see''
the previous positions (including $i$). Thus, at the last position,
we use $\AttHead_\mathsf{max}$ in the second encoder layer to
detect whether \emph{some} violation of the order has occurred.
Note that the neural network $\NN^{(2)}$ will just invert the
first two components so that we can apply $\NNout$ as in the
previous case.
\\[-1ex]

Note that the encoder layer $\EncLayer^{(2)}$ could also use
the non-masked self-attention semantics,
as we are only interested in the output at the very
last position.
\end{myexample}

\begin{figure}[t]
\centering
\includegraphics[scale=0.46]{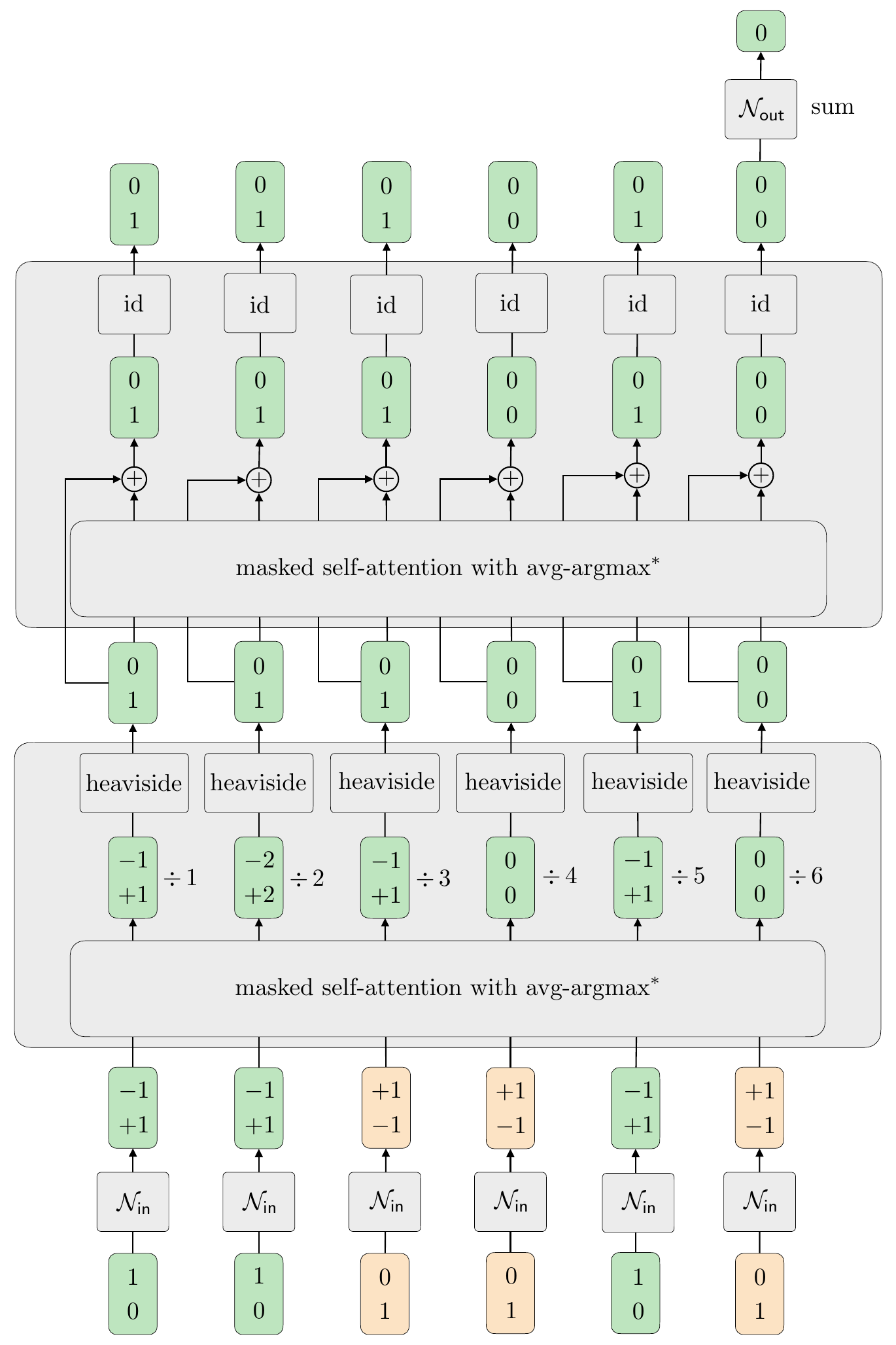}
\caption{Transformer recognizing well-formed bracket strings
\label{fig:dyck}}
\end{figure}

\begin{myexample}{Well-Formed Bracket Strings}
The last example, which is due to \cite{BhattamishraAG20},
is concerned with the language $L$ over the finite
alphabet $\Sigma = \{\langle\,, \rangle\}$ given by the following
grammar:
\[
A ::= \langle\, A\, \rangle ~\mid~ AA ~\mid~ \epsilon
\]
Thus, $L$ is the set of well-formed bracket strings.
Figure~\ref{fig:dyck} sketches an encoder-only transformer $\TF$
with two encoder layers such that $\rnnLangthr{\Sigma}{\TF}{=}{0} = L$.
Here, we assume $\onehot{\Sigma}(\,\langle\,) = (1, 0)^\top$ and
$\onehot{\Sigma}(\,\rangle\,) = (0, 1)^\top$.

The idea is that the first component of a global state checks
whether, in every prefix, there are at least as many opening
as closing brackets.
The second component checks whether, in the entire word, there are
as many closing as opening brackets.

A violation of the former property would result in some value 1
in the first component after the first encoder layer.
If not violated, the original string is valid if the second component
is zero, too. The second encoder layer checks zeroness in both cases.
\end{myexample}

A series of recent papers established a rather complete picture of the
language classes defined by transformer architectures,
both in terms of logic and circuit complexity. Examples include
\cite{abs-2311-00208,MerrillSS22,abs-2310-13897,abs-2310-03817,BhattamishraAG20}.
To the best of our knowledge, only a few works address the verification
of transformers. Exceptions are \cite{ShiZCHH20,LiaoCEK23}, which study
robustness verification. General positive decidability results for
transformers have yet to be explored.
Due to Turing completeness of the general architecture \cite{PerezBM21}, the challenge
lies in identifying architectures that allow for deciding
interesting properties. This represents a crucial area for future research,
particularly in the context of verification. A related question is what a useful
specification could be, for example in the spirit of $\NNSL$.

\bibliography{lit}
\bibliographystyle{plain}


\end{document}